\documentclass[11pt]{article}
\usepackage{eurosym}
\usepackage{amsfonts}
\usepackage{amssymb}
\usepackage{graphicx}
\usepackage{amsmath}
\usepackage{makeidx}
\usepackage{indentfirst}
\usepackage[T1]{fontenc}
\usepackage[utf8]{inputenc}

\newcounter{resultnum}[section]
\newcounter{conclusionnum}[section]
\newcounter{conditionnum}[section]
\newcounter{conjecturenum}[section]
\newcounter{examplenum}[section]
\newcounter{exercisenum}[section]
\newcounter{lemmanum}[section]
\newcounter{notationnum}[section]
\newcounter{theoremnum}[section]
\newcounter{definitionnum}[section]
\newcounter{corollarynum}[section]
\newcounter{remarknum}[section]
\newcounter{propositionnum}[section]
\newcounter{acknowledgementnum}[section]
\newcounter{algorithmnum}[section]
\newcounter{axiomnum}[section]
\newcounter{casenum}[section]
\newcounter{claimnum}[section]
\newcounter{summarynum}[section]
\newcounter{problemnum}[section]
\begin{document}

\title{Nonassociative black ellipsoids distorted by R-fluxes and\\
four dimensional thin locally anisotropic accretion disks}
\date{November 28, 2021}
\author{ \vspace{.1 in} {\textbf{Lauren\c{t}iu Bubuianu}}\thanks{%
email: laurentiu.bubuianu@tvr.ro} \\
{\small \textit{SRTV - Studioul TVR Ia\c{s}i} and \textit{University
Appolonia}, 2 Muzicii street, Ia\c{s}i, 700399, Romania} \vspace{.2 in} \\
\vspace{0.1in} \textbf{Sergiu I. Vacaru} \thanks{%
email: sergiu.vacaru@gmail.com \newline
\textit{Address for post correspondence in 2020-2021 as a visitor senior
researcher at YF CNU Ukraine is:\ } Yu. Gagarin street, 37-3, Chernivtsi,
Ukraine, 58008;\ authors are listed in alphabetic order when equal
contributions are assumed.} \\
{\small \textit{Physics Department, California State University at Fresno,
Fresno, CA 93740, USA; }} \\
{\small \textit{and Dep. Theoretical Physics and Computer Modelling,}} \\
{\small \textit{\ Yu. Fedkovych Chernivtsi National University}, 101
Storozhynetska street, Chernivtsi, 58029, Ukraine} \vspace{.2 in} \\
\vspace{.1 in} {\textbf{El\c{s}en Veli Veliev} } \thanks{%
email: elsen@kocaeli.edu.tr and elsenveli@hotmail.com} \\
{\small \textit{Department of Physics,\ Kocaeli University, 41380, Izmit,
Turkey}} }
\maketitle

\begin{abstract}
We construct nonassociative quasi-stationary solutions describing deformations of Schwarzschild black holes, BHs, to ellipsoid configurations, which can be black ellipsoids, BEs, and/or BHs with ellipsoidal accretion disks. Such solutions are defined by generic off-diagonal symmetric metrics and nonsymmetric components of metrics (which are zero on base four dimensional, 4-d, Lorentz manifold spacetimes but nontrivial in respective 8-d total (co) tangent bundles). Distorted nonassociative BH and BE solutions are found for effective real sources with terms proportional to $\hbar \kappa $ (for respective Planck and string constants). These sources and related effective nontrivial cosmological constants are determined by  nonlinear symmetries and deformations of the Ricci tensor by nonholonomic star products encoding R-flux contributions from string theory. To generate various classes of (non) associative / commutative distorted solutions we generalize and apply the anholonomic frame and connection deformation method for constructing exact and parametric solutions in modified gravity and/or general relativity theories. We study properties of locally anisotropic relativistic, optically thick, could and thin accretion disks around nonassociative distorted BHs, or BEs, when the effects due to the rotation are negligible. Such configurations describe angular anisotropic deformations of axially symmetric astrophysical models when the nonassociative distortions are related to the outer parts of the accretion disks.

\vskip3pt

\textbf{Keywords:}\ nonassociative geometry and gravity;\ R-flux
non-geometric background; nonholonomic star product deformations; symmetric
and nonsymmetric metrics; nonlinear connections; exact and parametric
solutions;\ distorted black holes; black ellipsoids; accretion disks.
\end{abstract}

\tableofcontents

\section{Introduction}
\subsection{Motivations for nonassociative geometry, physics and gravity}
Nonassociative algebras and nonassociative and noncommutative theories have a long and diverse history in mathematics and physics. We cite  \cite{shafer95,baez02,blumenhagen16,
aschieri17,szabo19} for introductions and reviews of results beginning  with the middle of the nineteenth century up to last few years containing various developments and applications in quantum field/gravity and string theories. For instance, the algebra octonions is an example of noncommutative Jordan algebras which provided the first example of appearance of nonassociativity in physics. Papers \cite{jordan32,jordan34} are for early approaches to nonassociative quantum mechanics, and \cite{kurdgelaidze,okubo,
castro1,castro2,mylonas12,mylonas13,kupriyanov15,kupriyanov18,gunaydin} are for further developments and references on noassociative algebras and mathematical particle physics.

In open string theories, nonassociative structures are present due to a non-vanishing background 2-form in the world volume of a D-brane. Such structures exist also for  closed strings, for instance, as consequences of flux compactification with non-trivial three-form and related non-geometric backgrounds. Various related and/or alternative approaches to noncommutative geometric flows, noncommutative and nonassociative gravity and gauge theories, membranes and double field theory, etc. were studied in \cite{bouwknegt04,alvarez06,luest10,blumenhagen10,mylonas12,condeescu13,blumenhagen13,
kupriyanov19,kupriyanov19a} and references therein. In this work, we do not attempt to review all subjects on nonassociative physics and gravity and do not  provide a comprehensive list of references, which can be found in  \cite{blumenhagen16,aschieri17,szabo19,partner01,partner02}.

A few years ago a geometric approach leading to nonassociative gravity, with
star product (in brief, $\star$) deformations determined by R-flux
backgrounds of string theory and modelled on a phase space $\mathcal{M}$
enabled with nonassociative geometric objects, has been provided and
elaborated in \cite{blumenhagen16,aschieri17}. A unique nonassociative Levi-Civita
connection (LC-connection $\nabla ^{\star }),$ which is torsionless and
compatible with respective $\star $--deformed symmetric and nonsymmetric
metric structures, was derived in \cite{aschieri17}. In that work the
nonassociative geometric constructions were performed in a form which is
covariant under the quasi-Hopf algebra \cite{drinf} generated by
infinitesimal diffeomorphisms on twisted nonassociative phase space. The
nonassociative vacuum gravitational equations,
\begin{equation}
Ric^{\star }[\nabla ^{\star }]=0,  \label{navacuum}
\end{equation}%
were introduced for star deformations of the standard Ricci tensor in
general relativity, GR, to a nonassociative $Ric^{\star }[\nabla ^{\star }]$%
, where $\nabla ^{\star }$ is defined and computed on $\mathcal{M}$. The
coefficient formulas were derived for Moyal-Weyl products formulated with
respect to tensor products involving coordinate bases extended on phase space.
Abstract definitions of fundamental geometric objects were provided in \cite%
{blumenhagen16,aschieri17} but without details on arbitrary frame and
coframe (dual frame) decompositions. Here we note that in the mentioned
approach the coefficient formulas for nonassociative geometric and physical objects on phase spaces  depend both on spacetime and momentum like coordinates (the
last ones are with respective multiplications on complex unity $i^{2}=-1$).

One of the original motivations for developing nonassociative geometric
models was to construct a nonassociative theory of gravity on a spacetime $V$
encoding ${\star}$--deformations via certain type R-flux contributions from
string/ brane theory, or other type theories. For physical applications, we
can consider $V$ as a four dimensional, 4-d, spacetime Lorentz manifold
using fundamental geometric objects defined as in GR. We can elaborate also
on various models of modified commutative and noncommutative gravity
theories, MGTs, which can be for 5-d, or 6-d, base spacetimes. The geometry
of a (total) nonassociative phase space $\mathcal{M}$ can be modelled on a
cotangent bundle $T^{\ast }V$ enabled with star product and R-flux
deformations. In \cite{aschieri17}, it was obtained also a very important
result for computing the Ricci tensor of nonassociative gravity in a form
with parametric decomposition on $\hbar ,$ the Planck constant, and $%
\kappa:=\ell _{s}^{3}/6\hbar $. In theories modelled on effective phase
spaces, $\kappa$ can be considered as a string constant to be determined by
experimental data or from some theoretic assumptions. Projecting
nonassociative geometric models from phase space to spacetime, $\mathcal{M}%
\rightarrow V$, with $\nabla ^{\star }\rightarrow \ ^{V}\nabla ,$ we can
compute star parametric deformations of the nonassociative Ricci tensor, $%
Ric^{\star}[\nabla ^{\star }]\rightarrow Ric^{\circ }[\ ^{V}\nabla ]$, when
\begin{equation}
Ric^{\circ }[\ ^{V}\nabla ]=Ric[\ ^{V}\nabla ]+\frac{\kappa }{2}\mathcal{R}%
^{jkm}B_{jkm}[\partial _{n},g_{no},\ ^{V}\nabla ]  \label{form1}
\end{equation}%
describes real star R-flux deformations induced on base spacetime $V$. In
these formulas, $Ric[\ ^{V}\nabla ]$ is the usual Ricci tensor of a
LC-connection $\ ^{V}\nabla $ for a pseudo-Riemannian metric $%
g=\{g_{no}(x^{k})\}$ of signature $(+++-)$ on a spacetime $V;$ coordinate
spacetime indices run values $i,j,k,...=1,2,3,4;$ partial derivatives $%
\partial _{j}=\partial /\partial x^{j}$ define a local coordinate
base/frame. In our notations, the anti-symmetric tensor $\mathcal{R}^{jkm}$
defines R-flux contributions via some nontrivial constant values with
respect to a chosen coordinate basis $\partial _{n}$ (such a tensor may
depend on space and cofiber coordinates for frame transforms on $\mathcal{M}$%
); and the anti-symmetric tensor $B_{jkm}$ is a functional of respective
operators and geometric objects.\footnote{%
We follow the Einstein convention rule with summation on up-low repeating
indices, when left up, or low labels will be used for stating that certain
geometric objects are defined on a some spaces, regions, for certain
conditions etc. Here we note that we elaborate a different system of
notation for coordinates, abstract and coordinate indices, and geometric
objects for nonassociative gravity. In coordinate form, the R-flux
deformation of the Ricci tensor (\ref{form1}) is given by formulas (5.90)
and (1.3) in \cite{aschieri17}. We shall provide in next section some
generalized formulas for nonholonomic frames and a canonical distinguished
connection distorting the LC-connection to certain configurations which
allow to construct exact and parametric solutions} The equations (\ref%
{navacuum}) and (\ref{form1}) present starting points for constructing
models of nonassociative gravity theory which encode the low-energy
effective dynamics of closed strings in so-called non-geometric backgrounds.

\subsection{Geometric  methods for constructing exact and parametric solutions in nonassociative gravity}
Geometric and physical models on (non) associative phase spaces are
formulated for geometric objects depending both on spacetime and phase
coordinates, for instance, with metrics $g_{\alpha \beta }(x^{i},p_{a}),$
where indices $\alpha ,\beta ,...=1,2,...8,$ are used as total phase space
ones; $i,j,k,..=1,2,3,4$ are used for spacetime indices and $%
a,b,c,...=5,6,7,8$ are cofiber ones. The nonassociative gravity models
elaborated in \cite{blumenhagen16,aschieri17} may be considered as certain
nonassociative generalizations and modifications (in some special cases with
LC-connections and coordinate frames) of (non) commutative
Finsler-Lagrange-Hamilton and higher dimension (super) string gravity
theories, various MGTs, studied in a series of our works
\cite{vacaru96b,vacaru03,vacaru09a}, see recent reviews and methods of constructing exact solutions in phase spaces in \cite{bubuianu18a,bubuianu19}.

In references \cite{blumenhagen16,aschieri17}, there were emphasized three tasks for further developments of nonassociative gravity which we approach in our research program (in this and partner works \cite{partner01,partner02}, and future ones):
 \newline  1) construct exact/ parametric solutions of nonassociative vacuum gravitational equations (\ref{navacuum});
\newline 2) study further generalizations for nontrivial matter sources; and
\newline 3)  elaborate on possible applications of such theories in modern particle physics, cosmology and astrophysics.
\newline
 The main technical difficulty in elaborating on above directions of research  is that all types of modified (non) associative/ commutative gravitational and geometric flow
theories are described by sophisticated nonlinear systems of partial
differential equations, PDEs, with coupling and mixing of indices of geometric
objects. To find exact and parametric solutions we have to use certain
coordinate and frame decompositions. Such (non) associative/ commutative
tensor and product coupling and/or mixing are consequences of various
conditions and constraints imposed on coefficients. For example, the star products
are defined via contractions with R-flux coefficients $\mathcal{R}^{jkm}$
and the metric compatibility and zero torsion conditions are
introduced for the LC-connection. In GR, a decoupling of Einstein equations
is possible, for instance, for diagonal metric ansatz depending on one space
variable (radial type). As a result, the vacuum and nonvacuum
gravitational equations stated as systems of nonlinear PDEs transform into
respective systems of decoupled ordinary differential equations, ODEs. For such assumptions, it is possible to construct black hole, BH, solutions; when there are considered additional spherical symmetry and asymptotic conditions resulting in the Newton gravitational potential of a point mass particle. This method can be generalized in certain forms for generating, for instance, BHs with rotation and nontrivial sources (we can consider a nontrivial cosmological constant); for wormholes; with nontrivial algebraic
structures, in cosmological models, when metrics depend on a time like
variable etc. The bulk of physically important exact and parametric
solutions described, for instance, in standard monographs \cite%
{misner,hawking73,wald82,kramer03}, are constructed for certain very special type ansatz (usually diagonalizable via coordinate transforms) transforming PDEs into ODEs with decoupling of tensor coefficients. There were studied also certain cases with solitonic gravitational metrics for various classes of nonlinear waves when PDEs are solved using other geometric and analytic techniques. Such methods can not be applied if our aim is to find exact solutions with generic off-diagonal metrics\footnote{which can not be diagonalized by coordinate transforms in a finite spacetime/ phase space region} of type $g_{kj}(x^{i})$ and/or $g_{\alpha \beta }(x^{i},p_{a})$ depending on some spacetime/phase space coordinates, generalized connections and/or LC-connections, even we project all geometric objects and equations on a GR background.

In a series of works \cite{bubuianu18a,bubuianu19,bubuianu17, bubuianu19a}, see
also references therein, we developed the so-called anholonomic frame and
connection deformation method, AFCDM (in previous works, we wrote AFDM) to a
level which allows us to prove some general decoupling and integrability
properties of modified gravitational field and geometric flow equations on
phase spaces. For nonassociative vacuum gravity models on star deformed
$V \rightarrow \mathcal{M}$, and with effective sources determined by R-fluxes of
type (\ref{form1}), we elaborated nonholonomic versions of nonassociative
geometry and gravity in two partner works \cite{partner01,partner02}. The
main idea of the AFCDM is to redefine the geometric constructions on $%
\mathcal{M}$ and $V$ in certain nonholonomic bases\footnote{%
i.e. non-integrable, equivalently anholonomic / nonholonomic; for details and definitions
see above cited works and next section}, when using  an auxiliary linear
connection $\widehat{\mathbf{D}}^{\star }=\nabla ^{\star }+\widehat{\mathbf{Z%
}}^{\star }$ we can prove certain general decoupling properties of physically important systems of PDEs. This way, we can generate certain very general classes of exact/parametric solutions determined by generic off-diagonal symmetric and nonsymmetric metrics, respectively, $\ _{\star}^{\shortparallel }\mathfrak{\check{g}}$ and $\ _{\star }^{\shortparallel }\mathfrak{a}$. Such metrics depend, in principle, on all phase space coordinates, with coefficients parameterized by off-diagonal matrices $g_{\alpha \beta }(x^{i},p_{a})$. In explicit form, the coordinate and parametric dependencies are determined by respective classes of generating functions and generating (effective) sources. For vacuum nonassociative gravitational equations, a general decoupling  property was proven in \cite{partner02} for a large class of quasi-stationary solutions with Killing symmetry on $\partial _{4}=\partial _{t}$ on the first two shells for nonholonomic dyadic decompositions of $\mathcal{M}$. The results of that paper will be used for further developments and applications of nonassociative geometric methods in modern gravity, cosmology and astrophysics, non-standard particle theories, etc., by constructing nonassociative locally anisotropic BH and cosmological solutions generalizing the constructions reviewed in \cite{bubuianu19}.

The nonassociative vacuum phase equations (\ref{navacuum}) and related
spacetime equations (\ref{form1}) can be considered as some 4-d curved
spacetime Einstein equations with an effective real source proportional to $%
\hbar \kappa $ encoding star deformations and R-fluxes from a generalized
co-fiber dynamics projected on $V.$ Re-defining such systems of nonlinear
PDEs in nonholonomic variables, we can construct generic off-diagonal
quasi-stationary solutions (when the coefficients of metrics and connections
depend only on space coordinates in certain adapted systems of reference).
There is a rigorous proof in \cite{partner02} that nonassociative vacuum
Einstein equations (\ref{navacuum}) can be decoupled in general form for
quasi-stationary phase spaces and a nontrivial cosmological constant. Using
such a decoupling property and respective classes of nonlinear symmetries,
we can encode the cofiber dynamics into real effective R-flux sources as in (%
\ref{form1}) and/or cosmological constants. As a result, we work
self-consistently with 4-d Einstein equations for nonassociative distortions
of off-diagonal components of metrics. Exact and parametric solutions
(locally anisotropic quasi-stationary and/or cosmological ones) can be
constructed applying the AFCDM method and results of \cite{bubuianu19} redefined in this work for real sources with R-fluxes. Using off-diagonal real nonholonomic vacuum configurations, we can compute also as parametric induced values respective nonsymmetric components of star-deformed metrics and complex parametric sources. This formalism can
be elaborated independently for the first two nonholonomic dyadic shells on
a 4-d spacetime. In general, there are 4 dyadic oriented shells on a 8-d
phase space and such methods are elaborated in \cite{partner02}. Those
formulas will be used in our further partner works in order to construct,
for instance, nonassociative BH and solitonic solutions on 8-d phase spaces
generalizing some respective classes of solutions from \cite%
{bubuianu17,bubuianu19a}.

\subsection{Main purposes of the paper}
In this work, we concentrate on a particular class of black hole, BH, and
black ellipsoid, BE, configurations with thin accretion disks determined by
nonassociative distorted solutions in 4-d modified gravity encoding
effective R-flux sources as in (\ref{form1}). The first goal is to construct
such generic off-diagonal parametric solutions in explicit form and consider
the conditions when quasi-stationary metrics describe BEs (which are
different from the Kerr metrics) and may transform into Schwarzschild like
BH configurations. It should be noted that the spacetime in the vicinity of
respective horizons with zero effective cosmological constants may remain
vacuum if we relax the condition of asymptotically flatness for
quasi-stationary R-flux distorted metrics. We proved that BEs can be stable
in various variants of modified (non) commutative gravity and geometric flow
theories (including associative gravity models with nonsymmetric metrics,
string gravity etc.) and in GR with spacetime and/or matter distortions
\cite{vacaru03be2,vacaru03,vacaru16b}. Those constructions were for nonholonomic generalizations of Chandrasekhar's equilibrium conditions and formulas on stability of BH solutions \cite{chandr02}. We can extend the geometric constructions and methods for solutions constructed in this paper because we use the same AFCDM but for different types of effective sources which can be stabilized by corresponding classes of
nonholonomic constraints.

The second goal of this paper is to study in brief the properties of
relativistic thin disks around nonassociative distorted Schwarzschild BHe
and BEs up to the quadrupole approximations. For associative and commutative
models, such constructions are reviewed in \cite{abram13,faraji20}. The
effective R-flux source can be treated similarly to an external mass type
distribution resulting in distortions with contributions of the outer parts
of the disk. Self-gravity and star product deformation effects play also
important roles. Using the nonassociative distorted geometry, we can also
study the inner part of locally anisotropic thin accretion disks. In this
paper, the anisotropy will be stated by dependencies on an angular type
space coordinate $x^{3}=\phi .$ For respective classes of generating
functions, we can describe stable ellipsoidal orbits encoding contributions
from effective R-flux sources.

The plan of the article work is as follows:\ In section \ref{sec2}, we outline
necessary results on nonassociative geometry and effective R-sources for
vacuum gravitational equations with nonholonomic 2+2 decoupling. Section \ref%
{sec3} is devoted to explicit constructions of quasi-stationary 4-d generic
off-diagonal and BH and BE solutions encoding nonassociative star
deformation and R-flux real effective sources. We study possible physical
effects of locally anisotropic thin accretion discs around nonassociative
BEs and BHs determined by nonassociative distortions in section \ref{sec4}.
Conclusions are provided in section \ref{sec5}.

\section{Nonholonomic 2+2 spacetime splitting \& nonassociative vacuum
Einstein equations}

\label{sec2} The geometric preliminaries on nonassociative star R-flux deformations presented in this section refer to a four dimensional, 4-d, nonholonomic spacetime projection of the nonholonomyc dyadic phase space geometry and vacuum gravity formulated for 8-d (co) tangent Lorentz bundles in \cite{partner02}. Such associative and commutative geometric methods for 6-10 dimensions are elaborated in details and reviewed in
\cite{vacaru16b,bubuianu19,bubuianu18a}. For a phase space $\mathcal{M}$ enabled with nonholonomic dyadic shell $s=1,2,3,4$ structure corresponding to a conventional nonholonomic (2+2)+(2+2) splitting (in brief, called a s-decomposition), it is possible to prove a general decoupling property of nonassociative vacuum gravitational equations
formulated for a canonical distinguished connection
$\widehat{\mathbf{D}}^{\star }=\nabla ^{\star }+\widehat{\mathbf{Z}}^{\star }$. Such a 
nonassociative linear distinguished connection, d--connection, can be
nonholonomically constrained for zero distortions, $\widehat{\mathbf{Z}}%
^{\star }=0,$ to $\nabla ^{\star }$. This can be used for generating solutions of the
nonassociative Einstein equations for a Levi Civita, LC, connection with
nontrivial cosmological constants and effective R-flux sources generalizing
the methods and solutions presented in \cite{bubuianu19}. On 4-d nonholonomic spacetimes $V$ with a corresponding nonholonomic (2+2)-splitting, the nonassociative vacuum equations (\ref{navacuum}) with real R-flux induced sources (\ref{form1}) can be solved using star
deformations of constructions from \cite{bubuianu17,bubuianu19a}, considered for shells $s=1,2.$

\subsection{Nonassociative nonholonomic star products from real R-flux
sources, metrics and connections}

In the partner works \cite{partner01,partner02}, the nonassociative vacuum
Einstein equations from \cite{blumenhagen16,aschieri17} were re-defined and generalized in
canonical nonholonomic variables on a phase space modeled as a cotangent
Lorentz bundle $\mathcal{M}=T_{\shortparallel }^{\ast }\mathbf{V.}$ Such
 variables are enabled with nonlinear connection, N-connection
structure $\ ^{\shortparallel }\mathbf{N}$ and a N-adapted star product
defined with respect to general frame structures. That allowed to develop
the AFCDM and  prove a general decoupling property and construct
exact and parametric solutions in nonassociative gravity when the geometric
constructions are adapted to a s-connection structure $\
_{s}^{\shortparallel }\mathbf{N}$ and a nonholonomic s-decomposed
nonassociative phase space $\mathcal{M}=T_{s\shortparallel }^{\ast }\mathbf{%
V.}$\footnote{\label{loccord} In this series of works on nonassociative
nonholonomic geometry and gravity, we follow such conventions and notations
for local real (spacetime and total phase space) and complex (co) fiber
coordinates:%
\begin{eqnarray*}
&\mbox{on}& \mathbf{V} \mbox{ and }\ _{s}\mathbf{V}:x=\{x^{i}\}=~_{s}x=%
\{x^{i_{s}}\}=(x^{i_{1}},x^{a_{2}}\rightarrow y^{a_{2}})=(x^{i_{2}}),%
\mbox{
with }x^{4}=t, \\
&&\mbox{ where }i,j,...=1,2,3,4;\mbox{ shells}:\ s=1:\mbox{ when }%
i_{1},j_{1},...=1,2;s=2,a_{2},b_{2},...=3,4; \\
&\mbox{on }&T\mathbf{V} \mbox{ and }T_{s}\mathbf{\mathbf{V}}:\
u=(x,y)=\{u^{\alpha }=(u^{k}=x^{k},\ u^{a}=y^{a})\}= \\
&&\ _{s}u =(~_{s}x,~_{s}y)=\{u^{\alpha _{s}}=(u^{k_{s}}=x^{k_{s}},\
u^{a_{s}}=y^{a_{s}})\}=(x^{i_{1}},x^{i_{2}},x^{a_{3}}\rightarrow
y^{a_{3}},x^{a_{4}}\rightarrow y^{a_{4}}),\mbox{ shells } \\
&&s=1,2,3,4,\mbox{ where }\alpha ,\beta
,...=1,2,...8;a,b,...=5,6,7,8;a_{3},b_{3},...=(5,6);\ a_{4},b_{4},..=(7,8);
\\
&\mbox{on }&T^{\ast }\mathbf{V} \mbox{ and } T_{s}^{\ast } \mathbf{V}: \ \
^{\shortmid }u=(x,\ ^{\shortmid }p)=\{\ u^{\alpha }=(u^{k}= x^{k},\ \
^{\shortmid }p_{a}=p_{a})\}  \notag \\
&&=(\ _{3}^{\shortmid }x,\ _{4}^{\shortmid }p)=\{\ ^{\shortmid }u^{\alpha
}=(\ ^{\shortmid }u^{k_{3}}=\ ^{\shortmid }x^{k_{3}},\ \
^{\shortmid}p_{a_{4}}=p_{a_{4}})\}=  \notag \\
&&\ _{s}^{\shortmid }u=( \ _{s}x,\ _{s}^{\shortmid }p) =\{\ ^{\shortmid
}u^{\alpha _{s}}=(x^{k_{s}}, \ ^{\shortmid
}p_{a_{s}}=p_{a_{s}})\}=(x^{i_{1}},x^{i_{2}}, \
^{\shortmid}p_{a_{3}}=p_{a_{3}},\ \ ^{\shortmid }p_{a_{4}}=p_{a_{4}}),
\notag \\
&&=(\ _{3}^{\shortmid }u~=\ _{3}^{\shortmid }x,\ _{4}^{\shortmid }p)= \{\
^{\shortmid }u^{\alpha _{3}}=(x^{i_{1}},x^{i_{2}},\ \ ^{\shortmid
}x^{i_{3}}\rightarrow ~\ ^{\shortmid }p_{a_{3}}),\ \ ^{\shortmid
}p_{a_{4}}\}, \mbox{ where }\ \ ^{\shortmid }x^{\alpha
_{3}}=(x^{i_{1}},x^{a_{2}},\ \ ^{\shortmid }p_{a_{3}}=p_{a_{3}}).  \notag \\
&\mbox{on}& T_{\shortparallel }^{\ast }\mathbf{V} \mbox{ and }
T_{\shortparallel s}^{\ast }\mathbf{V}: \ ^{\shortparallel }u=(x,\
^{\shortparallel }p)=\{\ ^{\shortparallel }u^{\alpha }=(u^{k}=x^{k},\
^{\shortparallel }p_{a}=(i\hbar )^{-1}p_{a})\}  \notag \\
&&= (\ _{3}^{\shortparallel }x,\ _{4}^{\shortparallel }p)=\{\
^{\shortparallel }u^{\alpha }=(^{\shortparallel }u^{k_{3}}=\
^{\shortparallel }x^{k_{3}},\ ^{\shortparallel }p_{a_{4}}=(i\hbar
)^{-1}p_{a_{4}})\}=  \notag \\
&& \ _{s}^{\shortparallel }u=(~_{s}x,\ _{s}^{\shortparallel }p) = \{\
^{\shortparallel }u^{\alpha _{s}}=(x^{k_{s}},\ \
^{\shortparallel}p_{a_{s}}=(i\hbar)^{-1}p_{a_{s}})\}=(x^{i_{1}},x^{i_{2}}, \
^{\shortparallel }p_{a_{3}}=(i\hbar )^{-1}p_{a_{3}},\ ^{\shortparallel
}p_{a_{4}}=(i\hbar )^{-1}p_{a_{4}}),  \notag \\
&&=(\ _{3}^{\shortparallel }u~=\ _{3}^{\shortparallel }x,\
_{4}^{\shortparallel }p)=\{\ ^{\shortparallel }u^{\alpha
_{3}}=(x^{i_{1}},x^{i_{2}},\ ^{\shortparallel }x^{i_{3}}\rightarrow
~^{\shortparallel }p_{a_{3}}),\ ^{\shortparallel }p_{a_{4}}\}, \mbox{ where }%
\ ^{\shortparallel }x^{\alpha _{3}}=(x^{i_{1}},x^{i_{2}},\ \
^{\shortparallel }p_{a_{3}}=(i\hbar )^{-1}p_{a_{3}}).  \notag
\end{eqnarray*}%
Boldface indices are used for spaces and geometric objects enabled with
(adapted to) N-connection structure. Our notations are different from the
spacetime coordinates with momentum like variables considered in \cite%
{blumenhagen16,aschieri17} and we consider a N- and/or s-adapting of
formulas for considering quasi-Hopf structures. An up (or low, on
convenience), label "$\ ^{\shortparallel }$" is used in our papers for
distinguishing coordinates with "complexified momenta" from real phase
coordinates $\ ^{\shortmid }u^{\alpha }=(x^{k},p_{a})$ on $T^{\ast }\mathbf{%
V,}$ see similar conventions in Finsler-Lagrange-Hamilton geometry \cite{bubuianu18a}. The formalism of N- and s-adapted labels and respective abstract or frame coefficient notations is elaborated in such a way that allows to use unified "symbolic" nonholonomic geometric calculus and many proofs by analogy.}

\subsubsection{Nonlinear connections with dyadic splitting and
nonassociative star product}

We outline for nonholonomic spacetime shells $s=1$ and 2 the definitions and
necessary formulas from section 2.1 of \cite{partner02}.

\paragraph{Nonholonomic dyadic decomposition (s-decomposition): \newline
}

We can define in global form (2+2)+(2+2) nonholonomic splitting of phase
spaces considering respective classes of N--connection structures
(s-connections, with $s=1,2$ when $s=3,4$ contribute indirectly via certain
effective sources):
\begin{eqnarray}
\ _{s}^{\shortmid }\mathbf{N}:\ \ _{s}T\mathbf{T}^{\ast }\mathbf{V} &=&\
^{1}hT^{\ast }V\oplus \ ^{2}vT^{\ast }V\oplus \ ^{3}cT^{\ast }V\oplus \
^{4}cT^{\ast }V,  \notag \\
\mbox{ when }\ _{2}^{\shortmid }\mathbf{N} &:&\ \ _{2}T\mathbf{T}^{\ast }%
\mathbf{V}=\ ^{1}hT^{\ast }V\oplus \ ^{2}vT^{\ast }V,\mbox{ and }  \notag \\
\ _{s}^{\shortparallel }\mathbf{N}:\ \ _{s}T\mathbf{T}_{\shortparallel
}^{\ast }\mathbf{V} &=&\ ^{1}hT_{\shortparallel }^{\ast }V\oplus \
^{2}vT_{\shortparallel }^{\ast }V\oplus \ ^{3}cT_{\shortparallel }^{\ast
}V\oplus \ ^{4}cT_{\shortparallel }^{\ast }V,\mbox{ when }  \label{ncon2} \\
\mbox{ when }\ _{2}^{\shortparallel }\mathbf{N} &:&\ \ _{2}T\mathbf{T}%
_{\shortparallel }^{\ast }\mathbf{V}=\ ^{1}hT_{\shortparallel }^{\ast
}V\oplus \ ^{2}vT_{\shortparallel }^{\ast }V.  \notag
\end{eqnarray}%
Such dyadic decompositions into conventional 2-dim nonholonomic
distributions of $TTV,$ $TT^{\ast }V$ and $TT_{\shortparallel }^{\ast }V$
involve dimensions $\dim (\ ^{1}hT^{\ast }V)=\dim (\ ^{2}vT^{\ast }V)=\dim
(\ ^{3}cT^{\ast }V)=(\ ^{4}cT^{\ast }V)=2$ and $\dim (\
^{1}hT_{\shortparallel }^{\ast }V)=\dim (\ ^{2}vT_{\shortparallel }^{\ast
}V)=\dim (\ ^{3}cT_{\shortparallel }^{\ast }V)=(\ ^{4}cT_{\shortparallel
}^{\ast }V)=2$, where, for instance, left up labels like $\ ^{1}h,\ ^{3}c$
etc. state that using nonholonomic (equivalently, anholonomic and/or
non-integrable) distributions we split respective 8-d total spaces into
oriented 2-d shells with numbers 1,2,3 and 4.

On a Lorentz spacetime manifold $\mathbf{V},$ nonholonomic dyadic splitting
with N-connections (\ref{ncon2}) are defined locally by coefficients $\
_{2}^{\shortparallel }\mathbf{N}=\{N_{i_{1}}^{i_{2}}(x^{i_{1}},x^{a_{2}})\}$
used for constructing N-elongated bases (N-/ s-adapted bases),
\begin{eqnarray}
\ ^{\shortparallel }\mathbf{e}_{\alpha _{s}} &=&(\ \ ^{\shortparallel }%
\mathbf{e}_{i_{s}}=\ \frac{\partial }{\partial x^{i_{s}}}-\ ^{\shortparallel
}N_{\ i_{s}a_{s}}\frac{\partial }{\partial \ ^{\shortparallel }p_{a_{s}}},\
\ ^{\shortparallel }e^{b_{s}}=\frac{\partial }{\partial \ ^{\shortparallel
}p_{b_{s}}})\mbox{ on }\ _{s}T\mathbf{T}_{\shortparallel }^{\ast }\mathbf{V,}%
\mbox{ for }s=1,2,3,4;  \notag \\
&& \mbox{ when for shells } s =1,2:\ ^{\shortparallel }\mathbf{e}_{\alpha
_{1}}=(\ \ \ ^{\shortparallel }e_{i_{1}}=\frac{\partial }{\partial x^{i_{1}}}%
=\partial _{i_{1}}),\mbox{
for }i_{1}=1,2;  \label{nadapbdsc} \\
\ ^{\shortparallel }\mathbf{e}_{\alpha _{2}} &=&(\ ^{\shortparallel }\mathbf{%
e}_{i_{1}}=\frac{\partial }{\partial x^{i_{1}}}-N_{i_{1}}^{a_{2}}\frac{%
\partial }{\partial x^{a_{2}}},\ ^{\shortparallel }e_{b_{2}}=\frac{\partial
}{\partial x^{b_{2}}})=(\ ^{\shortparallel }\mathbf{e}_{i_{1}}=\partial
_{i_{1}}-N_{i_{1}}^{a_{2}}\partial _{a_{2}},\ \ \ ^{\shortparallel
}e_{b_{2}}=\partial _{b_{2}}),\mbox{ for }b_{2}=3,4,  \notag
\end{eqnarray}%
where we follow the conventions for dyadic indices and coordinates stated in
footnote \ref{loccord}.

For dual shell s-adapted bases to (\ref{nadapbdsc}), s-cobases, we have%
\begin{eqnarray}
\ ^{\shortparallel }\mathbf{e}^{\alpha _{s}} &=&(\ ^{\shortparallel }\mathbf{%
e}^{i_{s}}=dx^{i_{s}},\ ^{\shortparallel }\mathbf{e}_{a_{s}}=d\
^{\shortparallel }p_{a_{s}}+\ ^{\shortparallel }N_{\ i_{s}a_{s}}dx^{i_{s}})%
\mbox{ on }\ \ _{s}T^{\ast }\mathbf{T}_{\shortparallel }^{\ast }\mathbf{V,}
\label{nadapbdss} \\
\mbox{ when on shells }s &=&1,2:~\ ^{\shortparallel }\mathbf{e}^{\alpha
_{1}}=(\mathbf{e}^{i_{1}}=dx^{i_{1}});\ ^{\shortparallel }\mathbf{e}^{\alpha
_{2}}=(\ \mathbf{e}^{i_{1}}=dx^{i_{1}},\ ^{\shortparallel }\mathbf{e}%
^{a_{2}}=d\ x^{a_{2}}+\ ^{\shortparallel }N_{\ i_{2}}^{a_{2}}dx^{i_{2}}),
\notag
\end{eqnarray}%
defined by the same N-connection coefficients and respective s-decomposition
which can be considered on spacetime and extended on phase spaces.

\subsubsection{Nonassociative star product and dyadic s-structures}

In our partner works \cite{partner01,partner02}, we generalized for
nonholonomic phase spaces endowed with N- /s--connection structure the
definition of nonassociative star product. For coordinate bases, such
formulas transform respectively into those presented is section 2 of \cite%
{blumenhagen16} and section 2 of \cite{aschieri17}. Here we provide
necessary definitions for shells $s=1,2$.

Considering a full phase space $\mathcal{M}$ containing a spacetime
direction $\ ^{\shortmid }\mathbf{e}_{i_{s}}$ and a momentum like cofiber
direction $\ ^{\shortmid }e^{b_{s}},$ see formulas (\ref{nadapbdsc}), for
any two functions $\ z(x,p)$ and $\ q(x,p),$ we define a nonholonomic
s-adapted star product $\star _{s}$:
\begin{eqnarray}
z\star _{s}q&:=& \cdot \lbrack \mathcal{F}_{s}^{-1}(z,q)]  \label{starpn} \\
&=&\cdot \lbrack \exp \left( -\frac{1}{2}i\hbar (\ ^{\shortmid }\mathbf{e}%
_{i_{s}}\otimes \ ^{\shortmid }e^{i_{s}}-\ ^{\shortmid }e^{i_{s}}\otimes \
^{\shortmid }\mathbf{e}_{i_{s}}) +\frac{i\mathit{\ell }_{s}^{4}}{%
12\hbar }R^{i_{s}j_{s}a_{s}}(p_{a_{s}}\ ^{\shortmid }\mathbf{e}%
_{i_{s}}\otimes \ ^{\shortmid }\mathbf{e}_{j_{a}}-\ ^{\shortmid }\mathbf{e}%
_{j_{s}}\otimes p_{a_{s}}\ ^{\shortmid }\mathbf{e}_{i_{s}})\right)]z\otimes q
\notag \\
&=&z\cdot q-\frac{i}{2}\hbar \lbrack (\ ^{\shortmid }\mathbf{e}_{i_{s}}z)(\
^{\shortmid }e^{i_{s}}q)-(\ ^{\shortmid }e^{i_{s}}z)(\ ^{\shortmid }\mathbf{e%
}_{i_{s}}q)]+\frac{i\mathit{\ell }_{s}^{4}}{6\hbar }%
R^{i_{s}j_{s}a_{s}}p_{a_{s}}(\ ^{\shortmid }\mathbf{e}_{i_{s}}z)(\
^{\shortmid }\mathbf{e}_{j_{s}}q)+\ldots .  \notag
\end{eqnarray}%
In these formulas, the constant $\mathit{\ell }$ defines the R-flux
contributions for a antisymmetric $R^{i_{s}j_{s}a_{s}}$ background in string
theory, with s-indices. We can restrict the definitions for any $s=1$ and $%
s=2,$ when the R-flux contracts with momentum like coordinates in an
un-stated explicit form if you consider only the base spacetime $V.$ In
these formulas, the tensor product $\otimes $ is used in a s-adapted form
indicating on which factor of $z\otimes q$ the s-adapted derivatives act
with the dot form (for many computations with small parametric
decompositions on $\hbar $ and $\kappa =\mathit{\ell }_{s}^{3}/6\hbar ,$ the
tensor products turn into usual multiplications).

\subsubsection{Symmetric and nonsymmetric spacetime s-metric structures and
star deformations}

For total phase space models, a (pseudo) Riemannian symmetric metric on
cotangent Lorentz bundle $T^{\ast }V$ is defined by a tensor $\
^{\shortparallel }g=\{\ ^{\shortparallel }g_{\alpha \beta }\}\in $ $TT^{\ast
}V\otimes TT^{\ast }V$ of local signature $(+,+,+,-;+,+,+,-).$ Such a metric
structure can be expressed in a nonholonomic dyadic form for shells $%
s=1,2,3,4$ as a s-metric on phase space $\mathcal{M}=\mathbf{T}%
_{\shortparallel }^{\ast }\mathbf{V.}$ We state nonholonomic 2+2
decompositions of geometric objects on the first two shells $s=1,2$ if we
consider symmetric tensor products of s-bases $\ ^{\shortparallel}\mathbf{\ e%
}^{\alpha _{2}}\in T_{2}^{\ast }\mathbf{T}_{\shortparallel }^{\ast }\mathbf{V%
}$ (\ref{nadapbdsc}),
\begin{eqnarray}
g = \ _{s}^{\shortparallel }\mathbf{g}&=&(h_{1}\ ^{\shortparallel }\mathbf{g}%
,~v_{2}^{\shortparallel }\mathbf{g},\ c_{3}\ ^{\shortparallel }\mathbf{g,}%
c_{4}\ ^{\shortparallel }\mathbf{g})\in T\mathbf{T}_{\shortparallel }^{\ast }%
\mathbf{V}\otimes _{\star N}T\mathbf{T}_{\shortparallel }^{\ast }\mathbf{V}
\label{sdm} \\
&=&\ ^{\shortparallel }\mathbf{g}_{\alpha _{s}\beta _{s}}(\
_{s}^{\shortparallel }u)\ ^{\shortparallel }\mathbf{e}^{\alpha _{s}}\otimes
_{\star s}\ ^{\shortparallel }\mathbf{e}^{\beta _{s}}=\{\ ^{\shortparallel }%
\mathbf{g}_{\alpha _{s}\beta _{s}}=(\ ^{\shortparallel }\mathbf{g}%
_{i_{1}j_{1}},\ ^{\shortparallel }\mathbf{g}_{a_{2}b_{2}},\ ^{\shortparallel
}\mathbf{g}^{a_{3}b_{3}},\ ^{\shortparallel }\mathbf{g}^{a_{4}b_{4}})\}
\notag \\
\rightarrow \ _{2}^{\shortparallel }\mathbf{g}&=&(h_{1}\ ^{\shortparallel }%
\mathbf{g},v_{2}\ ^{\shortparallel }\mathbf{g})\in h\mathbf{T}%
_{\shortparallel }^{\ast }\mathbf{V}\otimes _{\star N}v\mathbf{T}%
_{\shortparallel }^{\ast }\mathbf{V}  \notag \\
&=&\ ^{\shortparallel }\mathbf{g}_{\alpha _{2}\beta _{2}}(\
_{2}^{\shortparallel }u)\ ^{\shortparallel }\mathbf{e}^{\alpha _{2}}\otimes
_{\star 2}\ ^{\shortparallel }\mathbf{e}^{\beta _{2}}=\{\ ^{\shortparallel }%
\mathbf{g}_{\alpha _{2}\beta _{2}}=(\ ^{\shortparallel }\mathbf{g}%
_{i_{1}j_{1}},\ ^{\shortparallel }\mathbf{g}_{a_{2}b_{2}}\}.  \notag
\end{eqnarray}%
For star products and R-flux deformations to nonassociative geometry,
symmetric metrics transform, in general, into symmetric and nonsymmetric
ones \cite{blumenhagen16,aschieri17} (for nonholonomic N- / s-adapted
constructions, see respectively \cite{partner01,partner02}). We use such
s-adapted parameterizations (s-metrics)
\begin{eqnarray}
&&\mbox{ symmetric: }\ _{\star s}^{\shortparallel }\mathbf{g}=(h_{1}\
_{\star s}^{\shortparallel }\mathbf{g},v_{2}\ _{\star s}^{\shortparallel }%
\mathbf{g,}c_{3}\ _{\star s}^{\shortparallel }\mathbf{g,}c_{4}\ _{\star
s}^{\shortparallel }\mathbf{g})  \label{ssdm} \\
&=&\{\ _{\star }^{\shortparallel }\mathbf{g}_{\alpha _{s}\beta _{s}}=\
_{\star }^{\shortparallel }\mathbf{g}_{\beta _{s}\alpha _{s}}=(\ _{\star
}^{\shortparallel }\mathbf{g}_{i_{1}j_{1}}=\ _{\star }^{\shortparallel }%
\mathbf{g}_{j_{1}i_{1}},\ _{\star }^{\shortparallel }\mathbf{g}%
_{a_{2}b_{2}}=\ _{\star }^{\shortparallel }\mathbf{g}_{b_{2}a_{2}},\ _{\star
}^{\shortparallel }\mathbf{g}^{a_{3}b_{3}}=\ \ _{\star }^{\shortparallel }%
\mathbf{g}^{b_{3}a_{3}},\ _{\star }^{\shortparallel }\mathbf{g}%
^{a_{4}b_{4}}=\ _{\star }^{\shortparallel }\mathbf{g}^{b_{4}a_{4}})\}  \notag
\\
&\rightarrow &\ _{\star 2}^{\shortparallel }\mathbf{g}=(h_{1}\ _{\star
s}^{\shortparallel }\mathbf{g},v_{2}\ _{\star s}^{\shortparallel }\mathbf{g}%
)=\{\ _{\star }^{\shortparallel }\mathbf{g}_{\alpha _{2}\beta _{2}}=\
_{\star }^{\shortparallel }\mathbf{g}_{\beta _{2}\alpha _{2}}=(\ _{\star
}^{\shortparallel }\mathbf{g}_{i_{1}j_{1}}=\ _{\star }^{\shortparallel }%
\mathbf{g}_{j_{1}i_{1}},\ _{\star }^{\shortparallel }\mathbf{g}%
_{a_{2}b_{2}}=\ _{\star }^{\shortparallel }\mathbf{g}_{b_{2}a_{2}})\},
\notag \\
&&\mbox{ and nonsymmetric: }\ _{\star s}^{\shortparallel }\mathfrak{g}%
=(h_{1}\ _{\star s}^{\shortparallel }\mathfrak{g},v_{2}\ _{\star
s}^{\shortparallel }\mathfrak{g},c_{3}\ _{\star s}^{\shortparallel }%
\mathfrak{g,}c_{4}\ _{\star s}^{\shortparallel }\mathfrak{g})  \label{nssdm}
\\
&=&\{\ _{\star }^{\shortparallel }\mathfrak{g}_{\alpha _{s}\beta _{s}}=(\
_{\star }^{\shortparallel }\mathfrak{g}_{i_{1}j_{1}}\neq \ _{\star
}^{\shortparallel }\mathfrak{g}_{j_{1}i_{1}},\ \ _{\star }^{\shortparallel }%
\mathfrak{g}_{a_{2}b_{2}}\neq \ _{\star }^{\shortparallel }\mathfrak{g}%
_{b_{2}a_{2}}\ _{\star }^{\shortparallel }\mathfrak{g}^{a_{3}b_{3}}\neq \
_{\star }^{\shortparallel }\mathfrak{g}^{b_{3}a_{3}},\ _{\star
}^{\shortparallel }\mathfrak{g}^{a_{4}b_{4}}\neq \ _{\star }^{\shortparallel
}\mathfrak{g}^{b_{4}a_{4}})\neq \ _{\star }^{\shortparallel }\mathfrak{g}%
_{\beta _{s}\alpha _{s}}\}  \notag \\
&\rightarrow &\ _{\star 2}^{\shortparallel }\mathfrak{g}=(h_{1}\ _{\star
s}^{\shortparallel }\mathfrak{g},v_{2}\ _{\star s}^{\shortparallel }%
\mathfrak{g})=\{\ _{\star }^{\shortparallel }\mathfrak{g}_{\alpha _{2}\beta
_{2}}=(\ _{\star }^{\shortparallel }\mathfrak{g}_{i_{1}j_{1}}\neq \ _{\star
}^{\shortparallel }\mathfrak{g}_{j_{1}i_{1}},\ \ _{\star }^{\shortparallel }%
\mathfrak{g}_{a_{2}b_{2}}\neq \ _{\star }^{\shortparallel }\mathfrak{g}%
_{b_{2}a_{2}}\}.  \notag
\end{eqnarray}%
On spacetime $V$ and considering coordinate bases, when $\ ^{\shortparallel }%
\mathbf{e}^{\alpha _{2}}\rightarrow \ ^{\shortparallel }e^{\alpha _{2}}=d\
^{\shortparallel }u^{\alpha _{2}}\in hT_{\shortparallel }^{\ast }V$, we can
omit the shell/dyadic label $s.$ Respective star deformed metric structures
can be written in generic off-diagonal forms,
\begin{equation}
\mbox{ symmetric },\ _{\star 2}^{\shortparallel }g=\{\ _{\star
}^{\shortparallel }g_{\alpha _{2}\beta _{2}}\neq \ _{\star }^{\shortparallel
}g_{\beta _{2}\alpha _{2}}\},\mbox{ and nonsymmetric },\ _{\star
2}^{\shortparallel }\mathsf{G}=\{\ _{\star }^{\shortparallel }\mathsf{G}%
_{\alpha _{2}\beta _{2}}\neq \ _{\star }^{\shortparallel }\mathsf{G}_{\beta
_{2}\alpha _{2}}\}.  \label{offdns}
\end{equation}

The coefficients in above formulas are related via frame transforms,
\begin{eqnarray*}
\ ^{\shortparallel }\mathbf{g}_{\alpha _{s}\beta _{s}} &=&\ \
^{\shortparallel }e_{\ \alpha _{s}}^{\underline{\alpha }_{s}}\ \
^{\shortparallel }e_{\ \beta _{s}}^{\underline{\beta }_{s}}\ \
^{\shortparallel }g_{\underline{\alpha }_{s}\underline{\beta }_{s}}=\ \
^{\shortparallel }e_{\ \alpha _{s}}^{\underline{\alpha }}\ \
^{\shortparallel }e_{\ \beta _{s}}^{\underline{\beta }}\ \ ^{\shortparallel
}g_{\underline{\alpha }\underline{\beta }} \\
&\rightarrow &\ ^{\shortparallel }\mathbf{g}_{\alpha _{2}\beta _{2}}=\ \
^{\shortparallel }e_{\ \alpha _{2}}^{\underline{\alpha }_{s}}\ \
^{\shortparallel }e_{\ \beta _{2}}^{\underline{\beta }_{s}}\ \
^{\shortparallel }g_{\underline{\alpha }_{s}\underline{\beta }_{s}}=\ \
^{\shortparallel }e_{\ \alpha _{2}}^{\underline{\alpha }_{2}}\ \
^{\shortparallel }e_{\ \beta _{2}}^{\underline{\beta }_{2}}\ \
^{\shortparallel }g_{\underline{\alpha }_{2}\underline{\beta }_{2}},
\end{eqnarray*}%
relating respective dyadic decompositions with off-diagonal matrices.

\subsubsection{Canonical s-connections, (2+2)-splitting, and distortions of
LC-connections}

Choosing a s-metric $\ _{s}^{\shortparallel }\mathbf{g}$ $\rightarrow \ _{2}%
\mathbf{g}$ (\ref{sdm}), we can define two important linear connections and
respective phase space and spacetime projections:
\begin{equation}
(\ _{s}^{\shortparallel }\mathbf{g,\ _{s}^{\shortparallel }N})\rightarrow
\left\{
\begin{array}{cc}
\fbox{$%
\begin{array}{c}
\ ^{\shortparallel }\mathbf{\nabla :} \\
\downarrow \\
\ _{2}\mathbf{\nabla :}%
\end{array}%
$} & \fbox{$%
\begin{array}{c}
^{\shortparallel }\mathbf{\nabla }\ \ _{s}^{\shortparallel }\mathbf{g}=0;\
_{\nabla }^{\shortparallel }\mathcal{T}=0, \\
\downarrow \\
\ _{2}\mathbf{\nabla }\ \ _{2}\mathbf{g}=0;\ _{\nabla }^{2}\mathcal{T}=0%
\end{array}%
$}\ \mbox{\  LC--connection }; \\
\fbox{$%
\begin{array}{c}
\ _{s}^{\shortparallel }\widehat{\mathbf{D}}: \\
\downarrow \\
\ _{2}\widehat{\mathbf{D}}:%
\end{array}%
$} & \fbox{$%
\begin{array}{c}
\begin{array}{c}
\ _{s}^{\shortparallel }\widehat{\mathbf{D}}\ _{s}^{\shortparallel }\mathbf{g%
}=0;\ h_{1}\ ^{\shortparallel }\widehat{\mathcal{T}}=0,v_{2}\
^{\shortparallel }\widehat{\mathcal{T}}=0,c_{3}\ ^{\shortparallel }\widehat{%
\mathcal{T}}=0,c_{4}\ ^{\shortparallel }\widehat{\mathcal{T}}=0, \\
h_{1}v_{2}\ ^{\shortparallel }\widehat{\mathcal{T}}\neq 0,h_{1}c_{s}\
^{\shortparallel }\widehat{\mathcal{T}}\neq 0,v_{2}c_{s}\ ^{\shortparallel }%
\widehat{\mathcal{T}}\neq 0,c_{3}c_{4}\ ^{\shortparallel }\widehat{\mathcal{T%
}}\neq 0,%
\end{array}
\\
\downarrow \\
\ _{s}\widehat{\mathbf{D}}\ _{s}\mathbf{g}=0;\ h_{1}\ \widehat{\mathcal{T}}%
=0,v_{2}\ \widehat{\mathcal{T}}=0,h_{1}v_{2}\ \widehat{\mathcal{T}}\neq 0,%
\end{array}%
$}%
\begin{array}{c}
\mbox{ canonical } \\
\mbox{ s-connection  }.%
\end{array}%
\end{array}%
\right.  \label{twocon}
\end{equation}%
A s-operator $\ _{s}^{\shortparallel }\widehat{\mathbf{D}}=(h_{1}\
^{\shortparallel }\widehat{\mathbf{D}},\ v_{2}\ ^{\shortparallel }\widehat{%
\mathbf{D}},\ c_{3}\ ^{\shortparallel }\widehat{\mathbf{D}},\ c_{4}\
^{\shortparallel }\widehat{\mathbf{D}})$ acts on tangent spaces of phase
space, i.e. on $T\mathbf{T}_{\shortparallel }^{\ast }\mathbf{V},$ with
dyadic (co) vertical splitting, being a s-connection adapted to a
N--/s-connection structure $\ _{s}^{\shortparallel }\mathbf{N.}$ Such linear
(affine) connections and N-connections and their s-torsion are nonholonomic
when the torsions of type $\ ^{\shortparallel }\widehat{\mathcal{T}},$ etc.
are computed in standard form. We use "hat" labels in order to emphasize
that such s-adapted values are determined by a s-metric structure following
certain nonholonomic constraints involving (partial) zero torsion and metric
compatibility conditions. When such values are computed on the spacetime
shells $s=1$ and $2$ and trivial co-fibers $c_{3}$ and $c_{4},$ we can omit
the label \ $\ $"$^{\shortparallel \text{"}}$ and write $\ _{2}\widehat{%
\mathbf{D}}=(h_{1}\widehat{\mathbf{D}},\ v_{2}\widehat{\mathbf{D}}).$ The
LC-connection $\ ^{\shortparallel }\mathbf{\nabla =}$ $\
_{s}^{\shortparallel }\mathbf{\nabla }$ and its spacetime projection $\ _{2}%
\mathbf{\nabla }$ are not a d-/ or s-connections because such linear
connections do not preserve any N-/ s-connection splitting and/or dyadic
decompositions under parallel transports in respective phase space and
spacetime.

In \cite{partner01,partner02}, we provide all abstract and N- / s-adapted
formulas for canonical s-connections and their distortions on $T\mathbf{T}%
_{\shortparallel }^{\ast }\mathbf{V.}$ Here, we consider some important
nonholonomic formuls for 2+2 spacetime splitting and $\ _{2}^{\shortparallel
}\widehat{\mathbf{D}}$ which are important for our further considerations.
We can check by straightforward computations that the respective conditions
from (\ref{twocon}) are satisfied by such s-coefficients,
\begin{eqnarray}
\ _{2}^{\shortparallel }\widehat{\mathbf{D}}&=&\{\ ^{\shortparallel }%
\widehat{\mathbf{\Gamma }}_{\alpha _{2}\beta _{2}}^{\gamma _{2}}=(\widehat{L}%
_{j_{1}k_{1}}^{i_{1}},\widehat{L}_{b_{2}\ k_{1}}^{a_{2}},\widehat{C}_{\
j_{1}c_{2}}^{i_{1}\ },\widehat{C}_{b_{2}c_{2}}^{a_{2}})\}, \mbox{ where }
\notag \\
\widehat{L}_{j_{1}k_{1}}^{i_{1}} &=&\frac{1}{2}\ \ ^{\shortparallel
}g^{i_{1}r_{1}}(\ \ ^{\shortparallel }\mathbf{e}_{k_{1}}\ ^{\shortparallel
}g_{j_{1}r_{1}}+\ \ ^{\shortparallel }\mathbf{e}_{j_{1}}\ ^{\shortparallel
}g_{k_{1}r_{1}}-\ ^{\shortparallel }\mathbf{e}_{r_{1}}\ ^{\shortparallel
}g_{j_{1}k_{1}}),\   \label{canhcs} \\
\widehat{L}_{b_{2}\ k_{1}}^{a_{2}} &=&\ ^{\shortparallel }e_{b_{2}}(\
^{\shortparallel }N_{k_{1}}^{a_{2}})+\frac{1}{2}\ ^{\shortparallel
}g^{b_{2}c_{2}}(\ ^{\shortparallel }e_{k_{1}}\ ^{\shortparallel
}g_{b_{2}c_{2}}-\ ^{\shortparallel }g_{d_{2}c_{2}}\ ^{\shortparallel
}e_{b_{2}}\ ^{\shortparallel }N_{k_{1}}^{d_{2}}-\ ^{\shortparallel
}g_{d_{2}b_{2}}\ ^{\shortparallel }e_{c_{2}}\ ^{\shortparallel
}N_{k_{1}}^{d_{2}}),  \notag \\
\widehat{C}_{\ j_{1}c_{2}}^{i_{1}\ } &=&\frac{1}{2}\ \ ^{\shortparallel
}g^{ik}\ \ ^{\shortparallel }e_{c}\ \ ^{\shortparallel }g_{jk},\ \ \
\widehat{C}_{b_{2}c_{2}}^{a_{2}}=\frac{1}{2}\ \ ^{\shortparallel
}g^{a_{2}d_{2}}(\ \ ^{\shortparallel }e_{c_{2}}\ ^{\shortparallel
}g_{b_{2}d_{2}}+\ ^{\shortparallel }e_{b_{2}}\ ^{\shortparallel
}g_{c_{2}d_{2}}-\ ^{\shortparallel }e_{d_{2}}\ ^{\shortparallel
}g_{b_{2}c_{2}}).  \notag
\end{eqnarray}

We can compute a canonical distortion relation to a canonical s-connection,
or respective spacetime canonical d-connection,
\begin{equation}
\ _{s}^{\shortparallel }\widehat{\mathbf{D}}=\ ^{\shortparallel }\nabla +\
_{s}^{\shortparallel }\widehat{\mathbf{Z}},\mbox{ or }\ _{2}\widehat{\mathbf{%
D}}=\ _{2}\nabla +\ _{2}\widehat{\mathbf{Z}}.  \label{candistr}
\end{equation}%
For instance, the distortion s-tensor, $\ _{s}^{\shortparallel }\widehat{%
\mathbf{Z}}= \{\ ^{\shortparallel }\widehat{\mathbf{Z}}_{\ \beta _{s}\gamma
_{s}}^{\alpha _{s}} [\ ^{\shortparallel }\widehat{\mathbf{T}}_{\ \beta
_{s}\gamma _{s}}^{\alpha _{s}}]\}$ from (\ref{candistr}) is an algebraic
combination of the coefficients the canonical torsion s-tensor $\
_{s}^{\shortparallel }\widehat{\mathcal{T}}=\{\ ^{\shortparallel }\widehat{%
\mathbf{T}}_{\ \beta _{s}\gamma _{s}}^{\alpha _{s}}\}$ of $\
_{s}^{\shortparallel }\widehat{\mathbf{D}}.$ Similar algebraic combinations
can be defined for the nonholonomic spacetime projections with $s=1,2$.

\subsubsection{Canonical nonholonomic Ricci and Einstein d-tensors for
(2+2)-splitting}

For all types of linear connections defined in (\ref{twocon}), we can define
and compute in standard forms as in metric-affine geometry respective
torsions, see details on (non) associative formulations in \cite%
{partner01,partner02}. Here we provide necessary formulas without details on
computations of s-adapted coefficients of torsions and curvatures using
derivatives and contractions with s-connection and s-metric coefficients.
For a nonholonomic spacetime (2+2)-splitting, there are considered
fundamental geometric objects like torsions $\ _{\nabla }^{2}\mathcal{T}=0$
and $\ _{2}\widehat{\mathcal{T}},$ and curvature, $\ _{\nabla }^{2}\mathcal{R%
}=\{\ _{\nabla }R_{\ \beta _{2}\gamma _{2}\delta _{2}}^{\alpha _{2}}\}$ and $%
\ _{2}^{\shortparallel }\widehat{\mathcal{R}}=\{\ ^{\shortparallel }\widehat{%
\mathbf{R}}_{\ \beta _{2}\gamma _{2}\delta _{2}}^{\alpha _{2}}\},$ s-tensors.

The canonical Ricci s-tensor on the first two shells, $\ _{2}\widehat{%
\mathcal{R}}ic=\{\widehat{\mathbf{R}}_{\ \beta _{2}\gamma _{2}}:=\ \widehat{%
\mathbf{R}}_{\ \alpha _{2}\beta _{2}\gamma _{s}}^{\gamma _{2}}\}$, is
characterized by such a splitting of s-adapted coefficients,%
\begin{equation}
\widehat{\mathbf{R}}_{\ \beta _{2}\gamma _{2}}=\{\widehat{R}_{\
h_{1}j_{1}}=\ \widehat{\mathbf{R}}_{~h_{1}j_{1}i_{1}}^{i_{1}},\widehat{P}%
_{j_{1}a_{2}}^{\ }=-\ \widehat{\mathbf{R}}_{~\ j_{1}i_{1}a_{2}}^{i_{1}\quad
}\ ,\widehat{P}_{b_{2}k_{1}}=\widehat{\mathbf{R}}_{~b_{2}k_{1}c_{2}}^{c_{2}\
},\ ^{\shortparallel }\widehat{S}_{b_{2}c_{2}\ }=\widehat{\mathbf{R}}%
_{~b_{2}c_{2}a_{2}\ }^{a_{2}\ }\}.  \label{candricci}
\end{equation}%
Using two linear connections (\ref{twocon}), we define two different scalar
curvatures,
\begin{equation}
\ _{2}\widehat{\mathbf{R}}sc:=\mathbf{g}^{\alpha _{2}\beta _{2}}\
^{\shortparallel }\widehat{\mathbf{R}}_{\alpha _{2}\beta _{2}}=\
^{\shortparallel }g^{i_{1}j_{1}}\ ^{\shortparallel }\widehat{R}%
_{i_{1}j_{1}}+\ ^{\shortparallel }g^{a_{2}b_{2}}\ ^{\shortparallel }\widehat{%
R}_{a_{2}b_{2}}\mbox{ and }\ ^{\shortparallel }R:=\ ^{\shortparallel }%
\mathbf{g}^{\alpha _{2}\beta _{2}}\ ^{\shortparallel }R_{\alpha _{2}\beta
_{2}}.  \notag
\end{equation}%
The modified Einstein equations for $(\ _{2}\mathbf{g,}\ _{2}\widehat{%
\mathbf{D}})$ on $\mathbf{T}_{2\shortparallel }^{\ast }\mathbf{V}$ with a
nontrivial cosmological constant $\ ^{\shortparallel }\lambda $ can be
postulated using the same geometric principles as in GR. \footnote{%
we keep the label $\ $"$^{\shortparallel }"$ if this constant is taken for
the total phase space but the geometric objects are nonholonomically
constrained on shells $s=1,2$ and/or projected on a nonholonomic Lorentz base%
} In abstract geometric form (see, for instance, \cite{misner}), re-defining
the geometric constructions for nonholonomic Lorentz manifolds / (co)
bundles \cite{bubuianu18a}, we write
\begin{equation}
\ _{2}\widehat{E}n:=\ _{2}\widehat{\mathcal{R}}ic-\frac{1}{2}\ _{2}\widehat{%
\mathbf{R}}sc=\ ^{\shortparallel }\lambda \ _{2}\mathbf{g}.
\label{cnsveinst1}
\end{equation}%
The value $\ _{2}\widehat{E}n=\{\ \widehat{E}_{_{\alpha _{2}\beta _{2}}}\}$
is by definition the canonical Einstein d-tensor. The cosmological constant $%
\ ^{\shortparallel }\lambda $ encodes as an effective source certain
contributions from cofiber dynamics of phase space. If we completely ignore
such possible contributions, we should write $\ ^{\shortparallel }\lambda =$
$\lambda $ and work with nonholonomic 2+2 splitting and Einstein manifolds
for the canonical d-connection.

It should be emphasized that $\ _{2}\widehat{\mathbf{D}}(\ _{2}\widehat{E}%
n)\neq 0.$ This means that the conservation laws on nonholonomic manifolds
are subjected to additional non-integrable constraints (similar
constructions are considered in nonholonomic mechanics). Using distortions (%
\ref{candistr}), we can always nonholonomically deform such systems of
nonlinear PDEs into equivalent ones, $\ _{2}\widehat{E}n\rightarrow \
_{\nabla }En,$ when $\ _{2}\nabla (\ _{\nabla }En)=0.$ Such canonical
s-distortions are determined by respective distortions of the curvature and
Ricci tensors,
\begin{equation}
\ _{2}\widehat{\mathcal{R}}=\ \ _{\nabla }^{2}\mathcal{R+}\ \ \ _{2}\widehat{%
\mathcal{Z}},\ _{2}\widehat{\mathcal{R}}ic=\ \ _{\nabla }^{2}Ric+\ \ _{2}%
\widehat{\mathcal{Z}}ic,\ _{2}\widehat{\mathbf{R}}sc=\ \ _{\nabla }^{2}Rsc+\
\ _{2}\widehat{\mathcal{Z}}sc,\mbox{
and }\ _{2}\widehat{E}n=\ _{\nabla }^{2}En+\ \ _{2}\widehat{\mathcal{Z}}n.
\label{candriccidist}
\end{equation}%
In these formulas, the distortion s-tensors $\ _{\nabla }\widehat{\mathcal{Z}%
}$ and $\ _{\nabla }\widehat{\mathcal{Z}}ic$ are correspondingly determined
by $(\ _{2}\widehat{\mathbf{Z}},\ _{2}\mathbf{g)}$ on a (pseudo) Riemannian
phase background with LC-connection $\ ^{2}\nabla $ encoding the
N-connection coefficients for a nonholonomic 2+2 splitting. We omit abstract
and cumbersome s-adapted/ coordinate formulas for such s-objects and
respective geometric and physical systems of PDEs. Finally, we note that all
geometric constructions can be performed equivalently working with different
geometric data $\left( \ _{2}\mathbf{g},\ ^{2}\nabla \right) $ and/or $(\
_{2}\mathbf{g},\ _{2}\mathbf{N},\ _{2}\widehat{\mathbf{D}})$ if we prescribe
a zero, or non-zero cosmological constant of type $\ ^{\shortparallel
}\lambda $ or $\lambda .$

\subsection{Nonholonomic geometry of star deformed spacetime (non) symmetric
d-metrics}

In \cite{blumenhagen16,aschieri17}, nonassocitative geometric models
determined by R-flux star deformations are elaborated in coordinate frames
and beginning with a flat (co) fiber metric $\mathbf{\ ^{\shortparallel }}%
\eta .$ Star products $\star $ are defined in terms of coordinate bases $%
\mathbf{\ ^{\shortparallel }\partial }$ when nonassociative generalizations
of (pseudo) Riemann geometry are constructed with symmetric, $\ _{\star
}^{\shortparallel }g,$ and nonsymmetric, $\ _{\star }^{\shortparallel }%
\mathsf{G,}$ star-metric structures and a related nonassociative variant of
LC-connection $\mathbf{\ ^{\shortparallel }}\nabla ^{\star }.$
Conventionally, there are considered star-deformations of geometric
structures when $(\ ^{\shortparallel }\eta ,\ ^{\shortparallel }\partial ,\
^{\shortparallel }\nabla )\rightarrow (\star \ ,\ \mathcal{A}^{\star },\
_{\star }^{\shortparallel }g,\ _{\star }^{\shortparallel }\mathsf{G},\ \
^{\shortparallel }\partial \ ^{\shortparallel }\nabla ^{\star }),$ when the
constructions are adapted to quasi-Hopf algebras $\ \mathcal{A}^{\star }$,
or other type algebraic and geometric structures. In such an approach, it is
difficult technically to decouple physically important systems of nonlinear
PDEs.

\subsubsection{Convention 2 for nonassociative phases with nonholonomic
spacetime shells}

In our recent partner works \cite{partner01,partner02}, we followed 
Convention 2 on constructing nonassociative nonholonomic geometries which
allow the application of the AFCDM for constructing exact solutoins of
graviational and geometric flow equations. In this paper, we adapt those
assumptions for study nonholonomic vacuum Einstein equations with 2+2
splitting and effective sources encoding star and R-flux contributions form
the total 8-d phase space.

\vskip5pt \textbf{Convention 2 (extended)} on nonholonomic constraints from
nonassociative phase spaces to spacetime configurations:\ On phase spaces,
star products (\ref{starpn}) are defined via nonholonomic dyadic
decompositions on $\ ^{\shortparallel }\mathbf{e}_{\alpha _{s}}$ with R-flux
terms and there are computed respective star deformations of canonical
s-adapted geometric objects into nonassociative ones, with symmetric, $\
_{\star s}^{\shortparallel }\mathbf{g}$, and nonsymmetric, $\ _{\star
s}^{\shortparallel }\mathbf{\mathfrak{g}}$, star s-metrics and canonical
star s-connection $\ _{s}^{\shortparallel }\mathbf{D}^{\star },$ see
definitions below (details are provided in \cite{partner01,partner02}). The
nonholonomic dyadic star deformations of geometric canonical s-structures
and their projections on the spacetime shells $s=1,2$ are defined by such
star transforms of geometric data:%
\begin{equation}
\begin{array}{ccc}
\fbox{$%
\begin{array}{c}
(\star _{N},\ \ \mathcal{A}_{N}^{\star },\ _{\star }^{\shortparallel }%
\mathbf{g,\ _{\star }^{\shortparallel }\mathfrak{g,}\ \ ^{\shortparallel }N},%
\mathbf{\ \ ^{\shortparallel }e}_{\alpha }\mathbf{,\ \mathbf{\mathbf{\mathbf{%
\ ^{\shortparallel }}}}D}^{\star }) \\
\downarrow \\
(\star _{2},\ \ \mathcal{A}_{2}^{\star },\ _{\star }^{\shortparallel 2}%
\mathbf{g,\ _{\star }^{\shortparallel 2}\mathfrak{g,}}\ \ _{2}\mathbf{N},%
\mathbf{\ \ e}_{\alpha _{2}}\mathbf{,}\ \ _{2}^{\shortparallel }\mathbf{D}%
^{\star })%
\end{array}%
$} & \Leftrightarrow & \fbox{$%
\begin{array}{c}
(\star _{s},\ \ \mathcal{A}_{s}^{\star },\ _{\star s}^{\shortparallel }%
\mathbf{g,\ _{\star s}^{\shortparallel }\mathfrak{g,}}\ \
_{s}^{\shortparallel }\mathbf{N},\mathbf{\ \ ^{\shortparallel }e}_{\alpha
_{s}}\mathbf{,\ \mathbf{\mathbf{\mathbf{\ _{s}^{\shortparallel }}}}D}^{\star
}) \\
\downarrow \\
(\star _{2},\ \ \mathcal{A}_{2}^{\star },\ _{\star }^{\shortparallel 2}%
\mathbf{g,\ _{\star }^{\shortparallel 2}\mathfrak{g,}}\ \ _{2}\mathbf{N},%
\mathbf{\ \ e}_{\alpha _{2}}\mathbf{,}\ \ _{2}^{\shortparallel }\mathbf{D}%
^{\star })%
\end{array}%
$} \\
& \Uparrow &  \\
\fbox{$%
\begin{array}{c}
(\ \ ^{\shortparallel }\mathbf{g,\ \ ^{\shortparallel }N},\mathbf{\ \
^{\shortparallel }e}_{\alpha }\mathbf{,}\ ^{\shortparallel }\widehat{\mathbf{%
D}}) \\
\downarrow \\
(\ \ _{2}\mathbf{g,\ \ _{2}N},\mathbf{\ \ e}_{\alpha _{2}}\mathbf{,}\ _{2}%
\widehat{\mathbf{D}})%
\end{array}%
$} & \Leftrightarrow & \fbox{$%
\begin{array}{c}
(\ \ _{s}^{\shortparallel }\mathbf{g,\ \ _{s}^{\shortparallel }N},\mathbf{\
\ ^{\shortparallel }e}_{\alpha _{s}}\mathbf{,}\ _{s}^{\shortparallel }%
\widehat{\mathbf{D}}), \\
\downarrow \\
(\ \ _{2}\mathbf{g,\ \ _{2}N},\mathbf{\ \ e}_{\alpha _{2}}\mathbf{,}\ _{2}%
\widehat{\mathbf{D}})%
\end{array}%
$}%
\end{array}
\label{conv2s}
\end{equation}%
for $\ ^{\shortparallel }\mathbf{D}^{\star }=\ ^{\shortparallel }\nabla
^{\star }+\ ^{\shortparallel }\widehat{\mathbf{Z}}^{\star }$ and $\
_{s}^{\shortparallel }\mathbf{D}^{\star }=\ ^{\shortparallel }\nabla ^{\star
}+\ _{s}^{\shortparallel }\widehat{\mathbf{Z}}^{\star }$ with respective
spacetime 2+2 splitting $\ _{2}\mathbf{D}^{\star }=\ _{2}\nabla ^{\star }+\
_{2}\widehat{\mathbf{Z}}^{\star }.$ This convention with transforms (\ref%
{conv2s}) allows us to construct large classes of generic off-diagonal
solutions on 4-d spacetimes with effective sources encoding nonassociative
R-flux distortions.

\subsubsection{Nonsymmetric and symmetric 2+2 metrics and their inverses}

In \cite{partner02}, metric s-structures in nonassociative nonholonomic
differential geometry and related quasi-Hopf s-structures are studied on 8-d
phase spaces. A star metric symmetric s-tensor (\ref{ssdm}) for shells $%
s=1,2,3,4,$ on phase space, and $s=1,2$ on h-v-decompositions, with R-flux
induced terms on a Lorentz base manifold can be represented in the form
\begin{equation*}
\ _{\star s}^{\shortparallel }\mathbf{g}=\ _{\star }^{\shortparallel }%
\mathbf{g}_{\alpha _{s}\beta _{s}}\star _{s}(\ ^{\shortparallel }\mathbf{e}%
^{\alpha _{s}}\otimes _{\star s}\ ^{\shortparallel }\mathbf{e}^{\beta
_{s}})\in \Omega _{\star }^{1}\otimes _{\star s}\Omega _{\star }^{1};%
\mbox{
and }\ _{\star 2}^{\shortparallel }\mathbf{g}=\ _{\star }^{\shortparallel }%
\mathbf{g}_{\alpha _{2}\beta _{2}}\star _{2}(\ ^{\shortparallel }\mathbf{e}%
^{\alpha _{2}}\otimes _{\star 2}\ ^{\shortparallel }\mathbf{e}^{\beta
_{2}})\in \Omega _{\star }^{1}\otimes _{\star 2}\Omega _{\star }^{1},
\end{equation*}%
where the geometric objects on spacetime shells do not depend on co-fiber
coordinates only on some special s-adapted bases. In general, R-flux terms
connect and mix indices of all shells and we need additional assumptions on
how to extract spacetime geometric objects with dependencies only on $%
x^{\alpha _{2}}$ coordinates. We can work with background metrics with flat
cofiber components $c^{3}$ and $c^{4}$ when the s-connection structure is
prescribed to be nontrivial only for $s=1,2.$ In our approach, there are
considered real-valued s-adapted coefficients $\ _{\star }^{\shortparallel}%
\mathbf{g}(\ ^{\shortparallel}\mathbf{e}_{\alpha _s},\ ^{\shortparallel }%
\mathbf{e}_{\beta _s})= \ _{\star}^{\shortparallel}\mathbf{g}_{\alpha _s
\beta _s}= \ _{\star}^{\shortparallel}\mathbf{g}_{\beta _s \alpha _s}\in
\mathcal{A}_{s}^{\star }$, with respective restrictions for $s=1,2.$ Such a
s-metric and respective nonholonomic constraints for $\ _{\star
s}^{\shortparallel }\mathbf{g\rightarrow }\ _{\star 2}^{\shortparallel }%
\mathbf{g}$ are both compatible with a star s-connection $\
_{s}^{\shortparallel }\mathbf{D}^{\star }\rightarrow \ _{2}^{\shortparallel }%
\mathbf{D}^{\star }$ if the conditions
\begin{equation}
\ _{s}^{\shortparallel }\mathbf{D}^{\star }\ _{\star s}^{\shortparallel }%
\mathbf{g}=0\mbox{ and }\ _{2}^{\shortparallel }\mathbf{D}^{\star }\ _{\star
2}^{\shortparallel }\mathbf{g}=0  \label{mcompnas}
\end{equation}%
are satisfied. We shall consider below how star deformations of
s-connections and their shell projections are defined and computed, see
details in \cite{partner01,partner02}.

The models of nonassociative geometry and gravity formulated in \cite%
{blumenhagen16,aschieri17} involve "non-geometric" constructions when R-flux
deformations result in nonsymmetric metric structures. Such geometric
off-diagonal and/or d- / s-objects have to be considered additionally to the
symmetric metric. A nonsymmetric metric on phase space $\mathbf{T}%
_{\shortparallel }^{\ast }\mathbf{V}$ with restrictions to shells $s=1,2,$
can be defined in generic off-diagonal form with respect to local coordinate
bases when%
\begin{equation}
\ _{\star }^{\shortparallel }\mathsf{G}_{\alpha \beta }=\ _{\star
}^{\shortparallel }g_{\alpha \beta }-i\kappa \mathcal{R}_{\quad \alpha
}^{\tau \xi }\ \mathbf{^{\shortparallel }}\partial _{\xi }\ _{\star
}^{\shortparallel }g_{\beta \tau }\mbox{ and }\ _{\star }^{\shortparallel }%
\mathsf{G}_{\alpha _{2}\beta _{2}}=\ _{\star }^{\shortparallel }g_{\alpha
_{2}\beta _{2}}-i\kappa \mathcal{R}_{\quad \alpha _{2}}^{\tau \xi }\ \mathbf{%
^{\shortparallel }}\partial _{\xi }\ _{\star }^{\shortparallel }g_{\beta
_{2}\tau }.  \label{offdns1}
\end{equation}%
Such coefficients define, respectively, some nonsymmetric $8\times 8$ and $%
4\times 4$ matrices but on the spacetime shells the coefficients are
determined by $\mathcal{R}_{\quad \alpha _{2}}^{\tau \xi }$-coefficients
even $\ _{\star }^{\shortparallel }g_{\alpha _{2}\beta _{2}}$ are
constrained to depend only on spacetime coordinates $x^{\beta _{2}}$.

With respect to s-adapted bases $\ ^{\shortparallel }\mathbf{e}_{\xi
_{s}}\rightarrow \ ^{\shortparallel }\mathbf{e}_{\xi _{2}}$ and tensor
products of their dual s-bases, a nonsymmetric s-metric structure (\ref%
{nssdm}) can be parameterized in a $[(2\times 2)+(2\times 2)]+$ $[(2\times
2)+(2\times 2)] $ block form which can be nonholonomically restricted to a
matrix $[2\times 2]+$ $[2\times 2],$ when
\begin{equation}
\ _{\star }^{\shortparallel }\mathfrak{g}_{\alpha _{s}\beta _{s}}=\ _{\star
}^{\shortparallel }\mathbf{g}_{\alpha _{s}\beta _{s}}-i\kappa \overline{%
\mathcal{R}}_{\quad \alpha _{s}}^{\tau _{s}\xi _{s}}\ \mathbf{%
^{\shortparallel }e}_{\xi _{s}}\ _{\star }^{\shortparallel }\mathbf{g}%
_{\beta _{s}\tau _{s}}\rightarrow \ _{\star }^{\shortparallel }\mathfrak{g}%
_{\alpha _{2}\beta _{2}}=\ _{\star }^{\shortparallel }\mathbf{g}_{\alpha
_{2}\beta _{2}}-i\kappa \overline{\mathcal{R}}_{\quad \alpha _{2}}^{\tau
_{s}\xi _{s}}\ \mathbf{^{\shortparallel }e}_{\xi _{s}}\ _{\star
}^{\shortparallel }\mathbf{g}_{\beta _{2}\tau _{s}}.  \label{dmss1}
\end{equation}%
Such formulas are similar to (\ref{offdns1}) but we underline the R-flux
coefficients because they are re-written in s-adapted form $\overline{%
\mathcal{R}}_{\quad \alpha _{s}}^{\tau _{s}\xi _{s}}$ in order generate a
star nonsymmetric generalization of the commutative s-metric (\ref{sdm}). On
total phase space, we can write $\ _{\star s}^{\shortparallel }\mathfrak{g}%
=\ _{\star }^{\shortparallel }\mathfrak{g}_{\alpha _{s}\beta _{s}}\star
_{s}(\ ^{\shortparallel }\mathbf{e}^{\alpha _{s}}\otimes _{\star s}\
^{\shortparallel }\mathbf{e}^{\beta _{s}})=\ _{\star }^{\shortparallel }%
\mathsf{G}_{\alpha \beta }\star (\ d\ ^{\shortparallel }u^{\alpha }\otimes
_{\star }d\ ^{\shortparallel }u^{\beta })$, where $\ _{\star
}^{\shortparallel }\mathfrak{g}_{\alpha _{s}\beta _{s}}\neq \ _{\star
}^{\shortparallel }\mathfrak{g}_{\beta _{s}\alpha _{s}}$ and $\ _{\star
}^{\shortparallel }\mathsf{G}_{\alpha \beta }\neq \ _{\star
}^{\shortparallel }\mathsf{G}_{\beta \alpha }.$ We can also consider
nonholonomic constraints for $s=1,2,$%
\begin{equation}
\ _{\star 2}^{\shortparallel }\mathfrak{g}=\ _{\star }^{\shortparallel }%
\mathfrak{g}_{\alpha _{2}\beta _{2}}\star _{2}(\ ^{\shortparallel }\mathbf{e}%
^{\alpha _{2}}\otimes _{\star 2}\ ^{\shortparallel }\mathbf{e}^{\beta _{2}}),
\label{dmss1a}
\end{equation}%
from which we can find off-diagonal $\ _{\star }^{\shortparallel }\mathsf{G}%
_{\alpha _{2}\beta _{2}}$ but such coefficients in coordinate bases may
depend not only on $x^{\beta _{2}}$ but also on cofiber coordinates $%
p_{a_{3}}$ and $p_{a_{4}},$ for $s=3,4.$

In \cite{partner01,partner02}, we elaborated a nonholonomic and s-adapted
geometric formalism for quasi-Hopf s-structures, It generalized the
holonomic procedure for constructing inversions of matrices for a star
quasi-Hopf algebra $\mathcal{A}^{\star }$ formulated in \cite{aschieri17}.
So, working in coordinates, or s-adapted frames, we can compute, for
instance, the inverse matrix $\ ^{\shortparallel }\overline{\mathsf{G}}%
^{-1}=\{\ _{\star }^{\shortparallel }\mathsf{G}^{\alpha \beta }\}$ of a
matrix $\ ^{\shortparallel }\overline{\mathsf{G}}=\{\ _{\star
}^{\shortparallel }\mathsf{G}_{\alpha \beta }\}$ as a solution of algebraic
equations $\ _{\star }^{\shortparallel }\mathsf{G}^{\alpha \beta }\cdot \
_{\star }^{\shortparallel }\mathsf{G}_{\beta \gamma }=\ _{\star
}^{\shortparallel }\mathsf{G}_{\gamma \beta }\cdot $ $\ _{\star
}^{\shortparallel }\mathsf{G}^{\beta \alpha }=\delta _{\beta }^{\alpha }.$
In geometric series, such matrix formulas are of type
\begin{equation}
\ _{\star }^{\shortparallel }\mathsf{G}^{\alpha \beta }=\ ^{\shortparallel
}g^{\alpha \beta }-i\kappa \ ^{\shortparallel }g^{\alpha \tau }\mathcal{R}%
_{\quad \tau }^{\mu \nu }(\partial _{\mu }\ ^{\shortparallel }g_{\nu
\varepsilon })\ ^{\shortparallel }g^{\varepsilon \beta }+O(\kappa ^{2}),
\label{offdns1inv}
\end{equation}%
which $\mathcal{R}_{\quad \tau }^{\mu \nu }$ constructing via nonholonomic
transforms. We can write similar formulas in block s-forms for various
matrices constructed from s-adapted coefficients of a nonsymmetric star
s-metric $\ _{\star }^{\shortparallel }\mathfrak{g}_{\alpha _{s}\beta _{s}}$
and, respectively, of a symmetric star d-metric $\ _{\star }^{\shortparallel
}\mathbf{g}_{\alpha _{s}\beta _{s}}.$

\subsubsection{Star deformed LC-connections, canonical s-connections and
spacetime projections}

Using a star d-metric $\ _{\star }^{\shortparallel }\mathbf{g}$ (\ref{ssdm})
with spacetime shells $s=1,2$ nonholonomic constraints (projections) to $\
_{2\star }^{\shortparallel }\mathbf{g,}$ and applying tedious computations
with star product (\ref{starpn}), following Convention 2 (\ref{conv2s}), we
can define and compute star deformations of respective LC- and canonical
s-connections from (\ref{twocon}). Such details are contained for all shells
on phase spaces in our partner work \cite{partner02}. Here we present some
important formulas stated in additional forms for spacetime shells which
allow us to and work in equivalent forms with two different linear
connections both on total phase spaces and on spacetimes encoding R-fluxes
and star deformations: {\small
\begin{equation}
\begin{array}{c}
(\ _{\star s}^{\shortparallel }\mathbf{g,\ _{s}^{\shortparallel }N}) \\
\downarrow \\
(\ _{\star 2}^{\shortparallel }\mathbf{g,\ _{2}^{\shortparallel }N})%
\end{array}%
\rightarrow \left\{
\begin{array}{cc}
\begin{array}{c}
_{\star }^{\shortparallel }\mathbf{\nabla :} \\
\downarrow \\
_{\star }^{2}\mathbf{\nabla :}%
\end{array}%
\  &
\begin{array}{c}
\fbox{\ $_{\star }^{\shortparallel }\mathbf{\nabla }$\ $_{\star
s}^{\shortparallel }$\textbf{g}=0;\ $_{\nabla }^{\shortparallel }\mathcal{T}%
^{\star }$=0,} \\
\downarrow \\
\fbox{\ $_{\star }^{\shortparallel }\mathbf{\nabla }$\ $_{\star
s}^{\shortparallel }$\textbf{g}=0;\ $_{\nabla }^{\shortparallel }\mathcal{T}%
^{\star }$=0,}%
\end{array}%
\mbox{\
star LC-connection}; \\
\begin{array}{c}
\ _{s}^{\shortparallel }\widehat{\mathbf{D}}^{\star }: \\
\downarrow \\
\ _{2}\widehat{\mathbf{D}}^{\star }:%
\end{array}
&
\begin{array}{c}
\fbox{$%
\begin{array}{c}
\ _{s}^{\shortparallel }\widehat{\mathbf{D}}^{\star }\ _{\star
s}^{\shortparallel }\mathbf{g}=0;\ h_{1}\ ^{\shortparallel }\widehat{%
\mathcal{T}}^{\star }=0,v_{2}\ ^{\shortparallel }\widehat{\mathcal{T}}%
^{\star }=0,c_{3}\ ^{\shortparallel }\widehat{\mathcal{T}}^{\star }=0,c_{4}\
^{\shortparallel }\widehat{\mathcal{T}}^{\star }=0, \\
h_{1}v_{2}\ ^{\shortparallel }\widehat{\mathcal{T}}^{\star }\neq
0,h_{1}c_{s}\ ^{\shortparallel }\widehat{\mathcal{T}}^{\star }\neq
0,v_{2}c_{s}\ ^{\shortparallel }\widehat{\mathcal{T}}^{\star }\neq
0,c_{3}c_{4}\ ^{\shortparallel }\widehat{\mathcal{T}}^{\star }\neq 0,%
\end{array}%
$} \\
\downarrow \\
\fbox{\ $_{2}\widehat{\mathbf{D}}^{\star }$\ $_{\star 2}^{\shortparallel }$%
\textbf{g}=0;\ h$_{1}$\ $_{2}\widehat{\mathcal{T}}^{\star }$=0,v$_{2}$\ $%
\widehat{\mathcal{T}}^{\star }$=0,}%
\end{array}%
\mbox{canonic  s-connect}.%
\end{array}%
\right.  \label{twoconsstar}
\end{equation}%
} In these formulas, decompositions $\ _{s}^{\shortparallel }\widehat{%
\mathbf{D}}^{\star }=(h_{1}\ ^{\shortparallel }\widehat{\mathbf{D}}^{\star
},\ v_{2}\ ^{\shortparallel }\widehat{\mathbf{D}}^{\star },\ c_{3}\
^{\shortparallel }\widehat{\mathbf{D}}^{\star },\ c_{4}\ ^{\shortparallel }%
\widehat{\mathbf{D}}^{\star })$ and $\ _{2}\widehat{\mathbf{D}}^{\star
}=(h_{1}\ \widehat{\mathbf{D}}^{\star },\ v_{2}\ \widehat{\mathbf{D}}^{\star
})$ are considered for respective nonholonomic dyadic horizontal and (co)
vertical splitting, which define s-connections adapted to respective
nonlinear s--connection structures $\ _{s}^{\shortparallel }\mathbf{%
N\rightarrow \ _{2}\mathbf{N}.}$\footnote{%
In brief, such formulas are with respective h1-v2-c3-c4 and h1-v2
decompositions.} Such canonical s-connection and d-connection satisfy the
metricity conditions (\ref{mcompnas}) and involve some "mixing shells"
nontrivial torsion coefficients of $\ ^{\shortparallel }\widehat{\mathcal{T}}%
^{\star },$ and $\ \widehat{\mathcal{T}}^{\star }.$ Pure shell torsion
coefficients vanish in s-adapted (co) frames $\ ^{\shortparallel }\mathbf{e}%
_{\alpha _{s}}\rightarrow \mathbf{e}_{\alpha _{2}}$ (\ref{nadapbdsc}) and $\
^{\shortparallel }\mathbf{e}^{\alpha _{s}}\rightarrow \mathbf{e}^{\alpha
_{2}}$ (\ref{nadapbdss}). The torsion s-tensor of any s-connection can be
defined in standard form as for any affine (linear) \ connection and
subjected to respective star deformations (see below some necessary
formulas; more details are presented in \cite{partner01,partner02}).

The R-flux s-adapted contributions are proportional to $\kappa $, can be
computed both on total phase space with nontrivial projections on spacetime
shells using respective s-adapted frames and symmetric s-metric
coefficients,
\begin{eqnarray}
\ _{[1]}^{\shortparallel }\widehat{\mathbf{\Gamma }}_{\star \alpha _{s}\beta
_{s}\mu _{s}} &=&\frac{1}{2}\overline{\mathcal{R}}_{\quad \mu _{s}}^{\xi
_{s}\tau _{s}}\ (\ \mathbf{^{\shortparallel }e}_{\xi _{s}}\ \mathbf{%
^{\shortparallel }e}_{\alpha _{s}}\ _{\star }^{\shortparallel }\mathbf{g}%
_{\beta _{s}\tau _{s}}+\ \mathbf{^{\shortparallel }e}_{\xi _{s}}\mathbf{%
^{\shortparallel }e}_{\beta _{s}}\ _{\star }^{\shortparallel }\mathbf{g}%
_{\alpha _{s}\tau _{s}}),\mbox{ for }s=1,2,3,4;  \label{aux311} \\
\ _{[1]}^{\shortparallel }\widehat{\mathbf{\Gamma }}_{\star \alpha _{2}\beta
_{2}\mu _{2}} &=&\frac{1}{2}\overline{\mathcal{R}}_{\quad \mu _{2}}^{\xi
_{s}\tau _{s}}\ (\ \mathbf{^{\shortparallel }e}_{\xi _{s}}\ \mathbf{%
^{\shortparallel }e}_{\alpha _{2}}\ _{\star }^{\shortparallel }\mathbf{g}%
_{\beta _{2}\tau _{s}}+\ \mathbf{^{\shortparallel }e}_{\xi _{s}}\mathbf{%
^{\shortparallel }e}_{\beta _{2}}\ _{\star }^{\shortparallel }\mathbf{g}%
_{\alpha _{2}\tau _{s}}).  \notag
\end{eqnarray}%
We note that formulas (\ref{aux311}) involve higher shells (with $s=3,4$)
contributions on spacetime shells $s=1,2$ even we work in s-adapted frames.
Using (\ref{nadapbdss}), we can define and compute in complete parametric
form the s-adapted coefficients of the star canonical s-connection $\
_{s}^{\shortparallel }\widehat{\mathbf{D}}^{\star }=\{\ ^{\shortparallel }%
\widehat{\mathbf{\Gamma }}_{\star \alpha _{s}\beta _{s}}^{\gamma _{s}}\}$
and respective spacetime components. For simplicity, we provide only the
formulas for spacetime 2+2 decompositions if they do not encode certain
higher shell mixing, $\widehat{\mathbf{\Gamma }}_{\star \alpha _{2}\beta
_{2}}^{\gamma _{2}}=\ \ _{[0]}\widehat{\mathbf{\Gamma }}_{\star \alpha
_{2}\beta _{2}}^{\gamma _{2}}+i\kappa \ _{[1]}\widehat{\mathbf{\Gamma }}%
_{\star \alpha _{2}\beta _{2}}^{\gamma _{2}}=(\ \widehat{L}_{\star
j_{1}k_{1}}^{i_{1}},\ \ \widehat{L}_{\star b_{2}\ k_{1}}^{a_{2}},\ \widehat{C%
}_{\star \ j_{1}c_{2}}^{i_{1}\ },\widehat{C}_{\star b_{2}c_{2}}^{a_{2}})$.
The nonsymmetric s-metric $\ _{\star }^{\shortparallel }\mathfrak{g}_{\alpha
_{s}\beta _{s}}$ (\ref{dmss1}) is subjected to such conditions%
\begin{eqnarray}
\ \mathbf{^{\shortparallel }e}^{\tau _{s}}\star _{s}\ _{\star
}^{\shortparallel }\mathfrak{g}_{\tau _{s}\gamma _{s}}\star _{s}\
^{\shortparallel }\widehat{\mathbf{\Gamma }}_{\star \alpha _{s}\beta
_{s}}^{\gamma _{s}} &=&\ \mathbf{^{\shortparallel }e}^{\tau _{s}}\star
_{s}(\ _{[0]}^{\shortparallel }\widehat{\mathbf{\Gamma }}_{\star \tau
_{s}\alpha _{s}\beta _{s}}+i\kappa \ _{[1]}^{\shortparallel }\widehat{%
\mathbf{\Gamma }}_{\star \tau _{s}\alpha _{s}\beta _{s}}),  \label{eqnasdmdc}
\\
\ \mathbf{^{\shortparallel }e}^{\tau _{2}}\star _{2}\ _{\star
}^{\shortparallel }\mathfrak{g}_{\tau _{2}\gamma _{2}}\star _{2}\
^{\shortparallel }\widehat{\mathbf{\Gamma }}_{\star \alpha _{2}\beta
_{2}}^{\gamma _{2}} &=&\ \mathbf{^{\shortparallel }e}^{\tau _{2}}\star
_{2}(\ _{[0]}^{\shortparallel }\widehat{\mathbf{\Gamma }}_{\star \tau
_{2}\alpha _{2}\beta _{2}}+i\kappa \ _{[1]}^{\shortparallel }\widehat{%
\mathbf{\Gamma }}_{\star \tau _{2}\alpha _{2}\beta _{2}}).  \notag
\end{eqnarray}%
In these formulas, the s-components of $\ _{[0]}^{\shortparallel }\widehat{%
\mathbf{\Gamma }}_{\star \tau _{s}\alpha _{s}\beta _{s}}$ are determined by
formulas (\ref{canhcs}) using the respective (inverse) symmetric star
metric, when $\ _{[1]}^{\shortparallel }\widehat{\mathbf{\Gamma }}_{\star
\alpha _{s}\beta _{s}\mu _{s}}$ and $\ _{[1]}^{\shortparallel }\widehat{%
\mathbf{\Gamma }}_{\star \alpha _{2}\beta _{2}\mu _{2}}$ (\ref{aux311})
involve respective s-compositions of $\ _{\star }^{\shortparallel }\mathbf{g}%
_{\beta _{s}\tau _{s}}$ and $_{\star }^{\shortparallel }\mathbf{g}_{\beta
_{2}\tau _{2}}.$

All definitions and formulas from \cite{blumenhagen16,aschieri17} can be
rewritten in s-adapted form if $\ ^{\shortparallel }\partial _{\alpha
}\rightarrow \ ^{\shortparallel }\mathbf{e}_{\alpha _{s}}$ and $\
^{\shortparallel }\partial _{\alpha _{2}}\rightarrow \ ^{\shortparallel }%
\mathbf{e}_{\alpha _{2}}$ as we prove in \cite{partner01,partner02}.
Similarly to (\ref{candistr}), we can compute s-adapted coefficients of
nonassociative star deformed LC-connections to respective canonical
s-connections,
\begin{equation}
\ _{s}^{\shortparallel }\widehat{\mathbf{D}}^{\star }=\ ^{\shortparallel
}\nabla ^{\star }+\ _{\star s}^{\shortparallel }\widehat{\mathbf{Z}}%
\mbox{
and }\ _{2}\widehat{\mathbf{D}}^{\star }=\ _{2}\nabla ^{\star }+\ _{\star 2}%
\widehat{\mathbf{Z}}.  \label{candistrnas}
\end{equation}%
In explicit form, the s-adapted coefficients of distortion s-tensor are some
algebraic functionals on s-adapted torsion $\ _{2}\mathcal{T}^{\star }=\{\
^{\shortparallel }\widehat{\mathbf{T}}_{\ \star \beta _{2}\gamma
_{2}}^{\alpha _{2}}\}$, for instance, $\ _{2\star }^{\shortparallel }%
\widehat{\mathbf{Z}}=\{\ ^{\shortparallel }\widehat{\mathbf{Z}}_{\ \star
\beta _{2}\gamma _{2}}^{\alpha _{2}}[\ ^{\shortparallel }\widehat{\mathbf{T}}%
_{\ \star \beta _{2}\gamma _{2}}^{\alpha _{2}}]\}.$ We omit such formulas in
this work because they are not used for constructing exact and parametric
solutions.

Using nonholonomic dyadic decompositions, we can describe equivalently in
coordinated frames or s-adapted frames any model of nonassociative phase
geometry and gravity determined by a nonsymmetric s-metric structure and
spacetime projections $\ _{\star }^{\shortparallel }\mathfrak{g}_{\alpha
_{s}\beta _{s}}\rightarrow \ _{\star }\mathfrak{g}_{\alpha _{2}\beta _{2}}.$
For any given s-connection and R-flux s-coefficients data, we can consider
only the respective symmetric star s-metrics $\ _{\star }^{\shortparallel }%
\mathbf{g}_{\alpha _{s}\beta _{s}}\rightarrow \ _{\star }\mathbf{g}_{\alpha
_{2}\beta _{2}}.$ Such nonassociative geometric models with (non) symmetric
s-metrics can be described equivalently both in terms of the star
LC-connections $\ ^{\shortparallel }\nabla ^{\star }\rightarrow $ $\nabla
^{\star }$ and/or the star canonical s-connections $\ _{s}^{\shortparallel }%
\widehat{\mathbf{D}}^{\star }\rightarrow \ _{2}\widehat{\mathbf{D}}^{\star
}. $ The main idea of the AFCDM is that for constructing exact and
parametric solutions of physically important systems of PDEs is more
convenient to work with nonassociative nonholonomic dyadic canonical
geometric data $(\ _{\star s}^{\shortparallel }\mathfrak{g},\
_{s}^{\shortparallel }\widehat{\mathbf{D}}^{\star })\rightarrow (\ _{\star 2}%
\mathfrak{g},\ _{2}\widehat{\mathbf{D}}^{\star }).$ We can redefine the
constructions in terms of star deformed LC-configurations using
s-distortions (\ref{candistrnas}) and extract nonassociative
LC-configurations imposing respective zero s-torsion conditions%
\begin{eqnarray}
\ _{\star s}^{\shortparallel }\widehat{\mathbf{Z}} &=&0,%
\mbox{ which is
equivalent to }\ _{s}^{\shortparallel }\widehat{\mathbf{D}}_{\mid \
_{s}^{\shortparallel }\widehat{\mathbf{T}}=0}^{\star }=\ ^{\shortparallel
}\nabla ^{\star }\mbox{ and }  \label{lccondnonass} \\
\ _{\star 2}\widehat{\mathbf{Z}} &=&0,\mbox{ which is
equivalent to }\ _{2}\widehat{\mathbf{D}}_{\mid \ _{2}\widehat{\mathbf{T}}%
=0}^{\star }=\ _{2}\nabla ^{\star }.  \notag
\end{eqnarray}%
There are large classes of nontrivial solutions with zero torsion
coefficients ($\ ^{\shortparallel }\widehat{\mathbf{T}}_{\ \star \beta
_{s}\gamma _{s}}^{\alpha _{s}}=0$ and $\ \widehat{\mathbf{T}}_{\ \star \beta
_{2}\gamma _{2}}^{\alpha _{2}}=0),$ and the respective coefficients of
geometric s-objects solve the equations (\ref{lccondnonass}). In general,
such solutions contain certain nonzero anholonomy coefficients computed on
any (dual) s-shell and on spacetime shells. This states that respective
(non) symmetric s-metrics can be written in local coordinate forms as
generic off-diagonal matrices. We can not diagonalize such matrices by
coordinate transforms on a finite spacetime or phase space region if the
integrability conditions are not satisfied, i.e. if there are nonzero
anholonomy coefficients.

\subsection{Parametric decompositions of nonassociative Riemann \& Ricci
s-tensors}

We re-formulate for spacetime shells with h1-v2 decompositions the formulas
from section 3.3 of \cite{partner02}.

\subsubsection{Star and parametric (2+2)-deformed canonical curvature
d-tenors}

The nonassociative canonical Riemann s-tensor can be computed in standard
(non) associative \ forms for the (star) canonical s-connection coefficients
$\ ^{\shortparallel }\widehat{\mathbf{\Gamma }}_{\star \alpha _{s}\beta
_{s}}^{\gamma _{s}}$ (\ref{eqnasdmdc}). For shells $s=1,2$ and using "hat"
s-adapted variables with mixing on certain $s=3,4$ indices, we have
\begin{eqnarray}
\mathbf{\mathbf{\mathbf{\mathbf{\ ^{\shortparallel }}}}}\widehat{\mathcal{%
\Re }}_{\quad \alpha _{2}\beta _{2}\gamma _{2}}^{\star \mu _{2}} &=&\mathbf{%
\mathbf{\mathbf{\mathbf{\ _{1}^{\shortparallel }}}}}\widehat{\mathcal{\Re }}%
_{\quad \alpha _{2}\beta _{2}\gamma _{2}}^{\star \mu _{2}}+\mathbf{\mathbf{%
\mathbf{\mathbf{\ _{2}^{\shortparallel }}}}}\widehat{\mathcal{\Re }}_{\quad
\alpha _{2}\beta _{2}\gamma _{2}}^{\star \mu _{2}},\mbox{ where }
\label{nadriemhopfcan} \\
\mathbf{\mathbf{\mathbf{\mathbf{\ _{1}^{\shortparallel }}}}}\widehat{%
\mathcal{\Re }}_{\quad \alpha _{2}\beta _{2}\gamma _{2}}^{\star \mu _{2}}
&=&\ \mathbf{e}_{\gamma _{2}}\mathbf{\ }\widehat{\Gamma }_{\star \alpha
_{2}\beta _{2}}^{\mu _{2}}-\ \mathbf{e}_{\beta _{2}}\mathbf{\ }\widehat{%
\Gamma }_{\star \alpha _{2}\gamma _{2}}^{\mu _{2}}+\mathbf{\
^{\shortparallel }\ }\widehat{\Gamma }_{\star \nu _{s}\tau _{s}}^{\mu
_{2}}\star _{s}(\delta _{\ \gamma _{2}}^{\tau _{s}}\mathbf{\
^{\shortparallel }}\widehat{\Gamma }_{\star \alpha _{2}\beta _{2}}^{\nu
_{s}}-\delta _{\ \beta _{2}}^{\tau _{s}}\mathbf{\ ^{\shortparallel }}%
\widehat{\Gamma }_{\star \alpha _{2}\gamma _{2}}^{\nu _{s}})+\mathbf{\
^{\shortparallel }}w_{\beta _{2}\gamma _{2}}^{\tau _{s}}\star _{s}\mathbf{\
^{\shortparallel }}\widehat{\Gamma }_{\star \alpha _{2}\tau _{s}}^{\mu _{2}},
\notag \\
\ _{2}^{\shortparallel }\widehat{\mathcal{\Re }}_{\quad \alpha _{2}\beta
_{2}\gamma _{2}}^{\star \mu _{2}} &=&i\kappa \ ^{\shortparallel }\widehat{%
\Gamma }_{\star \nu _{s}\tau _{s}}^{\mu _{2}}\star _{s}(\mathcal{R}_{\quad
\gamma _{2}}^{\tau _{s}\xi _{s}}\ \mathbf{^{\shortparallel }e}_{\xi _{s}}%
\mathbf{\ ^{\shortparallel }}\widehat{\Gamma }_{\star \alpha _{2}\beta
_{2}}^{\nu _{s}}-\mathcal{R}_{\quad \beta _{2}}^{\tau _{s}\xi _{s}}\ \mathbf{%
^{\shortparallel }e}_{\xi _{s}}\mathbf{\ ^{\shortparallel }}\widehat{\Gamma }%
_{\star \alpha _{2}\gamma _{s}}^{\nu _{s}}).  \notag
\end{eqnarray}%
These formulas are h1-v2 decompositions with s-adapted "hat" coefficients of
formulas (A.15), (38) and (39) defined in \cite{partner02}.

Parametric decompositions of the star canonical s-connections can be written
in s-form on all phase space shells :
\begin{equation*}
\ ^{\shortparallel }\widehat{\mathbf{\Gamma }}_{\star \alpha _{s}\beta
_{s}}^{\nu _{s}}=\ _{[0]}^{\shortparallel }\widehat{\mathbf{\Gamma }}_{\star
\alpha _{s}\beta _{s}}^{\nu _{s}}+i\kappa \ _{[1]}^{\shortparallel }\widehat{%
\mathbf{\Gamma }}_{\star \alpha _{s}\beta _{s}}^{\nu _{s}}=\
_{[00]}^{\shortparallel }\widehat{\Gamma }_{\ast \alpha _{s}\beta _{s}}^{\nu
_{s}}+\ _{[01]}^{\shortparallel }\widehat{\Gamma }_{\ast \alpha _{s}\beta
_{s}}^{\nu _{s}}(\hbar )+\ _{[10]}^{\shortparallel }\widehat{\Gamma }_{\ast
\alpha _{s}\beta _{s}}^{\nu _{s}}(\kappa )+\ _{[11]}^{\shortparallel }%
\widehat{\Gamma }_{\ast \alpha _{s}\beta _{s}}^{\nu _{s}}(\hbar \kappa
)+O(\hbar ^{2},\kappa ^{2},...),
\end{equation*}%
with respective formulas for $s=1,2.$ Using such parametric formulas for (%
\ref{nadriemhopfcan}), we compute%
\begin{equation*}
\mathbf{\mathbf{\mathbf{\mathbf{\ ^{\shortparallel }}}}}\widehat{\mathcal{%
\Re }}_{\quad \alpha _{2}\beta _{2}\gamma _{2}}^{\star \mu _{2}}=\mathbf{%
\mathbf{\mathbf{\mathbf{\ }}}}\ _{[00]}^{\shortparallel }\widehat{\mathcal{%
\Re }}_{\quad \alpha _{2}\beta _{2}\gamma _{2}}^{\star \mu _{2}}+\mathbf{%
\mathbf{\mathbf{\mathbf{\ }}}}\ _{[01]}^{\shortparallel }\widehat{\mathcal{%
\Re }}_{\quad \alpha _{2}\beta _{2}\gamma _{2}}^{\star \mu _{2}}(\hbar )+%
\mathbf{\mathbf{\mathbf{\mathbf{\ }}}}\ _{[10]}^{\shortparallel }\widehat{%
\mathcal{\Re }}_{\quad \alpha _{2}\beta _{2}\gamma _{2}}^{\star \mu
_{2}}(\kappa )+\ _{[11]}^{\shortparallel }\widehat{\mathcal{\Re }}_{\quad
\alpha _{2}\beta _{2}\gamma _{2}}^{\star \mu _{2}}(\hbar \kappa )+O(\hbar
^{2},\kappa ^{2},...).
\end{equation*}%
Such nonholonomic dyadic decompositions of (non) associative/ commutative
canonical Riemann s-tensors and their star deformations on $\mathcal{M}=%
\mathbf{T}_{s\shortparallel }^{\ast }\mathbf{V}$ and $\mathbf{V}$ with $%
s=1,2,$ can be priscribed for sun nonholonomic configrations when values $\
_{[00]}^{\shortparallel }\widehat{\mathcal{\Re }}_{\quad \alpha _{s}\beta
_{s}\gamma _{s}}^{\star \mu _{s}}\rightarrow \ _{[00]}^{\shortparallel }%
\widehat{\mathcal{\Re }}_{\quad \alpha _{2}\beta _{2}\gamma _{2}}^{\star \mu
_{2}}$ define "hat" variants of "not star deformed" curvature s-tensors for
the respective canonical s-connections $\ ^{\shortparallel }\widehat{\Gamma }%
_{\star \alpha _{s}\tau _{s}}^{\mu _{s}}\rightarrow \ ^{\shortparallel }%
\widehat{\Gamma }_{\star \alpha _{2}\tau _{2}}^{\mu _{2}}$ (\ref{twocon}).

\subsubsection{Parametric decomposition of star canonical Ricci s-tensors
and spacetime distortions}

Contracting on the fist and forth indices in formulas (\ref{nadriemhopfcan})
and using formulas with$\ ^{\shortparallel }\widehat{\Gamma }_{\star \alpha
_{s}\gamma _{s}}^{\mu _{s}}\rightarrow \ ^{\shortparallel }\widehat{\Gamma }%
_{\star \alpha _{2}\gamma _{2}}^{\mu _{2}},$ we define and compute the h1-v2
s-components of the the nonassociative canonical Ricci s-tensor,
\begin{eqnarray}
\ _{2}^{\shortparallel }\widehat{\mathcal{\Re }}ic^{\star } &=&\mathbf{%
\mathbf{\mathbf{\mathbf{\ ^{\shortparallel }}}}}\widehat{\mathbf{\mathbf{%
\mathbf{\mathbf{R}}}}}ic_{\alpha _{2}\beta _{2}}^{\star }\star _{2}(\
\mathbf{^{\shortparallel }e}^{\alpha _{2}}\otimes _{\star 2}\ \mathbf{%
^{\shortparallel }e}^{\beta _{2}}),\mbox{ where }  \label{driccicanonstar} \\
&&\mathbf{\mathbf{\mathbf{\mathbf{\ ^{\shortparallel }}}}}\widehat{\mathbf{%
\mathbf{\mathbf{\mathbf{R}}}}}ic_{\alpha _{2}\beta _{2}}^{\star }:=\
_{2}^{\shortparallel }\widehat{\mathcal{\Re }}ic^{\star }(\mathbf{\ }\
^{\shortparallel }\mathbf{e}_{\alpha _{2}},\ ^{\shortparallel }\mathbf{e}%
_{\beta _{2}})=\mathbf{\langle }\ \mathbf{\mathbf{\mathbf{\mathbf{\
^{\shortparallel }}}}}\widehat{\mathbf{\mathbf{\mathbf{\mathbf{R}}}}}ic_{\mu
_{2}\nu _{2}}^{\star }\star _{2}(\ \mathbf{^{\shortparallel }e}^{\mu
_{2}}\otimes _{\star _{2}}\ \mathbf{^{\shortparallel }e}^{\nu _{2}}),\mathbf{%
\mathbf{\ }\ ^{\shortparallel }\mathbf{e}}_{\alpha _{2}}\mathbf{\otimes
_{\star 2}\ ^{\shortparallel }\mathbf{e}}_{\beta _{2}}\mathbf{\rangle }%
_{\star _{2}}.  \notag
\end{eqnarray}%
The parametric decompositions of such s-adapted coefficients are expressed:
\begin{eqnarray}
\ ^{\shortparallel}\widehat{\mathbf{R}}ic_{\alpha _{2}\beta _{2}}^{\star }
&=& \ ^{\shortparallel}\widehat{\mathcal{\Re }}_{\quad \alpha _{2}\beta
_{2}\mu _{2}}^{\star \mu _{2}}  \notag \\
&=& \ _{[00]}^{\shortparallel }\widehat{\mathbf{R}}ic_{\alpha _{2}\beta
_{2}}^{\star }+ \ _{[01]}^{\shortparallel }\widehat{\mathbf{R}}ic_{\alpha
_{2}\beta _{2}}^{\star }(\hbar )+ \ _{[10]}^{\shortparallel }\widehat{%
\mathbf{\mathbf{\mathbf{\mathbf{R}}}}}ic_{\alpha _{2}\beta _{2}}^{\star
}(\kappa )+\mathbf{\mathbf{\mathbf{\mathbf{\ }}}}_{[11]}^{\shortparallel }%
\widehat{\mathbf{\mathbf{\mathbf{\mathbf{R}}}}}ic_{\alpha _{2}\beta
_{2}}^{\star }(\hbar \kappa )+O(\hbar ^{2},\kappa ^{2},...),\mbox{ for}
\notag \\
&& \ _{[00]}^{\shortparallel }\widehat{\mathbf{R}}ic_{\alpha _{2}\beta
_{2}}^{\star } = \ _{[00]}^{\shortparallel }\widehat{\mathcal{\Re }}_{\quad
\alpha _{2}\beta _{2}\mu _{2}}^{\star \mu _{2}}, \ _{[01]}^{\shortparallel }%
\mathbf{\mathbf{\mathbf{\mathbf{\widehat{\mathbf{\mathbf{\mathbf{\mathbf{R}}}%
}}}}}}ic_{\alpha _{2}\beta _{2}}^{\star }=\ _{[01]}^{\shortparallel }\mathbf{%
\mathbf{\mathbf{\mathbf{\widehat{\mathcal{\Re }}}}}}_{\quad \alpha _{2}\beta
_{2}\mu _{2}}^{\star \mu _{2}},\ _{[10]}^{\shortparallel }\mathbf{\mathbf{%
\mathbf{\mathbf{\widehat{\mathbf{\mathbf{\mathbf{\mathbf{R}}}}}}}}}%
ic_{\alpha _{2}\beta _{2}}^{\star }=\ _{[10]}^{\shortparallel }\mathbf{%
\mathbf{\mathbf{\mathbf{\widehat{\mathcal{\Re }}}}}}_{\quad \alpha _{2}\beta
_{2}\mu _{2}}^{\star \mu _{2}},  \notag \\
&& \ _{[11]}^{\shortparallel }\widehat{\mathbf{\mathbf{\mathbf{\mathbf{R}}}}}%
ic_{\alpha _{2}\beta _{2}}^{\star }=\ _{[11]}^{\shortparallel }\mathbf{%
\mathbf{\mathbf{\mathbf{\widehat{\mathcal{\Re }}}}}}_{\quad \alpha _{2}\beta
_{2}\mu _{2}}^{\star \mu _{2}}.  \label{driccicanonstar1}
\end{eqnarray}%
In appendix A.3.3 of \cite{partner02}, it is shown how to compute parametric
[00],[01],[10], and [11] canonical Ricci s-coefficients for all
s-coefficients from (\ref{driccicanonstar1}).

We can nonholonomic dyadic distributions with $\ _{[00]}^{\shortparallel}
\widehat{\mathbf{R}}ic_{\alpha _{2}\beta _{2}}^{\star}= \widehat{\mathbf{R}}
_{\ \alpha _{2}\beta _{2}}$ (\ref{candricci}) but other parametric $%
[01,10,11]$ when $\left\lceil \hbar ,\kappa \right\rceil $ components
contain nonassociative and noncommutative contributions from star product
deformations on phase space and which can be real or complex ones. In
result, we can express the star s-deformed Ricci s-tensor in parametric
h1-v2- form
\begin{eqnarray}
\ _{2}^{\shortparallel }\widehat{\mathcal{R}}ic^{\star } &=&\{\
^{\shortparallel }\widehat{\mathbf{R}}_{\ \star \beta _{2}\gamma _{2}}\}=\
_{2}^{\shortparallel }\widehat{\mathcal{R}}ic+\ _{2}^{\shortparallel }%
\widehat{\mathcal{K}}ic\left\lceil \hbar ,\kappa \right\rceil =\{\widehat{%
\mathbf{R}}_{\ \beta _{2}\gamma _{2}}+\ ^{\shortparallel }\widehat{\mathbf{K}%
}_{\ \beta _{2}\gamma _{2}}\left\lceil \hbar ,\kappa \right\rceil \},%
\mbox{
where }  \label{paramsricci} \\
\ _{2}^{\shortparallel }\widehat{\mathcal{K}}ic &=&\{\ ^{\shortparallel }%
\widehat{\mathbf{K}}_{\ \beta _{2}\gamma _{2}}\left\lceil \hbar ,\kappa
\right\rceil =\mathbf{\mathbf{\mathbf{\mathbf{\ \ }}}}_{[01]}^{%
\shortparallel }\widehat{\mathbf{\mathbf{\mathbf{\mathbf{R}}}}}ic_{\beta
_{2}\gamma _{2}}^{\star }+\mathbf{\mathbf{\mathbf{\mathbf{\ \ }}}}%
_{[10]}^{\shortparallel }\widehat{\mathbf{\mathbf{\mathbf{\mathbf{R}}}}}%
ic_{\beta _{2}\gamma _{2}}^{\star }+\mathbf{\mathbf{\mathbf{\mathbf{\ \ }}}}%
_{[11]}^{\shortparallel }\widehat{\mathbf{\mathbf{\mathbf{\mathbf{R}}}}}%
ic_{\beta _{2}\gamma _{2}}^{\star }\}.  \notag
\end{eqnarray}%
Such formulas encode nonassociative parametric deformations of the canonical
Ricci s-tenor which, in general, are computed for all shells $s=1,2,3,4.$ We
can chose an associative and commutative solution $\ ^{\shortparallel }%
\mathbf{g}_{\alpha _{s}\beta _{s}}$ (\ref{sdm}) of nonholonomic dyadic
vacuum Einstein equations (\ref{cnsveinst1}). Then, we can compute $\
_{\star }^{\shortparallel }\mathbf{g}_{\alpha _{s}\beta _{s}}$ (\ref{ssdm})
\ and R-flux and star canonical s-deformations (\ref{dmss1}) to $\ _{\star
}^{\shortparallel }\mathfrak{g}_{\tau _{s}\gamma _{s}}$ (\ref{nssdm}).
Projections to h1-v2 shells can be considered for $\ _{\star
}^{\shortparallel }\mathfrak{g}_{\tau _{2}\gamma _{2}}$ (\ref{dmss1a}). For
such computations, we consider $\ _{2}^{\shortparallel }\widehat{\mathcal{K}}%
ic=\{\ ^{\shortparallel }\widehat{\mathbf{K}}_{\ \beta _{2}\gamma
_{2}}\left\lceil \hbar ,\kappa \right\rceil \},$ when nonassociative
distortions of Ricci s-tensors of type $\ ^{\shortparallel }\widehat{\mathbf{%
K}}_{\ \beta _{s}\gamma _{s}}\left\lceil \hbar ,\kappa \right\rceil $ for $%
s=3,4$ are encoded in (\ref{paramsricci}) using certain effective sources
and/or cosmological constants. This is possible because of nonlinear
symmetries relating generating functions and generating (effective) sources
on total phase spaces, see details in section 5.4 of \cite{partner02}. In
this work, we shall search for solutions of nonassociative vacuum Einstein
equations with R-flux deformations of Lorentz spacetimes involving
parametric decompositions of type (\ref{paramsricci}).

\subsection{Nonassociative vacuum Einstein equations with real R-flux source}

The goal of this subsection is to show how star deformations of nonholonomic
2+2 vacuum Einstein equations (\ref{cnsveinst1}) transform into systems of
nonlinear PDEs with effective R-flux sources computed in coordinate bases in
\cite{aschieri17}. We consider the nonholonomic dyadic formalism, elaborated
in \cite{partner02} for nonassociative geometry in s-adapted form for
R-flux, and star modifications of vacuum phase space gravitational equations
with (effective) cosmological constant.

\subsubsection{Nonassociative R-flux spacetime deformations with (non)
symmetric s-metrics}

A nonsymmetric star s-metric can be expressed in the form%
\begin{equation}
\ _{\star }^{\shortparallel }\mathfrak{g}_{\alpha _{s}\beta _{s}}=\ _{\star
}^{\shortparallel }\mathfrak{g}_{\alpha _{s}\beta _{s}}^{[0]}+\ _{\star
}^{\shortparallel }\mathfrak{g}_{\alpha _{s}\beta _{s}}^{[1]}(\kappa )=\
_{\star }^{\shortparallel }\mathfrak{\check{g}}_{\alpha _{s}\beta _{s}}+\
_{\star }^{\shortparallel }\mathfrak{a}_{\alpha _{s}\beta _{s}},
\label{aux40bb}
\end{equation}%
where $\ _{\star }^{\shortparallel }\mathfrak{\check{g}}_{\alpha _{2}\beta
_{2}}$ is the symmetric part and $\ _{\star }^{\shortparallel}\mathfrak{a}%
_{\alpha _{s}\beta _{s}}$ is the anti-symmetric part of star and R-flux
deformations of spacetime s-metrics for $s=1,2$ (for total phase space
computations, see section 4.1.1 and formulas (17), (A.20) in \cite{partner02}%
). We have
\begin{eqnarray}
\ _{\star }^{\shortparallel }\mathfrak{\check{g}}_{\alpha _{2}\beta _{2}}
&:=&\frac{1}{2}(\ _{\star }^{\shortparallel }\mathfrak{g}_{\alpha _{2}\beta
_{2}}+\ _{\star }^{\shortparallel }\mathfrak{g}_{\beta _{2}\alpha _{2}})=\
_{\star }^{\shortparallel }\mathbf{g}_{\alpha _{2}\beta _{2}}-\frac{i\kappa
}{2}\left( \overline{\mathcal{R}}_{\quad \beta _{2}}^{\tau _{s}\xi _{s}}\
\mathbf{^{\shortparallel }e}_{\xi _{s}}\ _{\star }^{\shortparallel }\mathbf{g%
}_{\tau _{s}\alpha _{2}}+\overline{\mathcal{R}}_{\quad \alpha _{2}}^{\tau
_{s}\xi _{s}}\ \mathbf{^{\shortparallel }e}_{\xi _{s}}\ _{\star
}^{\shortparallel }\mathbf{g}_{\beta _{2}\tau _{s}}\right)  \label{aux40b} \\
&=&\ _{\star }^{\shortparallel }\mathfrak{\check{g}}_{\alpha _{2}\beta
_{2}}^{[0]}+\ _{\star }^{\shortparallel }\mathfrak{\check{g}}_{\alpha
_{2}\beta _{2}}^{[1]}(\kappa ),  \notag \\
&&\mbox{ for }\ _{\star }^{\shortparallel }\mathfrak{\check{g}}_{\alpha
_{2}\beta _{s}}^{[0]}=\ _{\star }^{\shortparallel }\mathbf{g}_{\alpha
_{2}\beta _{2}}\mbox{ and }\ _{\star }^{\shortparallel }\mathfrak{\check{g}}%
_{\alpha _{2}\beta _{2}}^{[1]}(\kappa )=-\frac{i\kappa }{2}\left( \overline{%
\mathcal{R}}_{\quad \beta _{2}}^{\tau _{s}\xi _{s}}\ \mathbf{%
^{\shortparallel }e}_{\xi _{s}}\ _{\star }^{\shortparallel }\mathbf{g}_{\tau
_{s}\alpha _{2}}+\overline{\mathcal{R}}_{\quad \alpha _{2}}^{\tau _{s}\xi
_{s}}\ \mathbf{^{\shortparallel }e}_{\xi _{s}}\ _{\star }^{\shortparallel }%
\mathbf{g}_{\beta _{2}\tau _{s}}\right) ;  \notag \\
\ _{\star }^{\shortparallel }\mathfrak{a}_{\alpha _{2}\beta _{2}}&:= &\frac{1%
}{2}(\ _{\star }^{\shortparallel }\mathfrak{g}_{\alpha _{2}\beta _{2}}-\
_{\star }^{\shortparallel }\mathfrak{g}_{\beta _{2}\alpha _{2}})=\frac{%
i\kappa }{2}\left( \overline{\mathcal{R}}_{\quad \beta _{2}}^{\tau _{s}\xi
_{s}}\ \mathbf{^{\shortparallel }e}_{\xi _{s}}\ _{\star }^{\shortparallel }%
\mathbf{g}_{\tau _{s}\alpha _{2}}-\overline{\mathcal{R}}_{\quad \alpha
_{2}}^{\tau _{s}\xi _{s}}\ \mathbf{^{\shortparallel }e}_{\xi _{s}}\ _{\star
}^{\shortparallel }\mathbf{g}_{\beta _{2}\tau _{s}}\right)  \notag \\
&=&\ _{\star }^{\shortparallel }\mathfrak{a}_{\alpha _{2}\beta
_{2}}^{[1]}(\kappa )=\frac{1}{2}(\ _{\star }^{\shortparallel }\mathfrak{g}%
_{\alpha _{2}\beta _{2}}^{[1]}(\kappa )-\ _{\star }^{\shortparallel }%
\mathfrak{g}_{\beta _{2}\alpha _{2}}^{[1]}(\kappa )).  \label{aux40aa}
\end{eqnarray}%
We chose in such formulas $\ _{\star }^{\shortparallel }\mathfrak{a}_{\alpha
_{s}\beta _{s}}^{[0]}=0$ for nonassociative star deformations of commutative
theories with symmetric metrics. Here we note that for nonassociative
geometric models we have to apply a more sophisticate procedure for
computing inverse metrics and/or s-metrics, see respective details in \cite%
{aschieri17} and \cite{partner02}. The symmetric and nonsymmetric parts of a
nonsymmetric inverse s-metric with parametric decompositions can be written
similarly to
\begin{equation*}
\ _{\star }^{\shortparallel }\mathfrak{g}^{\alpha _{s}\beta _{s}}=\ _{\star
}^{\shortparallel }\mathfrak{\check{g}}^{\alpha _{s}\beta _{s}}+\ _{\star
}^{\shortparallel }\mathfrak{a}^{\alpha _{s}\beta _{s}},
\end{equation*}%
when, in general, $\ _{\star }^{\shortparallel }\mathfrak{\check{g}}^{\alpha
_{s}\beta _{s}}$ is not the inverse to $\ _{\star }^{\shortparallel }%
\mathfrak{\check{g}}_{\alpha _{s}\beta _{s}}$ and $\ _{\star
}^{\shortparallel }\mathfrak{a}^{\alpha _{s}\beta _{s}}$ is not inverse to $%
\ _{\star }^{\shortparallel }\mathfrak{a}_{\alpha _{s}\beta _{s}}$.

\subsubsection{Vacuum gravitational equations for star deformed canonical
s-connections}

In general, a (non) associative canonical Ricci s-tensor $\mathbf{\mathbf{%
\mathbf{\mathbf{\ ^{\shortparallel }}}}}\widehat{\mathbf{\mathbf{\mathbf{%
\mathbf{R}}}}}ic_{\mu _{s}\nu _{s}}^{\star }$ is not symmetric. We can
define and compute the nonassociative nonholonomic canonical Ricci scalar
curvature following formulas%
\begin{eqnarray}
\ _{s}^{\shortparallel }\widehat{\mathbf{R}}sc^{\star }:= &&\ _{\star
}^{\shortparallel }\mathfrak{g}^{\mu _{s}\nu _{s}}\mathbf{\mathbf{\mathbf{%
\mathbf{\ ^{\shortparallel }}}}}\widehat{\mathbf{\mathbf{\mathbf{\mathbf{R}}}%
}}ic_{\mu _{s}\nu _{s}}^{\star }=\left( \ _{\star }^{\shortparallel }%
\mathfrak{\check{g}}^{\mu _{s}\nu _{s}}+\ _{\star }^{\shortparallel }%
\mathfrak{a}^{\mu _{s}\nu _{s}}\right) \left( \mathbf{\mathbf{\mathbf{%
\mathbf{\ ^{\shortparallel }}}}}\widehat{\mathbf{\mathbf{\mathbf{\mathbf{R}}}%
}}ic_{(\mu _{s}\nu _{s})}^{\star }+\mathbf{\mathbf{\mathbf{\mathbf{\
^{\shortparallel }}}}}\widehat{\mathbf{\mathbf{\mathbf{\mathbf{R}}}}}%
ic_{[\mu _{s}\nu _{s}]}^{\star }\right) =\ _{s}^{\shortparallel }\widehat{%
\mathbf{\mathbf{\mathbf{\mathbf{R}}}}}ss^{\star }+\ _{s}^{\shortparallel }%
\widehat{\mathbf{\mathbf{\mathbf{\mathbf{R}}}}}sa^{\star },  \notag \\
&&\mbox{ where }\ _{s}^{\shortparallel }\widehat{\mathbf{\mathbf{\mathbf{%
\mathbf{R}}}}}ss^{\star }=:\ _{\star }^{\shortparallel }\mathfrak{\check{g}}%
^{\mu _{s}\nu _{s}}\mathbf{\mathbf{\mathbf{\mathbf{\ ^{\shortparallel }}}}}%
\widehat{\mathbf{\mathbf{\mathbf{\mathbf{R}}}}}ic_{(\mu _{s}\nu
_{s})}^{\star }\mbox{ and }\ _{s}^{\shortparallel }\widehat{\mathbf{\mathbf{%
\mathbf{\mathbf{R}}}}}sa^{\star }:=\ _{\star }^{\shortparallel }\mathfrak{a}%
^{\mu _{s}\nu _{s}}\mathbf{\mathbf{\mathbf{\mathbf{\ ^{\shortparallel }}}}}%
\widehat{\mathbf{\mathbf{\mathbf{\mathbf{R}}}}}ic_{[\mu _{s}\nu
_{s}]}^{\star },  \label{ricciscsymnonsym} \\
&&\mbox{ for }\ ^{\shortparallel }\widehat{\mathbf{R}}ic_{\mu _{s}\nu
_{s}}^{\star }=\mathbf{\mathbf{\mathbf{\mathbf{\ ^{\shortparallel }}}}}%
\widehat{\mathbf{\mathbf{\mathbf{\mathbf{R}}}}}ic_{(\mu _{s}\nu
_{s})}^{\star }+\mathbf{\mathbf{\mathbf{\mathbf{\ ^{\shortparallel }}}}}%
\widehat{\mathbf{\mathbf{\mathbf{\mathbf{R}}}}}ic_{[\mu _{s}\nu
_{s}]}^{\star },  \notag
\end{eqnarray}%
with respective symmetrization and anti-symmetrization with multiple $1/2.$

For total phase spaces, such constructions are provided in details in
section 4 of \cite{partner02} for the nonassociative vacuum Einstein
equations of a nonassociative star s-metric $\ _{\star }^{\shortparallel }%
\mathfrak{g}_{\alpha _{s}\beta _{s}}$ (\ref{aux40bb}) defined by s-adapted
star deformations using the convention (\ref{conv2s}) and formulas (\ref%
{twoconsstar}) for the nonholonomic associative Einstein equations (\ref%
{cnsveinst1}). Respective systems of nonlinear PDEs can be written in a form
emphasizing explicitly the symmetric and nonsymmetric components of
s-metric,
\begin{equation*}
\ _{\star }^{\shortparallel }\mathfrak{\check{g}}_{\mu _{s}\nu _{s}}(\
_{s}^{\shortparallel }\lambda +\frac{1}{2}\ _{s}^{\shortparallel }\widehat{%
\mathbf{\mathbf{\mathbf{\mathbf{R}}}}}sc^{\star })=\mathbf{\mathbf{\mathbf{%
\mathbf{\ ^{\shortparallel }}}}}\widehat{\mathbf{\mathbf{\mathbf{\mathbf{R}}}%
}}ic_{(\mu _{s}\nu _{s})}^{\star }\mbox{ and }\ _{\star }^{\shortparallel }%
\mathfrak{a}_{\mu _{s}\nu _{s}}(\ _{s}^{\shortparallel }\lambda +\frac{1}{2}%
\ _{s}^{\shortparallel }\widehat{\mathbf{\mathbf{\mathbf{\mathbf{R}}}}}%
sc^{\star })=\mathbf{\mathbf{\mathbf{\mathbf{\ ^{\shortparallel }}}}}%
\widehat{\mathbf{\mathbf{\mathbf{\mathbf{R}}}}}ic_{[\mu _{s}\nu
_{s}]}^{\star }.
\end{equation*}%
From the second system of equations, we observe that if we impose the
nonholonomic conditions that
\begin{equation}
(\ _{s}^{\shortparallel }\lambda +\frac{1}{2}\ _{s}^{\shortparallel }%
\widehat{\mathbf{\mathbf{\mathbf{\mathbf{R}}}}}sc^{\star })=0\mbox{ and }%
\mathbf{\mathbf{\mathbf{\mathbf{\ ^{\shortparallel }}}}}\widehat{\mathbf{%
\mathbf{\mathbf{\mathbf{R}}}}}ic_{[\mu _{s}\nu _{s}]}^{\star }=0
\label{zeroeffsource}
\end{equation}%
the s-adapted R-fluxes generate decoupled nonsymmetric components of
s--metrics $\ _{\star }^{\shortparallel }\mathfrak{a}_{\mu _{s}\nu _{s}}[%
\overline{\mathcal{R}}_{\quad \beta _{s}}^{\tau _{s}\xi _{s}}\ \mathbf{%
^{\shortparallel }e}_{\xi _{s}}\ _{\star }^{\shortparallel }\mathbf{g}_{\tau
_{s}\alpha _{s}}]$ (\ref{aux40aa}) for arbitrary symmetric solutions $\
_{\star }^{\shortparallel }\mathfrak{\check{g}}_{\mu _{s}\nu _{s}}=\ _{\star
}^{\shortparallel }\mathfrak{g}_{\mu _{s}\nu _{s}}[\ _{\star
}^{\shortparallel }\mathbf{g}_{\beta _{s}\tau _{s}},\overline{\mathcal{R}}%
_{\quad \beta _{s}}^{\tau _{s}\xi _{s}}\ \mathbf{^{\shortparallel }e}_{\xi
_{s}}\ _{\star }^{\shortparallel }\mathbf{g}_{\tau _{s}\alpha _{s}}]$ of (%
\ref{aux40b}), where respective s--coefficients are constructed as
functionals. The symmetric part can taken to be a solution of
\begin{equation}
\mathbf{\mathbf{\mathbf{\mathbf{\ ^{\shortparallel }}}}}\widehat{\mathbf{%
\mathbf{\mathbf{\mathbf{R}}}}}ic_{\mu _{s}\nu _{s}}^{\star }=\
_{s}^{\shortparallel }\lambda \mathbf{\mathbf{\mathbf{\mathbf{\ }}}}\
_{\star }^{\shortparallel }\mathfrak{g}_{\alpha _{s}\beta _{s}},
\label{nonassocdeinst2aa}
\end{equation}%
which is a star s-adapted deformation of (\ref{cnsveinst1}). The symmetric
solutions of such PDEs of zero order on parameters $\hbar $ and $\kappa $
can be used as prime ones for generating general (non) symmetric solutions
with parametric dependence in nonassociative gravity and geometric flow
theories, see next section.

We can use formulas (\ref{ricciscsymnonsym}) and impose nonholonomic
constraints
\begin{equation}
\mathbf{\mathbf{\mathbf{\mathbf{\ ^{\shortparallel }}}}}\lambda +\frac{1}{2}%
\mathbf{\mathbf{\mathbf{\mathbf{\ _{s}^{\shortparallel }}}}}\widehat{\mathbf{%
\mathbf{\mathbf{\mathbf{R}}}}}sc^{\star }=\mathbf{\mathbf{\mathbf{\mathbf{\
^{\shortparallel }}}}}\lambda +\ ^{\shortparallel }\mathbf{K}_{\ \alpha
_{s}}^{\alpha _{s}}[\hbar \kappa \overline{\mathcal{R}}]~=0,
\label{nonhconstr1a}
\end{equation}%
encoding the contributions from shells $s=3,4$ into an effective
cosmological constant $\ ^{\shortparallel }\lambda =\ _{s}^{\shortparallel
}\lambda \mathbf{\mathbf{\mathbf{\mathbf{\ }}}}$\ (for any $s),$ when there
are also nontrivial R-flux effective sources on shells $s=1,2;$ for
nontrivial quasi-stationary configurations with prescribed effective sources
\begin{eqnarray}
\ ^{\shortparallel }\mathbf{K}_{\ \beta _{s}}^{\alpha _{s}} &=&\{\
^{\shortparallel }\mathcal{K}_{\ j_{1}}^{i_{1}}=[\ \ _{1}^{\shortparallel
}\Upsilon (x^{k_{1}})+\ _{1}^{\shortparallel }\mathbf{K}(\kappa
,x^{k_{1}})]\delta _{j_{1}}^{i_{1}},\ ^{\shortparallel }\mathcal{K}_{\
j_{2}}^{i_{2}}=[\ \ _{2}^{\shortparallel }\Upsilon (x^{k_{1}},x^{3})+\
_{2}^{\shortparallel }\mathbf{K}(\kappa ,x^{k_{1}},x^{3})]\delta
_{b_{2}}^{a_{2}},  \label{ansatzsourchv} \\
&&\ ^{\shortparallel }\mathcal{K}_{\ a_{3}}^{b_{3}}=[\ \
_{3}^{\shortparallel }\Upsilon (x^{k_{2}},\ ^{\shortparallel }p_{6})+\
_{3}^{\shortparallel }\mathbf{K}(x^{k_{2}},\ ^{\shortparallel }p_{6})]\
\delta _{a_{3}}^{b_{3}},\ ^{\shortparallel }\mathcal{K}_{\ a_{4}}^{b_{4}}=[\
\ _{4}^{\shortparallel }\Upsilon (x^{k_{3}},\ ^{\shortparallel }p_{8})+\
_{4}^{\shortparallel }\mathbf{K}(x^{k_{3}},\ ^{\shortparallel }p_{8})]\delta
_{a_{4}}^{b_{4}}\},  \notag \\
&&\mbox{ where }\ ^{\shortparallel }\mathbf{K}_{\ j_{2}k_{2}}=-\
_{[11]}^{\shortparallel }\widehat{\mathbf{R}}ic_{j_{2}k_{2}}^{\star
}(x^{k_{1}},x^{3})\mbox{ as in formula (77) of \cite{partner02}  },  \notag
\\
&&\mbox{ for }\mathbf{g}_{j_{2}k_{2}}=%
\{g_{1}(x^{k_{1}}),g_{2}(x^{k_{1}}),g_{3}(x^{k_{1}},x^{3}),g_{4}(x^{k_{1}},x^{3})\}.
\notag
\end{eqnarray}%
In this work, we consider only real spacetime nonholonomic star R-flux
deformations characterize by star s-deformed Ricci s-tensor in parametric
h1-v2- form (\ref{paramsricci}). In such models, the higher shell $s=3,4$
dynamics with respective data for $\ \ _{s}^{\shortparallel }\Upsilon $ and $%
\ _{s}^{\shortparallel }\mathbf{K}$ are encoded into effective $\mathbf{%
\mathbf{\mathbf{\mathbf{\ ^{\shortparallel }}}}}\lambda $ from (\ref%
{nonhconstr1a}), which can be prescribed in order to have certain
compatibility with some experimental data or higher order theoretical
computations.

The system of real spacetime s-adapted spacetime projections of parametric
nonassociative gravitational equations (\ref{nonassocdeinst2aa}) with
sources (\ref{ansatzsourchv}) transforms into%
\begin{eqnarray}
\ ^{\shortparallel }\widehat{\mathbf{R}}_{\ i_{2}j_{2}}~ &=&\
^{\shortparallel }\mathbf{K}_{_{i_{2}j_{2}}}[\hbar \kappa \overline{\mathcal{%
R}}]~,\mbox{ where }  \label{cannonsymparamc2hv} \\
&&\ ^{\shortparallel }\mathbf{K}_{~i_{2}}^{j_{2}}=[~_{1}^{\shortparallel }%
\mathcal{K}(\kappa ,x^{k_{1}})\delta _{i_{1}}^{j_{1}},~_{2}^{\shortparallel }%
\mathcal{K}(\kappa ,x^{k_{1}},x^{3})\delta _{b_{2}}^{a_{2}}].  \notag
\end{eqnarray}%
To solve such systems of nonlinear PDEs with parametric sources we can apply
AFCDM method for constructing exact and parametric solutions of dimensions
2+2. For instance, we can construct (off-) diagonal black hole and
cosmological solutions following the technique summarized in \cite%
{bubuianu17}, see references therein. Prescribing any values for generating
sources $\ _{s}^{\shortparallel }\mathcal{K}=\left[ \ _{1}^{\shortparallel }%
\mathcal{K},\ _{2}^{\shortparallel }\mathcal{K}\right] ,$ for $s=1,2,$ in (%
\ref{cannonsymparamc2hv}), we can decouple and solve in certain general
forms such systems of nonlinear PDEs.

The main difference of this paper from our previous works on (non)
commutative MGTs and geometric flow models on Lorentz manifolds and
projected is that in this article we consider a special type of effective
parametric sources determined by star R-flux deformations with
\begin{eqnarray}
\ ^{\shortparallel }\mathbf{K}_{\ j_{2}k_{2}} &=&-\ _{[11]}^{\shortparallel }%
\widehat{\mathbf{R}}ic_{j_{2}k_{2}}^{\star }(x^{k_{1}},x^{3})=-\frac{\hbar
\kappa }{2}\overline{\mathcal{R}}_{\quad \quad \quad }^{n+o_{2}\ n+q_{2}\
~n+l_{2}}\{\ ^{\shortparallel }\mathbf{e}_{i_{2}}[(\ ^{\shortparallel }%
\mathbf{e}_{o_{2}}~^{\shortparallel }\mathbf{g}^{i_{2}m_{2}})(\
^{\shortparallel }\mathbf{e}_{q_{2}}\ ^{\shortparallel }\mathbf{g}%
_{m_{2}r_{2}})(\ ^{\shortparallel }\mathbf{e}_{l_{2}}\ ^{\shortparallel }%
\widehat{\mathbf{\Gamma }}_{\ j_{2}k_{2}}^{r_{2}})]  \notag \\
&&-\ ^{\shortparallel }\mathbf{e}_{k_{2}}[(\ ^{\shortparallel }\mathbf{e}%
_{o_{2}}~^{\shortparallel }\mathbf{g}^{i_{2}m_{2}})(\ ^{\shortparallel }%
\mathbf{e}_{q_{2}}\ ^{\shortparallel }\mathbf{g}_{m_{2}r_{2}})(\
^{\shortparallel }\mathbf{e}_{l_{2}}\ ^{\shortparallel }\widehat{\mathbf{%
\Gamma }}_{\ j_{2}i_{2}}^{r_{2}})]  \label{realrflux} \\
&&+(\ ^{\shortparallel }\mathbf{e}_{l_{2}}~^{\shortparallel }\mathbf{g}%
_{q_{2}r_{2}})[\ ^{\shortparallel }\mathbf{e}_{o_{2}}(\ ~^{\shortparallel }%
\mathbf{g}^{i_{2}q_{2}}\ ^{\shortparallel }\widehat{\mathbf{\Gamma }}_{\
i_{2}k_{2}}^{m_{2}})(\ ^{\shortparallel }\mathbf{e}_{q_{2}}\
^{\shortparallel }\widehat{\mathbf{\Gamma }}_{j_{2}m_{2}}^{r_{2}})-\
^{\shortparallel }\mathbf{e}_{o_{2}}(\ ~^{\shortparallel }\mathbf{g}%
^{i_{2}q_{2}}\ ^{\shortparallel }\widehat{\mathbf{\Gamma }}_{\
i_{2}m_{2}}^{m_{2}})(\ ^{\shortparallel }\mathbf{e}_{q_{2}}\
^{\shortparallel }\widehat{\mathbf{\Gamma }}_{j_{2}k_{2}}^{r_{2}})  \notag \\
&&+\left( \ ^{\shortparallel }\widehat{\mathbf{\Gamma }}_{\
j_{2}i_{2}}^{m_{2}}\ ^{\shortparallel }\mathbf{e}_{o_{2}}(\
~^{\shortparallel }\mathbf{g}^{i_{2}r_{2}})-\ ^{\shortparallel }\mathbf{e}%
_{o_{2}}(\ ^{\shortparallel }\widehat{\mathbf{\Gamma }}_{\
j_{2}i_{2}}^{m_{2}})\ ~^{\shortparallel }\mathbf{g}^{i_{2}r_{2}}\right) (\
^{\shortparallel }\mathbf{e}_{q_{2}}\ ^{\shortparallel }\widehat{\mathbf{%
\Gamma }}_{\ m_{2}k_{2}}^{r_{2}})  \notag \\
&&-\left( \ ^{\shortparallel }\widehat{\mathbf{\Gamma }}_{\
j_{2}k_{2}}^{m_{2}}\ ^{\shortparallel }\mathbf{e}_{o_{2}}(\
~^{\shortparallel }\mathbf{g}^{i_{2}r_{2}})-\ ^{\shortparallel }\mathbf{e}%
_{o_{2}}(\ ^{\shortparallel }\widehat{\mathbf{\Gamma }}_{\
j_{2}k_{2}}^{m_{2}})\ ~^{\shortparallel }\mathbf{g}^{i_{2}r_{2}}\right) (\
^{\shortparallel }\mathbf{e}_{q_{2}}\ ^{\shortparallel }\widehat{\mathbf{%
\Gamma }}_{\ m_{2}i_{2}}^{r_{2}})]\}.  \notag
\end{eqnarray}%
as we computed in formulas (77) of \cite{partner02}. In these formulas, $\
^{\shortparallel }\widehat{\mathbf{\Gamma }}_{\ j_{2}k_{2}}^{r_{2}}$ are
computed as in (\ref{canhcs}) for
\begin{eqnarray}
\ _{\star }^{\shortparallel }\mathbf{g}_{\beta _{s}\gamma _{s}} &\simeq &\
^{\shortparallel }\mathbf{g}_{\beta _{s}\gamma _{s}}=(\mathbf{g}%
_{i_{2}j_{2}}(x^{k_{2}}),1,1,1,-1);\ ^{\shortparallel }\mathbf{e}_{o_{2}}=(%
\mathbf{e}_{i_{2}},\ ^{\shortparallel }\mathbf{e}_{a_{3}}=\ ^{\shortparallel
}\mathbf{\partial }_{a_{3}},\ ^{\shortparallel }\mathbf{e}_{a_{4}}=\
^{\shortparallel }\mathbf{\partial }_{a_{4}},);  \label{aux40ss} \\
\ ^{\shortparallel }\widehat{\mathbf{R}}_{\ \beta _{s}\gamma _{s}}~ &=&\
_{[0]}^{\shortparallel }\Upsilon _{_{\beta _{s}\gamma _{s}}}=\mathbf{\mathbf{%
\mathbf{\mathbf{\ ^{\shortparallel }}}}}\lambda \ ^{\shortparallel }\mathbf{g%
}_{\beta _{s}\gamma _{s}}.  \notag
\end{eqnarray}

If we find an exact/ parametric solution $\mathbf{g}_{i_{2}j_{2}}(x^{k_{2}})$
of (\ref{cannonsymparamc2hv}) for any prescibed generating source data $%
\left[ \ _{1}^{\shortparallel }\mathcal{K},\ _{2}^{\shortparallel }\mathcal{K%
}\right] $ encoding an effective cosmological constant $\ ^{\shortparallel
}\lambda $ and parameters $\hbar $ and $\kappa ,$ we can compute in explicit
form the real R-flux sources (\ref{realrflux}) on a Lorentz spacetime base.
Such solutions are also characterized by associated complex values of the
Ricci s-tensor which can be computed but do not contribute as real spacetime
R-flux deformations. Such solutions are also charaterized by nontrivial
complex components of associated nonsymmetric metrics, i.e. by $\ _{\star
}^{\shortparallel }\mathfrak{\check{g}}_{\alpha _{2}\beta _{2}}$ (\ref%
{aux40b}) \ and $\ _{\star }^{\shortparallel }\mathfrak{a}_{\alpha _{2}\beta
_{2}}$ (\ref{aux40aa}) \ computed respectively for data (\ref{aux40ss}). In
our further partner works, we shall consider such solutions for total phase
spaces when complex configurations may play a substantial role. For
simplicity, in this article we restrict our considerations only for possible
4-d spacetime real R-flux deformations.

\section{Quasi-stationary configurations encoding nonassociative string
R-fluxes and gravity}

\label{sec3}

The main goal of this work is to extend the AFCDM to applications in
nonassociative vacuum gravity described by dyadic equations (\ref%
{cannonsymparamc2hv}) with effective sources (\ref{realrflux}) and prove a
general decoupling property of such nonlinear systems of PDEs. We shall
integrate corresponding equations in a general off-diagonal metric form for
quasi-stationary spacetime metrics encoding in parametrical form
nonassociative star R-flux deformations. In explicit form, we shall
construct a new class of black ellipsoid configurations determined by
nonassociative effective R-flux sources. We consider for spacetime shells $%
s=1,2$ the methods and results outlined in \cite{partner02}, section 5, and
\cite{bubuianu17} (see there sections on quasi-stationary off-diagonal
solutions and black hole, BH, configurations).

\subsection{Decoupling of nonassociative quasi-stationary solutions}

We can prove an important decoupling property of the system of modified
vacuum Einstein equations (\ref{cannonsymparamc2hv}) with effective sources
induced in parametric forms by nonassociative star R-flux deformations on
shells $s=1,2,$ for a class of generic off-diagonal metrics with Killing
symmetry on $\partial _{4}=\partial /\partial y^{4}= \partial
_{t}=\partial/\partial t$.\footnote{%
Similar decoupling properties can be proven for so-called locally
anisotropic cosmological solutions with Killing symmetry on $\partial
_{3}=\partial /\partial y^{3},$ for higher dimensions, and for more general
assumptions when generic off-diagonal spacetime and/or phase space s-metrics
depend on all coordinates, see details in \cite{bubuianu17,bubuianu19} and
references therein. For simplicity, we omit such geometric constructions in
this work.} For BH solutions and their nonassociative deformations, we can
use standard spacetime spherical coordinates with $\
^{\shortmid}u^{1}=x^{1}=r,\ ^{\shortmid }u^{2}=x^{2}=\theta , \ ^{\shortmid
}u^{3}=y^{3}=\varphi ,\ ^{\shortmid }u^{4}=y^{4}=t)$ when the phase space
dynamics is encoded in effective cosmological constants and prescribed $\
_{1}^{\shortparallel} \mathcal{K}$ and $\ _{2}^{\shortparallel }\mathcal{K}$
in $\ ^{\shortparallel}\mathbf{K}_{~j_{2}}^{i_{2}}=\ \delta
_{~j_{2}}^{i_{2}}\ _{s}^{\shortparallel }\mathcal{K}$ (\ref{ansatzsourchv})
constrained on spacetime shells. Those parametrization of (effective)
nonassociative and associative sources were stated in a quasi-stationary
form. In explicit form, $\ _{s}^{\shortparallel }\mathcal{K}$ should be
chosen a form which is compatible with some experimental and/o observational
data or computed for some models of string theory.

For an associative and commutative s-metric $\ _{s}^{\shortparallel }\mathbf{%
g}=\ ^{\shortparallel }\mathbf{g}_{\alpha _{s}\beta _{s}}\ ^{\shortparallel }%
\mathbf{e}^{\alpha _{s}}\otimes \ ^{\shortparallel }\mathbf{e}^{\beta _{s}}$
(\ref{aux40ss}) with spacetime projection to $\ _{2}^{\shortparallel }%
\mathbf{g}=\ ^{\shortparallel }\mathbf{g}_{\alpha _{2}\beta _{2}}\
^{\shortparallel }\mathbf{e}^{\alpha _{2}}\otimes \ ^{\shortparallel }%
\mathbf{e}^{\beta _{2}}$ (\ref{sdm}), we consider a quasi-stationary ansatz
for a linear quadratic element,
\begin{equation}
d\ ^{\shortparallel }\widehat{s}^{2}=\widehat{g}_{1}(r,\theta )dr^{2}+%
\widehat{g}_{2}(r,\theta )d\theta ^{2}+\widehat{g}_{3}(r,\theta ,\varphi
)\delta \varphi ^{2}+\widehat{g}_{4}(r,\theta ,\varphi )\delta t^{2},  \notag
\end{equation}%
where the s-adapted coefficients for the s-metric are parameterized in the
form
\begin{eqnarray*}
\widehat{g}_{i_{1}j_{i}} &=&diag[\widehat{g}_{i_{1}}(x^{k_{1}})],\mbox{ for }%
i_{1},j_{1}=1,2\mbox{ and }x^{k_{1}}=(x^{1}=r,x^{2}=\theta ); \\
\widehat{g}_{a_{2}b_{2}} &=&diag[\widehat{g}_{a_{2}}(x^{k_{1}},y^{3})],%
\mbox{ for }a_{2},b_{2}=3,4\mbox{ and }y^{3}=x^{3}=\varphi ,y^{4}=x^{4}=t;
\end{eqnarray*}%
Such coefficients will encode additional parametric dependencies on $\hbar $
and $\kappa $ which will be computed in next sections for respective classes
of exact/parametric solutions. We shall use "hat" labels on s-metric and
N-connection coefficients for a stationary ansatz which will be used for
computing the s-adapted coefficients of a respective canonical s-connection $%
\ _{2}^{\shortparallel }\widehat{\mathbf{D}}$ (\ref{twocon}) with
coefficients (\ref{canhcs}).

The N-adapted co-bases $\ \widehat{\mathbf{e}}^{\alpha _{2}}=(\mathbf{e}%
^{i_{1}}=dx^{i_{1}},\widehat{\mathbf{e}}^{a_{2}}=d\ y^{a_{2}}+\widehat{N}_{\
i_{1}}^{a_{2}}dx^{i_{1}})$ (\ref{nadapbdss}), with
\begin{equation*}
e^{1}=dr,e^{2}=d\theta ,\widehat{\mathbf{e}}^{3}=\delta \varphi =d\varphi
+w_{i_{1}}dx^{i_{1}},\widehat{\mathbf{e}}^{4}=\delta
t=dt+n_{i_{1}}dx^{i_{1}},
\end{equation*}%
are determined by N-connection coefficients $\widehat{N}_{j_{1}}^{3}=\
_{2}w_{j_{1}}=\ w_{j_{1}}(r,\theta ,\varphi ),\ N_{j_{1}}^{4}=\
_{2}n_{j_{1}}=\ n_{j_{1}}(r,\theta ,\varphi )$.

We can use the term "stationary" for metrics with coefficients which in
certain adapted coordinates do not depend on respective time like
coordinates but contain some off-diagonal terms, for instance, as for
rotating Kerr BH \cite{hawking73,misner,wald82,kramer03}. In our approach,
there are considered more general nonholonomic configurations (not only with
coordinate rotating frames). The h1-v2 part (i.e. the first 4 components for a
Lorentz manifold base) in (\ref{sdm}) is of stationary type. The term
\textbf{"quasi-stationary"} can be used for s-metric ansatz with associated N-connection
s-structure which is nonholonomic and encode in local anisotropic form
R-flux contributions from a phase space dynamics.

We shall use short notations for partial derivatives when, for instance, $%
\partial _{1}q=q^{\bullet },\partial _{2}q=q^{\prime},\partial
_{3}q=\partial _{\varphi }q=q^{\diamond }$ and construct quasi-stationary
configurations for $g_{4}^{\diamond }\neq 0.$\footnote{If such conditions are note imposed, we can fine more special classes of exact and parametric solutions with another type of nonlinear and nonholonomic structures, possible singularities etc. but the corresponding formulas are quite cumbersome and do not allow general explicit integration
of modified Einstein equations. We do not study such solutions in this work.}
Explicit computations of the canonical Ricci s-tensor (\ref{candricci}) for
$\ _{s}^{\shortparallel }\widehat{\mathbf{D}}$ (\ref{canhcs}) for a s-metric
(\ref{sdm}) allows to write the system of vacuum s-adapted
gravitational equations (\ref{cannonsymparamc2hv})
\begin{eqnarray}
\ \widehat{R}_{1}^{1} &=&\ \widehat{R}_{2}^{2}=\frac{1}{2g_{1}g_{2}}[\frac{%
g_{1}^{\bullet }g_{2}^{\bullet }}{2g_{1}}+\frac{(g_{2}^{\bullet })^{2}}{%
2g_{2}}-g_{2}^{\bullet \bullet }+\frac{g_{1}^{\prime }g_{2}^{\prime }}{2g_{2}%
}+\frac{\left( g_{1}^{\prime }\right) ^{2}}{2g_{1}}-g_{1}^{\prime \prime
}]=-\ \ _{1}^{\shortparallel }\mathcal{K}(\kappa ,r,\theta ),  \notag \\
\ \widehat{R}_{3}^{3} &=&\ \widehat{R}_{4}^{4}=\frac{1}{2g_{3}g_{4}}[\frac{%
\left( g_{4}^{\diamond }\right) ^{2}}{2g_{4}}+\frac{g_{3}^{\diamond
}g_{4}^{\diamond }}{2g_{3}}-g_{4}^{\diamond \diamond }]=-\
_{2}^{\shortparallel }\mathcal{K}(\kappa ,r,\theta ,\varphi ),
\label{riccist2} \\
\ ^{\shortmid }\widehat{R}_{3k_{1}} &=&\frac{\ w_{k_{1}}}{2g_{4}}%
[g_{4}^{\diamond \diamond }-\frac{\left( g_{4}^{\diamond }\right) ^{2}}{%
2g_{4}}-\frac{(g_{3}^{\diamond })(g_{4}^{\diamond })}{2g_{3}}]+\frac{%
g_{4}^{\diamond }}{4g_{4}}(\frac{\partial _{k_{1}}g_{3}}{g_{3}}+\frac{%
\partial _{k_{1}}g_{4}}{g_{4}})-\frac{\partial _{k_{1}}(g_{3}^{\diamond })}{%
2g_{3}}=0;  \notag \\
\ ^{\shortmid }\widehat{R}_{4k_{1}} &=&\frac{g_{4}}{2g_{3}}%
n_{k_{1}}^{\diamond \diamond }+\left( \frac{3}{2}g_{4}^{\diamond }-\frac{%
g_{4}}{g_{3}}g_{3}^{\diamond }\right) \frac{\ n_{k_{1}}^{\diamond }}{2g_{3}}%
=0.  \notag
\end{eqnarray}%
On shells $s=1$ and $s=2,$ with $i_{1},k_{1}...=1,2,$ the dependence on
string parameter $\kappa $ is determined by some formulas (\ref%
{ansatzsourchv}) related algebraically via some frame transforms to (\ref%
{realrflux}). For such two generating source ansatz $\ _{1}^{\shortparallel}%
\mathcal{K}$ and $\ _{2}^{\shortparallel }\mathcal{K},$ we restrict
nonholonomically the class of possible real R-flux deformations. This will
allow to find exact solutions in explicit form.

Using (\ref{riccist2}), (\ref{candricci}) and (\ref{cnsveinst1}), we find
such formulas for shell components of quasi-stationary phase spaces,
\begin{eqnarray}
\ \ _{1}\widehat{R}sc &=&2(\widehat{R}_{1}^{1}),\ _{2}\widehat{R}sc=2(%
\widehat{R}_{1}^{1}+\widehat{R}_{3}^{3}),\   \label{sourc1hv} \\
\ \widehat{R}_{1}^{1} &=&\widehat{R}_{2}^{2}=-\ \ \ _{1}^{\shortparallel }%
\mathcal{K};\ \widehat{R}_{3}^{3}=\widehat{R}_{4}^{4}=-\ \
_{2}^{\shortparallel }\mathcal{K}.  \notag
\end{eqnarray}

The nonassociative vacuum gravitational field equations (\ref%
{cannonsymparamc2hv}), written in geometric form as (\ref{sourc1hv}),
computed for a quasi-stationary s-metric ansatz (\ref{sdm}) decouple in a
2+2:
\begin{eqnarray}
s &=&1, \mbox{ with }g_{i_{1}}=e^{\psi (\hbar ,\kappa ;r,\theta )},i_{1}=1,2
\notag \\
&&\psi ^{\bullet \bullet }+\psi ^{\prime \prime }=2\ \ _{1}^{\shortparallel }%
\mathcal{K};  \label{eq1} \\
s &=&2, \mbox{ with }\left\{
\begin{array}{c}
\alpha _{i_{1}}=g_{4}^{\diamond }\partial _{i_{1}}(\ _{2}\varpi ),\
_{2}\beta =g_{4}^{\diamond }(\ _{2}\varpi )^{\diamond },\ _{2}\gamma =(\ln
\frac{|g_{4}|^{3/2}}{|g_{3}|})^{\diamond } \\
\mbox{ for }\ _{2}\Psi =\exp (\ _{2}\varpi ),\ _{2}\varpi =\ln
|g_{4}^{\diamond }/\sqrt{|g_{3}g_{4}}|,%
\end{array}%
\right.  \notag \\
&&(\ _{2}\varpi )^{\diamond }g_{4}^{\diamond }=2g_{3}g_{4}\ \
_{2}^{\shortparallel }\mathcal{K},  \notag \\
&&\ _{2}\beta \ w_{j_{1}}-\alpha _{j_{1}}=0,  \notag \\
&&\ n_{k_{1}}^{\diamond \diamond }+\ _{2}\gamma \ n_{k_{1}}^{\diamond }=0;
\notag
\end{eqnarray}%
These equations can be solved recurrently because possess an explicit
decoupling property on both shells $s=1$ and $2.$ The first equation in (\ref%
{eq1}) is a 2-d Poisson equation which can be solved in certain general
forms for any prescribed source $\ _{1}^{\shortparallel }\mathcal{K}$
encoding parametric R-flux contributions. Prescribing $\
_{2}^{\shortparallel }\mathcal{K}$ and $g_{3}$ (or $g_{4}$) on second shell $%
s=2,$ we can find integrating respective nonlinear equation relating such
coefficients and determine $g_{4}$ (or $g_{3}$). This allows us to compute
the coefficients $\alpha _{i_{1}},\ _{2}\beta $ and $\ _{2}\gamma $ and
determine the N-connection coefficients from respective linear algebraic
equations for $\ w_{j_{1}};$ and integrating two times in the last system of
equations in order to compute $n_{k_{1}}.$

\subsection{Integrability for nonassociative quasi-stationary vacuum
deformations}

\subsubsection{Quasi-stationary solutions with (non) associative induced
canonical s-torition}

The system (\ref{eq1}) can be integrated in general form by coefficients of
such a quasi-stationary quadratic linear form, {\small
\begin{eqnarray}
d\widehat{s}^{2} &=&e^{\psi (\hbar ,\kappa
;x^{k_{1}})}[(dx^{1})^{2}+(dx^{2})^{2}]+\frac{[\ _{2}\Psi ^{\diamond }(\hbar
,\kappa ;x^{k_{1}},y^{3})]^{2}}{4(\ _{2}^{\shortparallel }\mathcal{K}(\hbar
,\kappa ;x^{k_{1}},y^{3}))^{2}\{g_{4}^{[0]}-\int d\varphi \lbrack (\
_{2}\Psi (\hbar ,\kappa ;x^{k_{1}},y^{3}))^{2}]^{\diamond }/4(\
~_{2}^{\shortparallel }\mathcal{K}(\hbar ,\kappa ;x^{k_{1}},y^{3}))\}}
\notag \\
&&\{d\varphi +\frac{\partial _{i_{1}}(\ _{2}\Psi (\hbar ,\kappa
;x^{k_{1}},y^{3}))}{(\ _{2}\Psi (\hbar ,\kappa ;x^{k_{1}},y^{3}))^{\diamond }%
}dx^{i_{1}}\}^{2}+\{g_{4}^{[0]}(\hbar ,\kappa ;x^{k_{1}})-\int d\varphi
\frac{\lbrack (\ _{2}\Psi (\hbar ,\kappa ;x^{k_{1}},y^{3}))^{2}]^{\diamond }%
}{4(~_{2}^{\shortparallel }\mathcal{K}(\hbar ,\kappa ;x^{k_{1}},y^{3}))}%
\}\{dt+[\ _{1}n_{k_{1}}(\hbar ,\kappa ;x^{k_{1}})  \label{qeltorshv1} \\
&&\ +_{2}n_{k_{1}}(\hbar ,\kappa ;x^{k_{1}})\int d\varphi \frac{\lbrack (\
_{2}\Psi (\hbar ,\kappa ;x^{k_{1}},y^{3}))^{2}]^{\diamond }}{4(\
~_{2}^{\shortparallel }\mathcal{K}(\hbar ,\kappa
;x^{k_{1}},y^{3}))^{2}|g_{4}^{[0]}-\int d\varphi \lbrack (\ _{2}\Psi (\hbar
,\kappa ;x^{k_{1}},y^{3}))^{2}]^{\diamond }/4(~_{2}^{\shortparallel }%
\mathcal{K}(\hbar ,\kappa ;x^{k_{1}},y^{3}))|^{5/2}}]dx^{k_{1}}\}.  \notag
\end{eqnarray}%
}The coefficients of such an off-diagonal metric are determined by \
\begin{eqnarray*}
&&\mbox{generating functions: }\psi (\hbar ,\kappa ;x^{k_{1}});\ _{2}\Psi
(\hbar ,\kappa ;x^{k_{1}},y^{3}); \\
&&\mbox{generating sources:}\ ~_{1}^{\shortparallel }\mathcal{K}(\hbar
,\kappa ;x^{k_{1}});\ ~_{2}^{\shortparallel }\mathcal{K}(\hbar ,\kappa
;x^{k_{1}},y^{3}); \\
&&\mbox{integr. functions: }g_{4}^{[0]}(\hbar ,\kappa ;x^{k_{1}}),\
_{1}n_{k_{1}}(\hbar ,\kappa ;x^{j_{1}}),\ _{2}n_{k_{1}}(\hbar ,\kappa
;x^{j_{1}}).
\end{eqnarray*}%
Any such off-diagonal solution depends in parametric form on $\hbar ,\kappa $
for any R-flux nonassociative data encoded in $~_{s}^{\shortparallel }%
\mathcal{K}$. The canonical s-torsion structure $\ \widehat{\mathbf{T}}_{\
\alpha _{2}\beta _{2}}^{\gamma _{2}}$ of respective $\ _{2}\widehat{\mathbf{D%
}}=\ \nabla +\ _{2}\widehat{\mathbf{Z}}$ is not trivial for a general (\ref%
{qeltorshv1}), see similar details on computing such s-adapted coefficients
which are similar to those presented in \cite{bubuianu17,bubuianu19}.

Let us discuss the difference of quasi-stationary solutions (\ref{qeltorshv1}%
) for any class of stationary solutions (for black holes, wormholes etc.)
outlined in \cite{misner,hawking73,wald82,kramer03} etc. In those
monographs, the Einstein equations are transformed for certain special
diagonal ansatz into systems of nonlinear ODEs depending on one space like
variable (such ansatz are studied similarly in various MGTs). The general
solutions of such second order ODEs depend on two integration constants
which are defined from additional physical considerations, for instance, to
get some asymptotic limits to the Newton gravity potential, or to modify the
constructions for certain nontrivial cosmological constants etc.
Nonassociative R-flux deformations of the GR vacuum structures result in
more general classes of nonlinear PDEs which can not be transformed in a
general form to ODEs. We have to apply a generalized AFCDM which allow to
construct off-diagonal quasi-stationary solutions, when the coefficients of
d-metrics depend at least on three space coordinates via respective
integration functions (not only on integration constants) and on generating
functions and generating sources. There are parametric dependencies on
various (non) commutative / associative quantum/ string parameters etc. The
solutions are constructed as exact ones by for certain stated values of
parameters for corresponding effective sources. In such cases, we have to
prescribe integration / generation functions and effective sources in
certain explicit form in order to have compatibility with certain
experimental/ observational data. This is possible for such types of
nonholonomic constraints and N-adapted frames and canonical distortions of
(non) linear connections, which allow a self-consistent causal physical
descriptions of new classes of quasi-stationary solutions and their
nonholonomic deformations.

\subsubsection{Spacetime LC-configurations encoding nonassociative
parametric R-fluxes}

We have to impose additional nonholonomic constraints on generating
functions and effective sources in order to\ extract zero torsion
LC-configurations for a $\ ^{\shortparallel }\nabla $ generated constrained
on a spacetime encoding a nontrivial real R-flux structure. A class of
solutions (\ref{qeltorshv1}) can be constrained in such a form that up to
orders $\hbar ,\kappa $ and $\hbar \kappa ,$ for star deformations of
respective connections, there are satisfied the conditions $\
_{\star}^{\shortparallel 2}\widehat{\mathbf{Z}}=0,$ i. e. $\
_{2}^{\shortparallel }\widehat{\mathbf{D}}_{\mid \ _{s}^{\shortparallel }%
\widehat{\mathbf{T}}=0}^{\star }= \ ^{\shortparallel }\nabla ^{\star },$ see
(\ref{lccondnonass}). By straightforward computations, we can
verify that such equations are solved by such s-metric and N-connection
coefficients:
\begin{eqnarray}
\ w_{i_{1}}^{\diamond }(\hbar ,\kappa ;x^{k_{1}},y^{3}) &=&\mathbf{e}%
_{i_{1}}\ln \sqrt{|\ g_{3}|},\mathbf{e}_{i_{1}}\ln \sqrt{|\ g_{4}|}%
=0,\partial _{i_{1}}w_{j_{1}}=\partial _{j_{1}}w_{i_{1}};  \notag \\
n_{i_{1}}^{\diamond }&=&0 \mbox{ and } \partial
_{i_{1}}n_{j_{1}}(x^{k_{1}})=\partial _{j_{1}}n_{i_{1}}(x^{k_{1}});
\label{zerot}
\end{eqnarray}

Let us analyze the most important LC-conditions on respective classes of
generating functions and generating sources from (\ref{zerot}):\ If we
prescribe a shell generating function $\ _{2}\Psi =\ _{2}\check{\Psi}%
(\hbar,\kappa ,x^{i_{1}},y^{3}),$ for which $[\partial _{i_{1}}(\ _{2}\check{%
\Psi})]^{\diamond }=\partial _{i_{1}}(\ _{2}\check{\Psi})^{\diamond },$ we
can solve explicitly the conditions for $w_{j_{1}}$ in (\ref{zerot}) if $\
_{2}^{\shortparallel }\mathcal{K}=const,$ or for a functional $\
_{2}^{\shortparallel }\mathcal{K}(\hbar ,\kappa ,x^{i},y^{3})= \
_{2}^{\shortparallel }\mathcal{K}[\ _{2}\check{\Psi}].$ The third class of
conditions in (\ref{zerot}), $\partial _{i_{1}}w_{j_{1}}=\partial
_{j_{1}}w_{i_{1}},$ can be solved in parametric form for any generating
function $\ _{2}\check{A}=\ _{2}\check{A}(\hbar ,\kappa ,x^{k},y^{3})$ for
which $w_{i_{1}}=\check{w}_{i_{1}}=\partial _{i_{1}}\ _{2}\Psi /(\ _{2}\Psi
)^{\diamond }=\partial _{i_{1}}\ _{2}\check{A}.$ The forth class of values
can be solved by any $n_{i_{1}}(x^{k_{1}})=\partial _{i_{1}}n(x^{k_{1}}).$
We summarize these formulas in the form:
\begin{eqnarray}
&&\ _{2}\Psi =\ _{2}\check{\Psi}(\hbar ,\kappa ,x^{i_{1}},y^{3}),(\partial
_{i_{1}}\ _{2}\check{\Psi})^{\diamond }=\partial _{i_{1}}(\ _{2}\check{\Psi}%
)^{\diamond },\check{w}_{i_{1}}=\partial _{i_{1}}(\ _{2}\check{\Psi})/(\
_{2}^{\shortmid }\check{\Psi})^{\diamond }=\partial _{i_{1}}(\ _{2}\check{A}%
);  \notag \\
&&n_{i_{1}}=\partial _{i_{1}}[\ ^{2}n(\hbar ,\kappa ,x^{k_{1}})];\
~_{2}^{\shortparallel }\mathcal{K}(\hbar ,\kappa
,x^{i},y^{3})=~_{2}^{\shortparallel }\mathcal{K}[\ _{2}^{\shortmid }\check{%
\Psi}],\mbox{ or
}\ ~_{2}^{\shortparallel }\mathcal{K}=const;  \label{expconda}
\end{eqnarray}

For subclasses of s-coefficients determined by data (\ref{expconda}), the
quadratic line element (\ref{qeltorshv1}) transforms into
\begin{eqnarray}
d\widehat{s}_{LCst}^{2} &=&~^{\shortparallel }\widehat{\check{g}}%
_{i_{2}j_{2}}(\hbar ,\kappa )du^{i_{2}}du^{j_{2}}=e^{\psi (\hbar ,\kappa
,x^{k_{1}})}[(dx^{1})^{2}+(dx^{2})^{2}]+  \label{qellc} \\
&&\frac{[(\ _{2}\check{\Psi})^{\diamond }]^{2}}{4(~_{2}^{\shortparallel }%
\mathcal{K}[\ _{2}\check{\Psi}])^{2}\{g_{4}^{[0]}-\int d\varphi \lbrack (\
_{2}\check{\Psi})^{2}]^{\diamond }/4(\ ~_{2}^{\shortparallel }\mathcal{K})\}}%
\{d\varphi +[\partial _{i_{1}}(\ _{2}\check{A})]dx^{i_{1}}\}+  \notag \\
&&\{g_{4}^{[0]}-\int d\varphi \frac{\lbrack (\ _{2}\check{\Psi}%
)^{2}]^{\diamond }}{4(\ ~_{2}^{\shortparallel }\mathcal{K}[\ _{2}\check{\Psi}%
])}\}\{dt+\partial _{i_{1}}[\ ^{2}n(x^{k_{1}})]dx^{i_{1}}\}.  \notag
\end{eqnarray}

Any solution (\ref{qellc}) defines a LC-variant of (\ref{qeltorshv1}) for
quasi-stationary spacetime solutions with R-flux contributions encoded in $\
_{1}^{\shortparallel }\mathcal{K}$ and $\ _{2}^{\shortparallel }\mathcal{K}.$
This way, we generate exact solutions of the systems of nonassociative
vacuum Einstein equations (\ref{cannonsymparamc2hv}) for $\ ^{\shortparallel
}\nabla ^{\star }$ and reduced to (\ref{sourc1hv}) with spacetime $\
^{\shortparallel }\nabla ^{\star }\rightarrow \nabla ^{\star }.$

\subsubsection{Nonlinear symmetries for stationary generating functions and
R-flux sources}

We can change the spacetime nonassociative generating data with generating
functions and effective sources and a prescribed cosmological constant, $(\
_{2}\Psi (\hbar ,\kappa ,x^{i_{1}},y^{3}),\ ~_{2}^{\shortparallel }\mathcal{K%
}(\hbar ,\kappa ,x^{i_{1}},y^{3}))\leftrightarrow (\ _{2}\Phi (\hbar ,\kappa
,x^{i_{1}},y^{3}),\ _{2}\Lambda _{0})$, if there are used such nonlinear
transforms:%
\begin{eqnarray}
\frac{\lbrack (\ _{2}\Psi )^{2}]^{\diamond }}{\ ~_{2}^{\shortparallel }%
\mathcal{K}} &=&\frac{[(\ _{2}\Phi )^{2}]^{\diamond }}{\ _{2}\Lambda _{0}},%
\mbox{ which can be
integrated as  }  \label{nonltransf} \\
(_{2}\Phi )^{2} &=&\ _{2}\Lambda _{0}\int dx^{3}(~_{2}^{\shortparallel }%
\mathcal{K})^{-1}[(\ _{2}\Psi )^{2}]^{\diamond }\mbox{ and/or }(\ _{2}\Psi
)^{2}=(\ _{2}\Lambda _{0})^{-1}\int dx^{3}(~_{2}^{\shortparallel }\mathcal{K}%
)[(\ _{2}\Phi )^{2}]^{\diamond }.  \notag
\end{eqnarray}

Using nonlinear symmetries (\ref{nonltransf}), we can write the quadratic
element (\ref{qeltorshv1}) in an equivalent form encoding the nonlinear
symmetries of generating functions and generating sources of R-flux or a
respective effective cosmological constant,%
\begin{eqnarray}
d\widehat{s}^{2} &=&~^{\shortparallel }g_{\alpha _{s}\beta _{s}}(\hbar
,\kappa ,x^{k},y^{3},~_{2}^{\shortparallel }\Phi ,\ _{2}^{\shortparallel
}\Lambda _{0})d~^{\shortparallel }u^{\alpha _{s}}d~^{\shortparallel
}u^{\beta _{s}}=e^{\psi (\hbar ,\kappa
,x^{k_{1}})}[(dx^{1})^{2}+(dx^{2})^{2}]  \label{offdiagcosmcsh} \\
&&-\frac{(\ _{2}\Phi )^{2}[(\ _{2}\Phi )^{\diamond }]^{2}}{|\ _{2}\Lambda
_{0}\int dx^{3}(~_{2}^{\shortparallel }\mathcal{K})[(\ _{2}\Phi
)^{2}]^{\diamond }|[g_{4}^{[0]}-(\ _{2}\Phi )^{2}/4\ _{2}\Lambda _{0}]}%
\{dx^{3}+\frac{\partial _{i_{1}}\ \int dx^{3}(~_{2}^{\shortparallel }%
\mathcal{K})\ [(\ _{2}\Phi )^{2}]^{\diamond }}{(~_{2}^{\shortparallel }%
\mathcal{K})\ [(\ _{2}\Phi )^{2}]^{\diamond }}dx^{i_{1}}\}^{2}  \notag \\
&&-\{g_{4}^{[0]}-\frac{(\ _{2}\Phi )^{2}}{4\ _{2}\Lambda _{0}}\}\{dt+[\
_{1}n_{k_{1}}+\ _{2}n_{k_{1}}\int dy^{3}\frac{(\ _{2}\Phi )^{2}[(\ _{2}\Phi
)^{\diamond }]^{2}}{|\ _{2}\Lambda _{0}\int dy^{3}(~_{2}^{\shortparallel }%
\mathcal{K})[(\ _{2}\Phi )^{2}]^{\diamond }|[g_{4}^{[0]}-(\ _{2}\Phi
)^{2}/4\ _{2}\Lambda _{0}]^{5/2}}]\},  \notag
\end{eqnarray}%
for indices: $i_{1},j_{1},k_{1},...=1,2;i_{2},j_{2},k_{2},...=1,2,3,4$ and
\begin{eqnarray*}
&&\mbox{generating functions: }\psi (\hbar ,\kappa ,x^{k_{1}});\ _{2}\Phi
(\hbar ,\kappa ,x^{k_{1}}y^{3});\ \mbox{generating sources:}%
~_{1}^{\shortparallel }\mathcal{K}(\hbar ,\kappa ,x^{k_{1}});\
~_{2}^{\shortparallel }\mathcal{K}(\hbar ,\kappa ,x^{k_{1}},y^{3}); \\
&&\mbox{integration functions:}g_{4}^{[0]}(\hbar ,\kappa ,x^{k_{1}}),\
_{1}n_{k_{1}}(\hbar ,\kappa ,x^{j_{1}}),\ _{2}n_{k_{1}}(\hbar ,\kappa
,x^{j_{1}}).
\end{eqnarray*}

In a similar form, we can define nonlinear symmetries for generating
functions and R-flux sources and introduce effective cosmological constants $%
\ _{s}^{\shortparallel }\Lambda _{0}$ for LC-configurations and
quasi-stationary s-metrics (\ref{qellc}).

Finally, we note that nonlinear transforms (\ref{nonltransf}) induce
nonlinear symmetries for the symmetric part of the star deformed s-metric (%
\ref{aux40b}) and other geometric s-objects which may involve both real and
complex components. For the chosen class of quasi-stationary spacetime
s-metrics, the nonsymmetric part of star deformed s-metrics (\ref{aux40aa})
is still constrained to be zero for such re-definitions of generating
functions and sources but certain nontrivial components can be computed on
shells $s=3,4$ on (co) tangent Lorentz bundles.

\subsection{Nonholonomic dyadic deformations into parametric solutions}

In this section, we adapt for 4-d base spacetimes with shells $s=1,2$ dyadic
decompositions the results for 8-d phase space nonassociative gravity models
from Appendix B \cite{partner02}. Such extensions of the AFCDM are important for constructing exact and parametric locally anisotropic BH solutions of the system of nonlinear PDEs (\ref{eq1}) generated by nonassociative star nonholonomic deformations and
related effective R-flux sources. There are provided formulas for certain general quasi-stationary d-metrics encoding nonassociative deformations. We also derive formulas for such nonholonomic configurations which are defined as small parametric nonassociative distortions of Schwarzschild metrics via respective "gravitational polarization" functions. It is possible to prescribe such generating functions with ellipsoid symmetry when the generic off-diagonal solutions describe distorted black ellipsoids, BEs, and other type small nonholonomic deformations of static BHs in so-called Weyl coordinates.

\subsubsection{Using some d-metric coefficients as generating functions}

The nonlinear symmetries (\ref{nonltransf}) can be written in the form:%
\begin{equation*}
\lbrack (\ _{2}\Psi )^{2}]^{\diamond }=-\int dy^{3}(~_{2}^{\shortparallel }%
\mathcal{K})g_{4}^{\diamond }\mbox{ and/or }(\ _{2}\Phi )^{2}=-4\
_{2}\Lambda _{0}g_{4}.
\end{equation*}%
As a result, we conclude that the quadratic elements for quasi-periodic
solutions (\ref{qeltorshv1}) and/or (\ref{offdiagcosmcsh}) can be rewritten
equivalently in terms of generating data $(g_{4};\ _{1}^{\shortparallel }%
\mathcal{K},\ _{2}^{\shortparallel }\mathcal{K},\ _{s}^{\shortparallel
}\Lambda ),$
\begin{eqnarray}
d\ \widehat{s}^{2}&=&\ ~^{\shortparallel }g_{i_{2}j_{2}}(\hbar ,\kappa
,x^{k},y^{3}\ ;g_{4}\ )d\ u^{i_{2}}d\ u^{j_{2}}  \label{qeltorshv2} \\
&=&e^{\psi (\hbar ,\kappa ,x^{k_{1}})}[(dx^{1})^{2}+(dx^{2})^{2}]-\frac{%
(g_{4}^{\diamond })^{2}}{|\int dy^{3}[(~_{2}^{\shortparallel }\mathcal{K}%
)g_{4}]^{\diamond }|\ g_{4}}\{dy^{3}+\frac{\partial _{i_{1}}[\int
dy^{3}(~_{2}^{\shortparallel }\mathcal{K})\ g_{4}^{\diamond }]}{%
(~_{2}^{\shortparallel }\mathcal{K})\ g_{4}^{\diamond }}dx^{i_{1}}\}^{2}
\notag \\
&&+ g_{4}\{dt+[\ _{1}n_{k_{1}}+\ _{2}n_{k_{1}}\int dy^{3}\frac{%
(g_{4}^{\diamond })^{2}}{|\int dy^{3}[(~_{2}^{\shortparallel }\mathcal{K}%
)g_{4}]^{\diamond }|\ [g_{4}]^{5/2}}]dx^{\acute{k}_{1}}\}.  \notag
\end{eqnarray}%
The parametric dependence, signs and integration functions/constants in
above formulas have to be chosen in some forms which are compatible with
experimental/ observational data or corrections from R-fluxes in string
theories.

We can restrict the class generic off-diagonal solutions (\ref{qeltorshv2})
to LC-configurations if we use as a generating function any coefficient $%
\check{g}_{4}(\hbar ,\kappa ,x^{i_{1}},y^{3})$ and nonlinear symmetries
involving (\ref{expconda}) for quasi-stationary solutions (\ref{qellc}). For
such zero torsion solutions, the R-flux contributions of effective sources $%
\ _{2}^{\shortparallel }\mathcal{K}$ are encoded correspondingly in the
N-connection coefficients $\ _{2}^{\shortparallel }\check{A}.$

\subsubsection{Polarization functions for nonassociative prime and target
spacetime d-metrics}

We can elaborate on associative/commutative models of gravity on a
nonholonomic Lorentz manifold enabled with nonholonomic dyadic structure, $\
^{\shortparallel }\mathbf{e}^{\alpha _{2}}\in T_{2}^{\ast }\mathbf{T}^{\ast }%
\mathbf{V}$ (\ref{nadapbdsc}) and defined by a \textbf{prime } d-metric $\
_{2}\mathbf{\mathring{g}}$ structure (\ref{sdm}), with possible star
deformations to respective symmetric and nonsymmetric s-metrics of type (\ref%
{ssdm}) and (\ref{nssdm}). For such constructions, we can consider trivial
shells $s=3,4$ enabled with (co) fiber Minkowski metric structure. In
N-adapted form, we parameterize such a spacetime d-metric for $s=1,2$ in the
form
\begin{eqnarray}
\ _{2}\mathbf{\mathring{g}} &=&\ \mathring{g}%
_{i_{2}j_{2}}(x^{k_{1}},y^{a_{2}})d\ u^{i_{2}}\otimes d\ u^{j_{2}}=\mathbf{%
\mathring{g}}_{\alpha _{2}\beta _{2}}(\ _{2}u)\ ~^{\shortparallel }\mathbf{%
\mathbf{\mathring{e}}}^{\alpha _{2}}\mathbf{\otimes \ ~^{\shortparallel }%
\mathbf{\mathring{e}}}^{\beta _{2}}  \label{primedm} \\
&=&~^{\shortparallel }\mathring{g}_{i_{1}j_{1}}(x^{k_{1}})~e^{i_{1}}\otimes
e^{j_{1}}+\ ~^{\shortparallel }\mathbf{\mathring{g}}%
^{a_{2}b_{2}}(x^{i_{1}},y^{a_{2}})\ \mathbf{\mathring{e}}_{a_{2}}\otimes \
~^{\shortparallel }\mathbf{\mathring{e}}_{b_{2}},\mbox{ for }  \notag \\
&&\mathbf{\mathring{e}}_{\alpha _{2}} =(\mathbf{\mathring{e}}%
_{i_{1}}=\partial _{i_{1}}-\mathring{N}_{i_{1}}^{b_{2}}(~x,y)\partial
_{b_{2}},\ ~{e}_{a_{2}}=\partial _{a_{2}})\mbox{ and } \ ^{\shortparallel }%
\mathbf{\mathring{e}}^{\alpha _{2}} =(dx^{i_{1}},\mathbf{\mathring{e}}%
^{a_{2}}=dy^{a_{2}}+\mathring{N}_{i_{1}}^{a_{2}}(x,y)dx^{i_{1}}).  \notag
\end{eqnarray}%
To label coefficients of prime d-metrics and related geometric d-objects, we
shall use left/right/up labels with a small circle. In general, a prime $\
_{2}\mathbf{\mathring{g}}$ (\ref{primedm}) my be, or not, a solution of
certain gravitational field equations in a MGT or GR. In this work, we
consider that $\ _{2}\mathbf{\mathring{g}}$ is a Schwarzschild metric
written in some adapted coordinates which allow to apply the AFCDM and
construct generic off-diagonal solutions for some target metrics. We can
always consider phase space extensions of prime d-metrics resulting in
parametric R-flux deformations to nonassociative configurations following
formulas $\ _{\star }^{\shortparallel }\mathfrak{\check{g}}_{\mu _{s}\nu
_{s}}^{\circ }=(\ _{\star }^{\shortparallel }\mathfrak{\check{g}}%
_{i_{1}j_{1}}^{\circ },\ _{\star }^{\shortparallel }\mathfrak{\check{g}}%
_{a_{2}b_{2}}^{\circ },\ _{\star }^{\shortparallel }\mathfrak{\check{g}}%
_{\circ }^{a_{3}b_{3}},\ _{\star }^{\shortparallel }\mathfrak{\check{g}}%
_{\circ }^{a_{4}b_{4}})$ (\ref{aux40b}) and $\ _{\star }^{\shortparallel }%
\mathfrak{a}_{\mu _{s}\nu _{s}}^{\circ }=(0,0,\ _{\star }^{\shortparallel }%
\mathfrak{a}_{c_{3}b_{3}}^{\circ },\ _{\star }^{\shortparallel }\mathfrak{a}%
_{c_{4}b_{4}}^{\circ })$ (\ref{aux40aa}), when the star-metrics are
parametric extensions of some associative and commutative s-metrics.

For nontrivial R-flux and star deformations, we can study nonassociative
parametric nonholonomic deformations of a prime metric $\ _{2}\mathbf{%
\mathring{g}}$ (\ref{primedm}) to nonlinear quadratic elements determined by
\textbf{target} quasi-stationary d-metrics $\ _{2}^{\shortparallel }\mathbf{%
g,}$ which can be parameterized in any necessary form (\ref{qeltorshv1}), (%
\ref{offdiagcosmcsh}), or (\ref{qeltorshv2}). Such a target $\
_{2}^{\shortparallel }\mathbf{g}$ is a solution of the nonassociative vacuum
Einstein equations represented in any form (\ref{cannonsymparamc2hv}), (\ref%
{riccist2}), or (\ref{eq1}). Nonholonomic star s-deformations of type "prime
to target" s-metrics can be described in terms of \textbf{gravitational
polarization} ($\eta - $\textbf{polarization}) functions,
\begin{equation*}
\ ~_{2}^{\shortparallel }\mathbf{\mathring{g}}\rightarrow \
_{2}^{\shortparallel }\mathbf{g}=[~^{\shortparallel }g_{\alpha _{2}}=\
^{\shortparallel }\eta _{\alpha _{2}}\ ~^{\shortparallel }\mathring{g}%
_{\alpha _{2}},\ ^{\shortparallel }N_{i_{1}}^{a_{2}}=\ ~^{\shortparallel
}\eta _{i_{1}}^{a_{2}}\ ~^{\shortparallel }\mathring{N}_{i_{1}}^{a_{2}}],
\end{equation*}%
when the target d-metrics are parameterized in the form
\begin{eqnarray}
\ _{2}^{\shortparallel }\mathbf{g} &=&g_{i_{1}}(\hbar ,\kappa
,x^{k_{1}})dx^{i_{1}}\otimes dx^{i_{1}}+g_{a_{2}}(\hbar ,\kappa
,x^{i_{1}},y^{b_{2}})\mathbf{e}^{a_{2}}\otimes \mathbf{e}^{a_{2}}
\label{dmpolariz} \\
&=&\eta _{i_{1}}(\hbar ,\kappa ,x^{i_{1}},y^{a_{2}})\ ~^{\shortparallel }%
\mathring{g}_{i_{1}}(\hbar ,\kappa ,x^{i_{1}},y^{a_{2}})dx^{i_{1}}\otimes
dx^{i_{1}}  \notag \\
&&+\eta _{b_{2}}(\hbar ,\kappa ,x^{i_{1}},y^{a_{2}})\ ~^{\shortparallel }%
\mathring{g}_{b_{2}}(\hbar ,\kappa ,x^{i_{1}},y^{a_{2}})\ ~^{\shortparallel }%
\mathbf{e}^{b_{2}}[\eta ]\otimes \ ~^{\shortparallel }\mathbf{e}%
^{b_{2}}[\eta ], \mbox{ for }  \notag \\
\ ^{\shortparallel }\mathbf{e}^{\alpha _{2}}[\eta ] &=& (dx^{i_{1}},\mathbf{e%
}^{a_{2}}=dy^{a_{2}}+\eta _{i_{1}}^{a_{2}}(\hbar ,\kappa
,x^{i_{1}},y^{a_{2}})\ \mathring{N}_{i_{1}}^{a_{2}}(\hbar ,\kappa
,x^{i_{1}},y^{a_{2}})dx^{i_{1}}).  \notag
\end{eqnarray}

We emphasize that any multiple $\eta \ \mathring{g}$ in (\ref{dmpolariz})
may depend on mixed (higher) shell coordinates and various parameters,
sources, etc. To apply the AFCDM for constructing solutions we have to use
products subjected to the condition that the target d-metrics, for instance,
is of type (\ref{qeltorshv2}) or any equivalent form. The term "gravitational polarization" is used because for $\eta $-deformations with a small parameter, we can generate new classes of exact/ parametric solutions, for instance, of black hole/ ellipsoid type but with effective polarization of fundamental physical constants. Such diagonal and off-diagonal solutions were constructed in (non) commutative MGTs, see
\cite{vacaru03,vacaru09a,bubuianu17,bubuianu18a,bubuianu19} and reference therein. By straightforward computations, we can check that any solution (\ref{qeltorshv2}) can be parameterized in a form (\ref{dmpolariz}) and described by a nonlinear quadratic element with explicit dependence on $\eta $-polarizations,
\begin{eqnarray}
&&d\ ^{\shortparallel }\widehat{s}^{2}=\ ^{\shortparallel }g_{\alpha
_{s}\beta _{s}}(\hbar ,\kappa ,x^{k},y^{3};\ g_{4},\ \ _{s}^{\shortparallel
}\Lambda _{0};\ _{2}^{\shortparallel }\mathcal{K})d~^{\shortparallel
}u^{\alpha _{s}}d~^{\shortparallel }u^{\beta _{s}}  \label{offdiagpolf} \\
&&=e^{\psi (\hbar ,\kappa ,x^{k_{1}})}[(dx^{1})^{2}+(dx^{2})^{2}]-\frac{[(\
^{\shortparallel }\eta _{4}\ ^{\shortparallel }\mathring{g}_{4})^{\diamond
}]^{2}}{|\int dy^{3}(~_{2}^{\shortparallel }\mathcal{K})(\ ^{\shortparallel
}\eta _{4}\ ^{\shortparallel }\mathring{g}_{4})^{\diamond }|\ (\
^{\shortparallel }\eta _{4}\ ^{\shortparallel }\mathring{g}_{4})}\{dy^{3}+%
\frac{\partial _{i_{1}}[\int dy^{3}(\ _{2}^{\shortparallel }\mathcal{K})\
(~^{\shortparallel }\eta _{4}\ ^{\shortparallel }\mathring{g}_{4})^{\diamond
}]}{(~_{2}^{\shortparallel }\mathcal{K})(\ ^{\shortparallel }\eta _{4}\
^{\shortparallel }\mathring{g}_{4})^{\diamond }}dx^{i_{1}}\}^{2}  \notag \\
&&+(\ ^{\shortparallel }\eta _{4}\ ^{\shortparallel }\mathring{g}%
_{4})\{dt+[\ _{1}n_{k_{1}}+\ _{2}n_{k_{1}}\int dy^{3}\frac{[(\ \
^{\shortparallel }\eta _{4}\ ~^{\shortparallel }\mathring{g}_{4})^{\diamond
}]^{2}}{|\int dy^{3}(\ _{2}^{\shortparallel }\mathcal{K})(\ ^{\shortparallel
}\eta _{4}\ ^{\shortparallel }\mathring{g}_{4})^{\diamond }|\ (\
~^{\shortparallel }\eta _{4}\ ~^{\shortparallel }\mathring{g}_{4})^{5/2}}%
]dx^{\acute{k}_{1}}\},  \notag
\end{eqnarray}%
where the polarization functions are determined by generating data $\psi
(\hbar ,\kappa ,x^{i_{1}})$ and $^{\shortparallel }\eta _{4}(\hbar ,\kappa
,x^{i_{1}},y^{3}).$ We keep left labels "$\ ^{\shortparallel }"$ for such
spacetime configurations in order to emphasize that the coefficients encode
certain R-flux deformations in the (co) tangent bundle.

\subsubsection{Parametric nonassociative transforms to quasi-stationary
spacetime metrics}

The nonassociative vacuum gravitational field equations can be projected on
spacetime background, in instance, in the form (\ref{sourc1hv}). Such
nonlinear system of PDEs involve in parametrical form an effective R-flux
source (\ref{realrflux}). Similar geometric constructions were considered in
our previous works \cite{vacaru09a,bubuianu17,bubuianu18a,bubuianu19} for (other types) effective sources with parametric deformations on a small parameter $\varepsilon ,0\leq \varepsilon <1.$ In \cite{partner02}, we proved that the same AFCDM can be applied for a
quasi-stationary s-metric ansatz on total phase spaces when as a small
parameter it is considered the string constant, $\varepsilon \rightarrow
\kappa .$ \ For the goals of this paper, we adapt those constructions for
generating parametric spacetime solutions with gravitational polarizations
depending in N-adapted form only on space coordinates.

Let us consider parametric $\kappa $--decompositions of the $\eta $%
-polarization functions in a d-metric (\ref{offdiagpolf}) resulting in
quasi-stationary solutions of type (\ref{qeltorshv1}) and/or (\ref%
{offdiagcosmcsh}):
\begin{eqnarray*}
\ ^{\shortparallel }g_{i_{1}}(\kappa ,x^{k_{1}}) &=&~^{\shortparallel }\eta
_{i_{i}}\ ~^{\shortparallel }\mathring{g}_{i_{1}}=~^{\shortparallel }\zeta
_{i_{1}}(x^{i_{1}},y^{a_{2}})[1+\kappa ~^{\shortparallel }\chi
_{i_{1}}(x^{i_{1}},y^{a_{2}})]\ ~^{\shortparallel }\mathring{g}%
_{i_{1}}(x^{i_{1}},y^{a_{2}}), \\
\ ~^{\shortparallel }g_{b_{2}}(\kappa ,x^{i_{1}},y^{3}) &=&\
~^{\shortparallel }\eta _{b_{2}}\ ~^{\shortparallel }\mathring{g}%
_{b_{1}}=~^{\shortparallel }\zeta _{b_{2}}(x^{i_{1}},y^{a_{2}})[1+\kappa \
~^{\shortparallel }\chi _{b_{2}}(x^{i_{1}},y^{a_{2}})]\ ~^{\shortparallel }%
\mathring{g}_{b_{1}}(x^{i_{1}},y^{a_{2}}), \\
\ ^{\shortparallel }N_{i_{1}}^{a_{2}}(\kappa ,x^{k_{1}},y^{3})
&=&~^{\shortparallel }\eta _{i_{1}}^{a_{2}}\ ~^{\shortparallel }\mathring{N}%
_{i_{1}}^{a_{2}}=~^{\shortparallel }\zeta
_{i_{1}}^{a_{2}}(x^{i_{1}},y^{b_{2}})[1+\kappa \ ~^{\shortparallel }\chi
_{i_{1}}^{a_{2}}(x^{i_{1}},y^{b_{2}})]\ ~^{\shortparallel }\mathring{N}%
_{i_{1}}^{a_{2}}(x^{i_{1}},y^{b_{2}}).
\end{eqnarray*}%
These formulas for star parametric deformations of a d-metric and
N-connection structure on $_{2}\mathbf{T}_{\shortparallel }^{\ast }\mathbf{V}
$ can be written in the form
\begin{equation}
\ _{2}^{\shortparallel }\mathbf{\mathring{g}}\rightarrow \
_{2}^{\shortparallel \kappa }\mathbf{g}=[\ ^{\shortparallel }g_{\alpha
_{2}}=\ ^{\shortparallel }\zeta _{\alpha _{2}}(1+\kappa \ ^{\shortparallel
}\chi _{\alpha _{2}})\ \ ^{\shortparallel }\mathring{g}_{\alpha _{2}},\ \
^{\shortparallel }N_{i_{1}}^{a_{2}}=\ ^{\shortparallel }\zeta
_{i_{1}}^{a_{2}}(1+\kappa \ \ ^{\shortparallel }\chi _{i_{1}}^{a_{2}})\ \
^{\shortparallel }\mathring{N}_{i_{1}}^{a_{2}}],  \label{epstargsm}
\end{equation}%
where $\zeta $- and $\chi $-coefficients for deformations (\ref{epstargsm})
are generated by nonholonomic deformation data
\begin{equation*}
\ \ \ ^{\shortparallel }\eta _{2}=\ ^{\shortparallel }\zeta _{2}(1+\kappa \
^{\shortparallel }\chi _{2}),\ \ \ ^{\shortparallel }\eta _{4}=\
^{\shortparallel }\zeta _{4}(1+\kappa \ ^{\shortparallel }\chi _{4}).
\end{equation*}%
For quasi-stationary configurations, $\ ^{\shortparallel }\eta _{2}$ and $\
^{\shortparallel }\eta _{4}$ are considered as generating functions. In a
more general context, we can consider$\ ^{\shortparallel }g_{2}$ and $\
^{\shortparallel }g_{4}$ as generating functions and, for small parametric
deformations, we can take $\ ^{\shortparallel }\chi _{2}$ and $\
^{\shortparallel }\chi _{4}$ as generating functions.

We can compute parametric deformations following such a procedure: \ For $%
s=1,$ we consider $\ \ ^{\shortparallel }\zeta _{i_{1}}=(\ ^{\shortparallel }%
\mathring{g}_{i_{1}})^{-1}e^{\psi _{0}(x^{k_{1}})}$ and $\ ^{\shortparallel
}\chi _{i_{1}}=(\ ^{\shortparallel }\mathring{g}_{i_{1}})^{-1}\ ^{\psi }\
^{\shortparallel }\chi (x^{k_{1}})$, \ where%
\begin{equation*}
\ ^{\shortparallel }\zeta _{i_{1}}(1+\kappa \ ^{\shortparallel }\chi
_{i_{1}})\ \ ^{\shortparallel }\mathring{g}_{i_{1}}=e^{\psi
(x^{k_{1}})}\approx e^{\psi _{0}(x^{k_{1}})(1+\kappa \ ^{\psi }\chi
(x^{k_{1}}))}\approx e^{\psi _{0}(x^{k_{1}})}(1+\kappa \ ^{\psi }\
^{\shortparallel }\chi )
\end{equation*}%
for $\psi _{0}(x^{k_{1}})$ and $\ ^{\shortparallel }\chi (x^{k_{1}})$
defined by a solution of a 2-d Poisson equation (\ref{eq1}). \ For $s=2$ we
have generating functions $\ ^{\shortparallel }\zeta _{4}$ and $\
^{\shortparallel }\chi _{4};$ generating source and cosmological constant,
respectively,$\ ~_{2}^{\shortparallel }\mathcal{K}$ and $\
_{2}^{\shortparallel }\Lambda _{0};$ integration functions$\
_{1}^{\shortparallel }n_{k_{1}}$ and $\ _{2}^{\shortparallel }n_{k_{1}};$
and certain prescribed data for a prime s-metric, $(\ ^{\shortparallel }%
\mathring{g}_{3},\ ^{\shortparallel }\mathring{g}_{4};\ ^{\shortparallel }%
\mathring{N}_{i_{1}}^{3},\ ^{\shortparallel }\mathring{N}_{k_{1}}^{4}).$
After tedious computations, we express
\begin{eqnarray}
\ ^{\shortparallel }\zeta _{3} &=&-\frac{4}{\ \ ^{\shortparallel }\mathring{g%
}_{3}}\frac{[(|\ \ ^{\shortparallel }\zeta _{4}\ \ ^{\shortparallel }%
\mathring{g}_{4}|^{1/2})^{\diamond }]^{2}}{|\int
dy^{3}\{(~_{2}^{\shortparallel }\mathcal{K})(\ ^{\shortparallel }\zeta _{4}\
^{\shortparallel }\mathring{g}_{4})^{\diamond }\}|}\mbox{ and }
\label{paramdef} \\
\ ^{\shortparallel }\chi _{3} &=&\frac{(\ ^{\shortparallel }\chi _{4}|\ \
^{\shortparallel }\zeta _{4}\ \ ^{\shortparallel }\mathring{g}%
_{4}|^{1/2})^{\diamond }}{4(|\ \ ^{\shortparallel }\zeta _{4}\ \
^{\shortparallel }\mathring{g}_{4}|^{1/2})^{\diamond }}-\frac{\int
dy^{3}\{[(~_{2}^{\shortparallel }\mathcal{K})\ (\ ^{\shortparallel }\zeta
_{4}\ ^{\shortparallel }\mathring{g}_{4})\ ^{\shortparallel }\chi
_{4}]^{\diamond }\}}{\int dy^{3}\{(~_{2}^{\shortparallel }\mathcal{K})(\
^{\shortparallel }\zeta _{4}\ \ ^{\shortparallel }\mathring{g}%
_{4})^{\diamond }\}},  \notag
\end{eqnarray}%
\begin{eqnarray*}
\ ^{\shortparallel }\zeta _{i_{1}}^{3} &=&\frac{\partial _{i_{1}}\ \int
dy^{3}(~_{2}^{\shortparallel }\mathcal{K})\ (\ ^{\shortparallel }\zeta
_{4})^{\diamond }}{(\ ^{\shortparallel }\mathring{N}_{i_{1}}^{3})(~_{2}^{%
\shortparallel }\mathcal{K})(\ ^{\shortparallel }\zeta _{4})^{\diamond }}%
\mbox{ and } \\
\ ^{\shortparallel }\chi _{i_{1}}^{3} &=&\frac{\partial _{i_{1}}[\int
dy^{3}(~_{2}^{\shortparallel }\mathcal{K})(\ ^{\shortparallel }\zeta _{4}\
^{\shortparallel }\chi _{4})^{\diamond }]}{\partial _{i_{1}}\ [\int
dy^{3}(~_{2}^{\shortparallel }\mathcal{K})(\ ^{\shortparallel }\zeta
_{4})^{\diamond }]}-\frac{(\ ^{\shortparallel }\zeta _{4}\ ^{\shortparallel
}\chi _{4})^{\diamond }}{(\ ^{\shortparallel }\zeta _{4})^{\diamond }},
\end{eqnarray*}%
\begin{eqnarray*}
\ ^{\shortparallel }\zeta _{k_{1}}^{4} &=&\ (\ \ ^{\shortparallel }\mathring{%
N}_{k_{1}}^{4})^{-1}[\ _{1}^{\shortparallel }n_{k_{1}}+16\
_{2}^{\shortparallel }n_{k_{1}}[\int dy^{3}\{\frac{\left( [(\
^{\shortparallel }\zeta _{4}\ \ ^{\shortparallel }\mathring{g}%
_{4})^{-1/4}]^{\diamond }\right) ^{2}}{|\int dy^{3}(~_{2}^{\shortparallel }%
\mathcal{K})(\ ^{\shortparallel }\zeta _{4}\ \ ^{\shortparallel }\mathring{g}%
_{4})^{\diamond }|}]\mbox{ and } \\
\ ^{\shortparallel }\chi _{k_{1}}^{4} &=&\ -\frac{16\ _{2}^{\shortparallel
}n_{k_{1}}\int dy^{3}\frac{\left( [(\ \ ^{\shortparallel }\zeta _{4}\ \
^{\shortparallel }\mathring{g}_{4})^{-1/4}]^{\diamond }\right) ^{2}}{|\int
dy^{3}(~_{2}^{\shortparallel }\mathcal{K})[(\ \ ^{\shortparallel }\zeta
_{4}\ ^{\shortmid }\mathring{g}_{4})]^{\diamond }|}(\frac{[(\ \
^{\shortparallel }\zeta _{4}\ \ ^{\shortparallel }\mathring{g}_{4})^{-1/4}\
^{\shortparallel }\chi _{4})]^{\diamond }}{2[(\ \ ^{\shortparallel }\zeta
_{4}\ \ ^{\shortparallel }\mathring{g}_{4})^{-1/4}]^{\diamond }}+\frac{\int
dy^{3}[(~_{2}^{\shortparallel }\mathcal{K})(\ ^{\shortparallel }\zeta _{4}\
^{\shortparallel }\chi _{4}\ \ ^{\shortparallel }\mathring{g}%
_{4})]^{\diamond }}{\int dy^{3}(~_{2}^{\shortparallel }\mathcal{K})(\ \
^{\shortparallel }\zeta _{4}\ \ ^{\shortparallel }\mathring{g}%
_{4})^{\diamond }})}{\ _{1}^{\shortparallel }n_{k_{1}}+16\
_{2}^{\shortparallel }n_{k_{1}}[\int dy^{3}\frac{\left( [(\ \
^{\shortparallel }\zeta _{4}\ \ ^{\shortparallel }\mathring{g}%
_{4})^{-1/4}]^{\diamond }\right) ^{2}}{|\int dy^{3}(~_{2}^{\shortparallel }%
\mathcal{K})[(\ ^{\shortparallel }\zeta _{4}\ \ ^{\shortparallel }\mathring{g%
}_{4})]^{\diamond }|}].}.
\end{eqnarray*}

Introducing above coefficients with $\kappa $-decomposition instead of $\eta
$-coefficients of (\ref{offdiagpolf}), we obtain respective nonlinear
quadratic elements for quasi-stationary solutions encoding nonassociative
star R-flux deformations. In next subsection, there are provided such
formulas for such distortions of Schwarzschild BHs. This subclass of
quasi-stationary solutions consist a restriction on shells $s=1,2$ of the
phase space solutions from Appendix B, formulas (B.7) in \cite{partner02}
(in that works, the solutions are with $\eta $-polarization functions
depending also on phase space coordinates).

\subsection{Nonassociative distorted black holes and black ellipsoids}

Our goal is to construct exact quasi-stationary solutions encoding in
parametric form nonassociative nonholonomic (ellipsoid) deformations of the
Schwarzschild spacetime. Such solutions are obtained by assuming that
existence of a R-flux effective matter source (\ref{realrflux}).

\subsubsection{Prime metrics as a distorted Schwarzschild BH}

Let us introduce on spacetime $\mathbf{V}$ local (prolate spheroidal)
coordinates
\begin{equation}
u^{1}=x^{1}=x\in (1,+\infty );u^{2}=x^{2}=y\in \lbrack
-1,1];u^{3}=x^{3}=y^{3}=\phi \in \lbrack 0,2\pi ];u^{4}=x^{4}=y^{4}=t\in
(-\infty ,+\infty ),  \label{wcoord}
\end{equation}%
and consider a prime d-metric $\ _{2}\mathbf{\mathring{g}}^{W}= \left( \ \
\mathring{g}_{i_{1}}^{W},\ \ \mathring{g}_{a_{2}}^{W},\mathring{N}%
_{i_{1}}^{Wa_{2}}\right) $ (\ref{primedm}) (with a label W referring to Weyl
coordinates (\ref{wcoord})) with such parameterizations of nontrivial
coefficients: {\small
\begin{eqnarray}
\ \ \mathring{g}_{1}^{W} &=&\mathring{g}_{11}(x^{i_{1}})=M_{0}^{2}(x+1)^{2}%
\frac{\exp [2(\mathring{\gamma}(x^{i_{1}})-\mathring{\psi}(x^{i_{1}}))]}{%
x^{2}-1},\ \ \mathring{g}_{2}^{W}=\mathring{g}%
_{22}(x^{i_{1}})=M_{0}^{2}(x+1)^{2}\frac{\exp [2(\mathring{\gamma}%
(x^{i_{1}})-\mathring{\psi}(x^{i_{1}}))]}{1-y^{2}},  \notag \\
\ \ \mathring{g}_{3}^{W} &=&\ \mathring{g}%
_{33}(x^{i_{1}})=M_{0}^{2}(x+1)^{2}(1-y^{2})\exp [-2\mathring{\psi}%
(x^{i_{1}})],\ \mathring{g}_{4}^{W}=\ \mathring{g}_{44}(x^{i_{1}})=-\frac{x-1%
}{x+1}\exp [2\mathring{\psi}(x^{i_{1}})],  \notag \\
\mathring{N}_{i_{1}}^{a_{2}}(u^{\beta _{2}}) &\neq &0%
\mbox{  are defined by
a fixed system of local coordinates }.  \label{prime1a}
\end{eqnarray}%
} To avoid coordinate singularities and non-compatible constraints for
nonholonomic and/ or off-diagonal deformation in d-metrics of type (\ref%
{epstargsm}), we can consider such coordinate transforms $u^{\alpha
_{2}^{\prime }}\rightarrow u^{\alpha _{2}^{\prime }}(u^{\alpha _{2}}),$ when
$\mathring{N}_{i_{1}^{\prime }}^{a_{2}^{\prime }}(u^{\beta _{2}^{\prime
}})=0 $ transform into some coefficients $\mathring{N}_{i_{1}}^{a_{2}}(u^{%
\beta _{2}})\neq 0$ of necessary smooth class, preserving a respective (2+2)
splitting for the same d-metric possessing a horizon for $x=1$ and
singularity for $x=-1.$ In (\ref{prime1a}), the parameter $M_{0}$ can be
identified as the BH mass. We use also distortion functions which can be
expressed in terms of Legendre polynomials $P_{l}$ (of the first kind, see
details in \cite{chandr02,breton97,faraji20}),%
\begin{eqnarray}
\mathring{\psi}(x^{i_{1}}) &=&\sum\limits_{l>0}a_{l}\check{\rho}^{l}P_{l}%
\mbox{  and }  \label{legpoliy} \\
\mathring{\gamma}(x^{i_{1}}) &=&\sum\limits_{l>0}a_{l}\sum\limits_{l^{\prime
}=0}^{l-1}[(-1)^{l-l^{\prime }+1}(x+y)-x+y]\check{\rho}^{l^{\prime
}}P_{l^{\prime }}+\sum\limits_{l,l^{\prime }=1}\frac{ll^{\prime
}a_{l}a_{l^{\prime }}}{l+l^{\prime }}\check{\rho}^{l+l^{\prime
}}(P_{l}P_{l^{\prime }}-P_{l-1}P_{l^{\prime }-1}),  \notag
\end{eqnarray}%
where $P_{l}:=P_{l}(xy/\check{\rho})$ for $\check{\rho}=\sqrt{x^{2}+y^{2}-1}$%
. In these formulas $a_{l}\in \mathbb{R}$ are called multipole moments
defining a distortion due, in our case, to an effective R-flux source. The
Schwarzschild spacetime is recovered for $a_{l}=0,$ $a_{1}=1$ is considered
as the dipole moment, $a_{2}:=\mathring{q}$ is the quadrupole moment, etc.
for higher momenta which describe deviations from the spherically symmetric
shape of a central compact object.\footnote{%
We consider that in the presence of an external static and axially symmetric
matter distribution the exterior of a BH is described as a distorted
Schwarzschild solution as in \cite{geroch82}. In such a case, a prime metric
(\ref{prime1a}) defines a static vacuum solution with a regular event
horizon but such a metric is not asymptotically flat \cite{chandr02,faraji20}%
. We can consider Schwarzschild coordinates $x=r/M_{0}-1$ and $y=\cos \theta
$ and related them to the Weyl coordinates $\rho =\sqrt{r(r-2M_{0})}\sin
\theta $ and $z=(r-M_{0})\cos \theta .$ In such variables, a prime metric (%
\ref{prime1a}) can be written as a Weyl metric
\begin{equation*}
d\mathring{s}^{2}=e^{2[\gamma (\rho ,z)-\psi (\rho ,z)]}(d\rho
^{2}+dz^{2})+e^{-2\psi (\rho ,z)}\rho ^{2}d\phi ^{2}-e^{2\psi (\rho
,z)}dt^{2}.
\end{equation*}%
For a Schwarzschild solution, we have $\mathring{\psi}=\frac{1}{2}\ln \frac{%
x-1}{x+1}$ and $\ \mathring{\gamma}=\frac{1}{2}\ln \frac{x^{2}-1}{x^{2}-y^{2}%
},$ when $\rho =M_{0}\sqrt{(x^{2}-1)(1-y^{2})}$ and $z=M_{0}xy$, see details
in \cite{quevedo90} and section II of \cite{faraji20}.}

\subsubsection{Target d-metrics for distorted nonassociative black ellipsoids%
}

For small $\kappa $-ellipsoidal deformations with
\begin{equation*}
\exp [\kappa \ ^{\shortparallel }\chi (x^{i_{1}})]\simeq 1+\kappa \
^{\shortparallel }\chi (x^{i_{1}})\text{\mbox{ and }}\exp [\kappa \
^{\shortparallel }\chi _{3}(x^{i_{1}},\phi )]\simeq 1+\kappa \
^{\shortparallel }\chi _{3}(x^{i_{1}},\phi ),
\end{equation*}%
for $\ ^{\shortparallel }\chi _{1}=\ ^{\shortparallel }\chi _{2}=\
^{\shortparallel }\chi (x^{i_{1}}),$ we can define values
\begin{equation}
M_{0}^{2}\rightarrow \check{M}^{2}(x^{i_{1}},\phi )=M_{0}^{2}\exp [\kappa \
^{\shortparallel }\chi _{3}(x^{i_{1}},\phi )],\mbox{ and }\check{\gamma}%
(x^{i_{1}},\phi )=\mathring{\gamma}(x^{i_{1}})+\frac{\kappa }{2}[\
^{\shortparallel }\chi (x^{i_{1}})-\ ^{\shortparallel }\chi
_{3}(x^{i_{1}},\phi )].  \label{deflegpoly}
\end{equation}%
The coefficients (in Weyl coordinates) of a target d-metric $\ _{2}\mathbf{g}%
^{W}\mathbf{=}\left( \ \ g_{i_{1}}^{W},\ \
g_{a_{2}}^{W},N_{i_{1}}^{Wa_{2}}\right) $ with geometric data (\ref{prime1a}%
) for a nonholonomic deformation for $\ _{2}\mathbf{\mathring{g}=}\left( \ \
\mathring{g}_{i_{1}}^{W},\ \ \mathring{g}_{a_{2}}^{W},\mathring{N}%
_{i_{1}}^{Wa_{2}}\right) $ (\ref{prime1a}) can be written in the form
\begin{eqnarray*}
\ \ g_{1}^{W} &=&\ \ ^{W}\zeta _{1}(1+\kappa \ ^{W}\chi _{1})\ \mathring{g}%
_{1}^{W}=\mathring{g}_{11}(x^{i_{1}})=\ ^{W}\zeta _{1}M_{0}^{2}\exp [\kappa
\ ^{W}\chi _{3}](x+1)^{2}\frac{\exp [2(\check{\gamma}(x^{i_{1}},\phi )-%
\mathring{\psi}(x^{i_{1}}))]}{x^{2}-1},\ \  \\
g_{2}^{W} &=&\ \ ^{W}\zeta _{2}(1+\kappa \ ^{W}\chi _{2})\ \ \mathring{g}%
_{2}^{W}=\mathring{g}_{22}(x^{i_{1}})=\ \ ^{W}\zeta _{2}M_{0}^{2}\exp
[\kappa \ ^{W}\chi _{3}](x+1)^{2}\frac{\exp [2(\check{\gamma}(x^{i_{1}},\phi
)-\mathring{\psi}(x^{i_{1}}))]}{1-y^{2}}, \\
\ \ g_{3}^{W} &=&\ \ ^{W}\zeta _{3}(1+\kappa \ ^{W}\chi _{3})\ \mathring{g}%
_{3}^{W}=\ \mathring{g}_{33}(x^{i_{1}})=\ \ \ ^{W}\zeta _{3}M_{0}^{2}\exp
[\kappa \ ^{W}\chi _{3}](x+1)^{2}(1-y^{2})\exp [-2\mathring{\psi}%
(x^{i_{1}})], \\
g_{4}^{W} &=&\ \ ^{W}\zeta _{4}(1+\kappa \ ^{W}\chi _{4})\ \mathring{g}%
_{4}^{W}=\mathring{g}_{44}(x^{i_{1}})=-\frac{x-(1+\kappa \ ^{W}\chi _{4})}{%
x+1}\ \ ^{W}\zeta _{4}\exp [2\mathring{\psi}(x^{i_{1}})], \\
\ ^{W}N_{i_{1}}^{a_{2}} &=&\ ^{W}\zeta _{i_{1}}^{a_{2}}(1+\kappa \ \
^{W}\chi _{i_{1}}^{a_{2}})\ \mathring{N}_{i_{1}}^{a_{2}}(u^{\beta _{2}})\neq
0\mbox{  are defined by
a fixed system of local coordinates },
\end{eqnarray*}%
where we changed labels from (\ref{epstargsm}) as \ $\ \ ^{\shortparallel
}\zeta _{\alpha }\rightarrow \ ^{W}\zeta _{\alpha },\ ^{\shortparallel
}\zeta _{i_{1}}^{a_{2}}\rightarrow \ ^{W}\zeta _{i_{1}}^{a_{2}},\
^{\shortparallel }\chi _{4}\rightarrow \ ^{W}\chi _{4},$ etc. in order to
emphasize that we use Weyl coordinates and begin deformations of a prime
Weyl metric.

In result, we generate a family of such quasi-stationary off-diagonal
solutions%
\begin{eqnarray}
&& d\ ^{\shortparallel }\widehat{s}_{W}^{2}=M_{0}^{2}\exp [\kappa \ ^{W}\chi
_{3}(x,y,\phi )](x+1)^{2}\exp [2(\check{\gamma}(x,y,\phi )-\mathring{\psi}%
(x,y)] (\frac{\ ^{W}\zeta _{1}(x,y)}{x^{2}-1}dx^{2}+\frac{\ ^{W}\zeta
_{2}(x,y)}{1-y^{2}}dy^{2}) +  \notag \\
&& \ ^{W}\zeta _{3}M_{0}^{2}\exp [\kappa \ ^{W}\chi _{3}(x,y,\phi
)](x+1)^{2}(1-y^{2})\exp [-2\mathring{\psi}(x,y)]  \notag \\
&&\{d\phi +\ ^{W}\zeta _{i_{1}}^{3}(x,y,\phi )[1+\kappa \ ^{W}\chi
_{i_{1}}^{3}(x,y,\phi )]\ ^{\shortparallel }\mathring{N}_{i_{1}}^{3}(x,y,%
\phi )dx^{i_{1}}\}^{2}-  \label{solut1a} \\
&&\frac{x-(1+\kappa \ ^{W}\chi _{4}(x,y,\phi ))}{x+1}\ ^{W}\zeta _{4}\exp [2%
\mathring{\psi}(x,y)]\{dt+\ ^{W}\zeta _{i_{1}}^{3}(x,y,\phi ) [1+\kappa \
^{W}\chi _{i_{1}}^{4}(x,y,\phi )] \ ^{\shortparallel }\mathring{N}%
_{i_{1}}^{4}(x,y,\phi )dx^{i_{1}}\}^{2}.  \notag
\end{eqnarray}%
In this quadratic element, the parametric deformations (\ref{paramdef}) are
computed for the prime metric (\ref{prime1a}), when the effective R-sources
are computed as functionals $~_{s}^{W}\mathcal{K=}~_{s}^{\shortparallel }%
\mathcal{K}[\ _{2}\mathbf{g}^{W},\ _{W}^{\shortparallel }\widehat{\mathbf{%
\Gamma }}_{\ i_{2}k_{2}}^{m_{2}},\ _{s}^{\shortparallel }\Lambda ,\hbar
\kappa \overline{\mathcal{R}}]$ following formulas (\ref{cannonsymparamc2hv}%
) with effective sources (\ref{realrflux}) determined by $\ ^{W}\widehat{%
\mathbf{\Gamma }}_{\ i_{2}k_{2}}^{m_{2}}=\ ^{\shortparallel }\widehat{%
\mathbf{\Gamma }}_{\ i_{2}k_{2}}^{m_{2}}[\ _{2}\mathbf{g}^{W}],$ \ $\ _{2}%
\mathbf{g=}\ _{2}^{W}\mathbf{g}$ and $\ _{2}^{W}\mathbf{\mathring{g}}=(\
\mathring{g}_{\alpha }^{W})$,
\begin{eqnarray}
\ ^{W}\zeta _{3} &=&-\frac{4}{\ \ \mathring{g}_{3}^{W}}\frac{[(|\ \
^{W}\zeta _{4}\ \mathring{g}_{4}^{W}|^{1/2})^{\diamond }]^{2}}{|\int d\phi
\{(~_{2}^{W}\mathcal{K})(\ ^{W}\zeta _{4}\ \mathring{g}_{4}^{W})^{\diamond
}\}|}\mbox{ and }  \label{paramdefw} \\
\ ^{W}\chi _{3} &=&\frac{(\ ^{W}\chi _{4}|\ \ ^{W}\zeta _{4}\ \mathring{g}%
_{4}^{W}|^{1/2})^{\diamond }}{4(|\ ^{W}\zeta _{4}\ \mathring{g}%
_{4}^{W}|^{1/2})^{\diamond }}-\frac{\int d\phi \{[(~_{2}^{W}\mathcal{K})\ (\
^{W}\zeta _{4}\ ^{\shortparallel }\mathring{g}_{4}^{W})\ ^{W}\chi
_{4}]^{\diamond }\}}{\int d\phi \{(~_{2}^{W}\mathcal{K})(\ ^{W}\zeta _{4}\
\mathring{g}_{4}^{W})^{\diamond }\}},  \notag \\
\ ^{W}\zeta _{i_{1}}^{3} &=&\frac{\partial _{i_{1}}\ \int d\phi (~_{2}^{W}%
\mathcal{K})\ (\ ^{W}\zeta _{4})^{\diamond }}{(\ ^{\shortparallel }\mathring{%
N}_{i_{1}}^{3})(~_{2}^{W}\mathcal{K})(\ ^{W}\zeta _{4})^{\diamond }}%
\mbox{
and } \\
\ ^{W}\chi _{i_{1}}^{3} &=&\frac{\partial _{i_{1}}[\int d\phi (~_{2}^{W}%
\mathcal{K})(\ ^{W}\zeta _{4}\ ^{W}\chi _{4})^{\diamond }]}{\partial
_{i_{1}}\ [\int d\phi (~_{2}^{W}\mathcal{K})(\ ^{W}\zeta _{4})^{\diamond }]}-%
\frac{(\ ^{W}\zeta _{4}\ ^{W}\chi _{4})^{\diamond }}{(\ ^{W}\zeta
_{4})^{\diamond }},  \notag
\end{eqnarray}%
\begin{eqnarray*}
\ ^{W}\zeta _{k_{1}}^{4} &=&\ (\ ^{\shortparallel }\mathring{N}%
_{k_{1}}^{4})^{-1}[\ _{1}^{\shortparallel }n_{k_{1}}+16\
_{2}^{\shortparallel }n_{k_{1}}[\int d\phi \{\frac{\left( [(\ ^{W}\zeta
_{4}\ \mathring{g}_{4}^{W})^{-1/4}]^{\diamond }\right) ^{2}}{|\int d\phi
(~_{2}^{W}\mathcal{K})(\ ^{W}\zeta _{4}\ \mathring{g}_{4}^{W})^{\diamond }|}]%
\mbox{ and } \\
\ ^{W}\chi _{k_{1}}^{4} &=&\ -\frac{16\ _{2}^{\shortparallel }n_{k_{1}}\int
d\phi \frac{\left( \lbrack (\ ^{W}\zeta _{4}\ \mathring{g}%
_{4}^{W})^{-1/4}]^{\diamond }\right) ^{2}}{|\int dy^{3}(~_{2}^{W}\mathcal{K}%
)[(\ \ ^{W}\zeta _{4}\ \mathring{g}_{4}^{W})]^{\diamond }|}(\frac{[(\ \
^{W}\zeta _{4}\ \mathring{g}_{4}^{W})^{-1/4}\ ^{W}\chi _{4})]^{\diamond }}{%
2[(\ \ ^{W}\zeta _{4}\ \ \mathring{g}_{4}^{W})^{-1/4}]^{\diamond }}+\frac{%
\int dy^{3}[(~_{2}^{W}\mathcal{K})(\ ^{W}\zeta _{4}\ ^{W}\chi _{4}\
\mathring{g}_{4}^{W})]^{\diamond }}{\int dy^{3}(~_{2}^{W}\mathcal{K})(\
^{W}\zeta _{4}\ \ \mathring{g}_{4}^{W})^{\diamond }})}{\
_{1}^{\shortparallel }n_{k_{1}}+16\ _{2}^{\shortparallel }n_{k_{1}}[\int
d\phi \frac{\left( \lbrack (\ \ ^{W}\zeta _{4}\ \ \mathring{g}%
_{4}^{W})^{-1/4}]^{\diamond }\right) ^{2}}{|\int dy^{3}(~_{2}^{W}\mathcal{K}%
)[(\ ^{W}\zeta _{4}\ \ \mathring{g}_{4}^{W})]^{\diamond }|}].}.
\end{eqnarray*}

Putting together above coefficients (\ref{paramdefw}), we express the
quadratic linear element (\ref{solut1a}) for nonassociative R-flux distorted
Schwarzschild BH in the form:
\begin{eqnarray*}
&&d\ ^{\shortparallel }\widehat{s}_{W}^{2}=g_{\alpha _{2}\beta
_{2}}^{W}(x,y,\phi )du^{\alpha _{2}}du^{\beta _{2}}=e^{\psi
_{0}(x,y)}[1+\kappa \ ^{\psi }\ ^{\shortparallel }\chi (x,y,\phi
)][(dx)^{2}+(dy)^{2}] \\
&&-\{\frac{4[(|\ ^{W}\zeta _{4}\ \mathring{g}_{4}^{W}|^{1/2})^{\diamond
}]^{2}}{\ ^{\shortparallel }\mathring{g}_{3}^{W}|\int d\phi \{(~_{2}^{W}%
\mathcal{K})(\ ^{W}\zeta _{4}\ \mathring{g}_{4}^{W})^{\diamond }\}|}-\kappa
\lbrack \frac{(\ ^{W}\chi _{4}|\ ^{W}\zeta _{4}\ \mathring{g}%
_{4}^{W}|^{1/2})^{\diamond }}{4(|\ ^{W}\zeta _{4}\ \mathring{g}%
_{4}^{W}|^{1/2})^{\diamond }}-\frac{\int d\phi \{(~_{2}^{W}\mathcal{K}%
)[(~^{W}\zeta _{4}\ \mathring{g}_{4}^{W})\ ^{W}\chi _{4}]^{\diamond }\}}{%
\int d\phi \{(~_{2}^{W}\mathcal{K})(\ ^{W}\zeta _{4}\ \mathring{g}%
_{4}^{W})^{\diamond }\}}]\}\mathring{g}_{3}^{W} \\
&&+\{d\phi +[\frac{\partial _{i_{1}}\ \int d\phi (~_{2}^{W}\mathcal{K})\ (\
^{W}\zeta _{4})^{\diamond }}{(\ ^{\shortparallel }\mathring{N}%
_{i_{1}}^{3})(~_{2}^{W}\mathcal{K})(\ ^{W}\zeta _{4})^{\diamond }}+\kappa (%
\frac{\partial _{i_{1}}[\int d\phi (~_{2}^{W}\mathcal{K})(\ ^{W}\zeta _{4}\
^{W}\chi _{4})^{\diamond }]}{\partial _{i_{1}}\ [\int d\phi (~_{2}^{W}%
\mathcal{K})(\ ^{W}\zeta _{4})^{\diamond }]}-\frac{(\ ^{W}\zeta _{4}\
^{W}\chi _{4})^{\diamond }}{(\ ^{W}\zeta _{4})^{\diamond }})](\
^{\shortparallel }\mathring{N}_{i_{1}}^{3})dx^{i_{1}}\}^{2}
\end{eqnarray*}%
\begin{eqnarray}
&&+\ ^{W}\zeta _{4}(1+\kappa \ ^{W}\chi _{4})\ \mathring{g}_{4}^{W}\{dt+[(\
^{\shortparallel }\mathring{N}_{k_{1}}^{4})^{-1}[\ _{1}^{\shortparallel
}n_{k_{1}}+16\ _{2}^{\shortparallel }n_{k_{1}}[\int d\phi \{\frac{\left( [(\
^{W}\zeta _{4}\ \mathring{g}_{4}^{W})^{-1/4}]^{\diamond }\right) ^{2}}{|\int
dy^{3}[(~_{2}^{W}\mathcal{K})(\ ^{W}\zeta _{4}\ \ \mathring{g}%
_{4}^{W})]^{\diamond }|}]  \label{solut1b} \\
&&-\kappa \frac{16\ _{2}^{\shortparallel }n_{k_{1}}\int d\phi \frac{\left(
\lbrack (\ ^{W}\ \zeta _{4}\ \ \mathring{g}_{4}^{W})^{-1/4}]^{\diamond
}\right) ^{2}}{|\int dy^{3}[(~_{2}^{W}\mathcal{K})(\ ^{W}\zeta _{4}\
\mathring{g}_{4}^{w})]^{\diamond }|}(\frac{[(\ ^{w}\zeta _{4}\ \ \mathring{g}%
_{4}^{w})^{-1/4}\ ^{W}\chi _{4})]^{\diamond }}{2[(\ ^{W}\zeta _{4}\
\mathring{g}_{4}^{W})^{-1/4}]^{\diamond }}+\frac{\int dy^{3}[(~_{2}^{W}%
\mathcal{K})(\ ^{W}\zeta _{4}\ ^{W}\chi _{4}\ \ \mathring{g}%
_{4}^{W})]^{\diamond }}{\int dy^{3}[(~_{2}^{W}\mathcal{K})(\ ^{W}\zeta _{4}\
\ \mathring{g}_{4}^{W})]^{\diamond }})}{\ _{1}^{\shortparallel
}n_{k_{1}}+16\ _{2}^{\shortparallel }n_{k_{1}}[\int d\phi \frac{\left(
\lbrack (\ ^{W}\zeta _{4}\ \ \mathring{g}_{4}^{W})^{-1/4}]^{\diamond
}\right) ^{2}}{|\int dy^{3}[(~_{2}^{W}\mathcal{K})(\ ^{W}\zeta _{4}\
\mathring{g}_{4}^{W})]^{\diamond }|}]}](\ ^{\shortparallel }\mathring{N}%
_{k_{1}}^{4})dx_{1}^{k}\}.  \notag
\end{eqnarray}

We can prescribe solutions (\ref{solut1b}) with ellipsoidal configurations
for generating functions of type
\begin{equation}
\ ^{W}\chi _{4}=\ ^{e}\chi _{4}(x,y,\phi )=2\underline{\chi }(x,y)\sin
(\omega _{0}\phi +\phi _{0}),  \label{ellipsconf}
\end{equation}%
for a smooth function $\underline{\chi }(x,y)$ (in particular, $\underline{%
\chi }$ can be a constant) and constants $\omega _{0}$ and $\phi _{0}.$ Such
d-metrics have an ellipsoidal horizon with eccentricity $\kappa $\ stated by
the equation  $x=1+\kappa \ ^{W}\chi _{4}(x,y,\phi )$ 
of zero horizon when the coefficients before 
$\ ^{W}\zeta _{4}\ \exp [2\mathring{\psi}(x,y)]\{dt+...dx^{i}\}^{2}$ in (\ref{solut1a}). The integration and generating functions and generating source for such
d-metrics are defined as in (\ref{offdiagcosmcsh}) but for the case when the
term $\ ^{W}\zeta _{4}(1+\kappa \ ^{e}\chi _{4})$ is considered as a
generating function as in (\ref{qeltorshv2}) (see also respective nonlinear
symmetries involving (\ref{expconda}) for quasi-stationary solutions (\ref%
{qellc})). \ We can restrict the class generic off-diagonal solutions of
type (\ref{solut1b}) (equivalently (\ref{solut1a})), in particular, with
ellipsoidal horizons (\ref{ellipsconf}) in order to extract to
LC-configurations with zero torsion, when the R-flux contributions of
effective sources $\ _{2}^{W}\mathcal{K}$ are encoded correspondingly in the
N-connection coefficients of type $\ _{2}^{\shortparallel }\check{A}.$

Finally, we note that black ellipsoid, BE, solutions were studied in details in a series of our former works, for instance, see  \cite{vacaru03} and \cite{vacaru16b}. Such BE distorted configurations can be prescribed to obey well-defined stability conditions
as in \cite{vacaru03be2}. The stability and instability of some small parametric deformations depend on the type generating and integration functions are used for constructing respective exact/ parametric solutions.

\section{Thin accretion disks around nonassociative black ellipsoids}

\label{sec4}A series of important astrophysical phenomena (for active
galactic nuclei and/or X-ray binaries) are related to accretion of matter
onto BHs. This topic has been discussed intensively in modern literature,
see reviews of results in \cite{abram13,faraji20,geroch82,breton97}. In
addition to numerical simulations, the approach to finding analytical
solutions to accretion disk models and for BH - disk systems in GR and
various MGTs is essential for study and understanding properties of such
gravitational and astrophysical models. In this section, we concentrate on
accretion onto compact objects for a particular class of BH and BE like
solutions in nonassociative gravity theories with star R-flux distortions.
Such nonholonomic configurations are defined in the presence of external
effective matter sources defined for axially symmetric prescribed
constraints (\ref{cannonsymparamc2hv}) on effective R-flux sources (\ref%
{realrflux}). This is analogous to external axially symmetric distributions
of some effective mass outside the horizon. To study such models we use
classes of solutions (\ref{solut1a}) and (\ref{solut1b}) of the
nonassociative nonholonomic and parametric deformed vacuum Einstein
equations (\ref{eq1}). For such solutions, we can consider generating
functions for black ellipsoid configurations (\ref{ellipsconf}) and impose
zero torsion conditions (\ref{zerot}) via generating data (\ref{expconda})
resulting in metrics of type (\ref{qellc}). All constructions simplify to
the Schwarzschild solution for holonomic (diagonalizable, with trivial
N-connection structure) configurations and if the R-flux source is zero. A
quasi-stationary nonassociative star R-flux spacetime in the vicinity of an
(ellipsoidal, or other type smooth distorted) horizon is described by an
off-diagonal vacuum metric with the cost of relaxing the assumption of
asymptotically flatness for respective nonholonomic modifications of the
Einstein equations.

\subsection{Approximations and equations for locally anisotropic thin disk
models}

The structure of thin accretion disks and respective equations can be
defined in a simple form using prolate coordinates (\ref{wcoord}) which can
easily transformed into distorted Schwarzschild coordinates easily. We can
follow all assumptions and formulas from section III of \cite{faraji20} but
using locally anisotropic Weyl coefficients (\ref{deflegpoly}) as
nonholonomic deformations of Legendre polynomical formulas (\ref{legpoliy}),
when
\begin{equation*}
\lbrack M_{0},\mathring{\psi}(x,y),\mathring{\gamma}(x,y)]\rightarrow
\lbrack \check{M}(x,y,\phi ),\mathring{\psi}(x,y),\check{\gamma}(x,y,\phi )]
\end{equation*}%
determined $\kappa $--parametrically by generating functions $\
^{\shortparallel }\chi (x^{i_{1}})$ and $^{\shortparallel}\chi
_{3}(x^{i_{1}},\phi ).$ We can elaborate on standard thin disk models and
consider small nonholonomic and/or off-diagonal $\kappa $-deformations. This
results in geometrically thin, optically thick, and cold accretion disks
which be described as certain ellipsoid like quasi-stationary
configurations. Effectively, such locally anisotropic accretion effects can
modelled as standard ones in GR but (in well defined cases)
self-consistently embedded in a nonassociative $\kappa $-distorted vacuum,
when all formulas are written with respect to N-adapted bases.

Let us analyze three fundamental equations encoding nonassociative
contributions and governing the radial structure of thin disk models. We
adopt the coordinates with $c=1,$ $G=1$ and $M_{0}=1$. The first equation
(in N-adapted bases) for the particle number conservation are R-distorted%
\begin{equation}
\nabla _{\alpha _{2}}(\rho \mathbf{u}^{\alpha _{2}})=0\ \Longrightarrow \
\widehat{\mathbf{D}}_{\alpha _{2}}(\rho \mathbf{u}^{\alpha _{2}})=\ \widehat{%
\mathbf{Z}}_{\alpha _{2}}(\rho \mathbf{u}^{\alpha _{2}}),  \label{cfeq1}
\end{equation}%
where the distortion d-tensor $\widehat{\mathbf{Z}}_{\alpha _{2}}$ defines
real canonical s-deformations of the LC connection by respective
nonassociative R-deformations as in formulas (\ref{candistrnas}). In this
formula, $\mathbf{u}^{\alpha _{2}}$ is the velocity d-vector of a fluid and $%
\rho $ is its mass density. Such a (nonholonomic) conservation law means
that we expect that the mass accretion rate is constant in certain N-adapted
frames.

We can introduce the second fundamental equation (for the radial momentum)
as a component of the relativistic Navier-Stokes equations and their
canonical N-adapted deformation,%
\begin{equation}
\widehat{h}_{\alpha _{2}\beta _{2}}\nabla _{\gamma _{2}}(T^{\beta _{2}\gamma
_{2}})=0\ \Longrightarrow \ \widehat{\mathbf{h}}_{\alpha _{2}\beta _{2}}%
\widehat{\mathbf{D}}_{\gamma _{2}}(\mathbf{T}^{\beta _{2}\gamma _{2}})=%
\widehat{\mathbf{h}}_{\alpha _{2}\beta _{2}}\widehat{\mathbf{Z}}_{\gamma
_{2}}(\mathbf{T}^{\beta _{2}\gamma _{2}})  \label{cfeq2}
\end{equation}%
where $\widehat{\mathbf{h}}^{\alpha _{2}\beta _{2}}:=\mathbf{u}^{\alpha _{2}}%
\mathbf{u}^{\beta _{2}}+\widehat{\mathbf{g}}^{\alpha _{2}\beta _{2}}$ is the
projection d-tensor defining the spacial d-metric which is normal to $%
\mathbf{u}_{\alpha _{2}}.$ Such quasi-stationary values can be defined if $%
\widehat{\mathbf{g}}^{\alpha _{2}\beta _{2}}$ is determined, for instance,
by a vacuum solution (\ref{solut1a}) encoding nonassociative star
deformations. The stress-energy d-tensor $\mathbf{T}^{\beta _{2}\gamma _{2}}$
is for the accreting fluid type matter which is different from the effective
nonassociative R-flux source (\ref{realrflux}).

The third energy conservation equation is
\begin{equation}
\mathbf{u}_{\beta _{2}}\nabla _{\gamma _{2}}(T^{\beta _{2}\gamma _{2}})=0\
\Longrightarrow \ \mathbf{u}_{\beta _{2}}\widehat{\mathbf{D}}_{\gamma _{2}}(%
\mathbf{T}^{\beta _{2}\gamma _{2}})=\mathbf{u}_{\beta _{2}}\widehat{\mathbf{Z%
}}_{\gamma _{2}}(\mathbf{T}^{\beta _{2}\gamma _{2}}),  \label{cfeq3}
\end{equation}%
where the stress-energy d-tensor is parameterized in N-adapted form as%
\begin{equation*}
\mathbf{T}^{\beta _{2}\gamma _{2}}=h\mathbf{u}^{\beta _{2}}\mathbf{u}%
^{\gamma _{2}}-P\widehat{\mathbf{g}}^{\beta _{2}\gamma _{2}}+\mathbf{q}%
^{\beta _{2}}\mathbf{u}^{\gamma _{2}}+\mathbf{u}^{\beta _{2}}\mathbf{q}%
^{\gamma _{2}}+\mathbf{S}^{\beta _{2}\gamma _{2}}.
\end{equation*}%
In this formula, $h$ is the enthalpy density (defined as the sum of internal
energy per unit proper volume and the pressure over the rest mass density); $%
P$ is the pressure; the d-vector $\mathbf{q}^{\beta _{2}}$ describes the
transverse energy flux; and the viscous stress energy tensor $\mathbf{S}%
^{\beta _{2}\gamma _{2}}=-2\widetilde{\lambda }\mathbf{\sigma }^{\beta
_{2}\gamma _{2}}$ is taken in a relativistic form without no bulk viscosity,
where $\widetilde{\lambda }$ is the dynamical viscosity and $\mathbf{\sigma }%
^{\beta _{2}\gamma _{2}}$ is the shear d-tensor. In the thin disk
approximation and with respect to N-adapted frames using prolate coordinates
with $x^{1}=r$ and $x^{3}=\phi $, one approximates
\begin{equation*}
\mathbf{\sigma }_{13}=\frac{1}{2}[(\widehat{\mathbf{D}}_{\gamma _{2}}\mathbf{%
u}_{1})\widehat{h}_{\ 3}^{\beta _{2}}+(\widehat{\mathbf{D}}_{\gamma _{2}}%
\mathbf{u}_{1})\widehat{h}_{\ 3}^{\beta _{2}}]-\frac{1}{3}\widehat{h}_{\ 13}(%
\widehat{\mathbf{D}}_{\gamma _{2}}\mathbf{u}^{\gamma _{2}}).
\end{equation*}%
In coordinate frames and for LC--configurations, we obtain the formula (24)
from \cite{faraji20}. In next subsection, we summarize certain important
formulas and results from sections III.B and IV of that work using
Convention 2 extended in the form (\ref{conv2s}), which allows us to
transform (non) associative geometric constructions from coordinate bases to
N-adapted ones, and inversely.

\subsection{Thin accretion disk around nonassociative distorted
quasi-stationary BEs}

The system of fundamental equations (\ref{cfeq1}), (\ref{cfeq2}), and (\ref%
{cfeq3}) and respective N-adapted energy transport law, the equation of
state and opacity, allow us to derive in dyadic variables the system of
nonlinear algebraic equations for the thin disk model \cite%
{novikov73,compere17}. Let us introduce such important values compute per
unit mass of geodesic circular motion in equatorial planes (when $\theta=\pi
/2,$ or $y=0$ in Weyl coordinates (\ref{wcoord})):%
\begin{equation*}
\begin{array}{cccc}
\mathbf{E}=-\mathbf{u}_{4} & \Longrightarrow & \mathring{E}=-u_{4}=-u_{t}, & %
\mbox{ energy  }; \\
\mathbf{L}=\mathbf{u}_{3} & \Longrightarrow & \mathring{L}=u_{3}=u_{\phi },
& \mbox{ angular  momentum }; \\
\mathbf{\Omega =u}^{3}/\mathbf{u}^{4} & \Longrightarrow & \mathring{\Omega}%
=u^{3}/u^{4}, & \mbox{ angular velocity }.%
\end{array}%
\end{equation*}%
Above boldface values are defined and computed in N-adapted prolated frames
for any solution (\ref{solut1a}) and/or (\ref{solut1b}) encoding
nonassociative R-flux effects but circle values are in prolated coordinate
frames for prime data $(\ _{2}\mathbf{\mathring{g}}^{W},\mathring{\nabla})$ (%
\ref{prime1a}).

We omit long N-adapted calculations which are similar to the coordinate ones
in \cite{faraji20} (see there all assumptions on thermodynamic models for
thin disk accretion) and provide such results for most important physical
quantities appearing in the equations of the thin disk model:
\begin{equation}
\begin{array}{ccc}
\mathbf{E}=\left( \frac{(x-1)^{2}[\mathring{q}x^{2}+\mathring{q}x+1]}{%
(x+1)e^{\mathring{q}(x^{2}+3)}[2x^{3}\mathring{q}+x(1-2\mathring{q})-2]}%
\right) ^{1/2} & =\mathring{E}; &  \\
\mathbf{L}=\check{M}\left( \frac{-(x+1)e^{\mathring{q}(x^{2}+3)}[-\mathring{q%
}x^{3}+\mathring{q}x+1]}{2x^{3}\mathring{q}+x(1-2\mathring{q})-2}\right)
^{1/2} & \Longrightarrow \mathring{L}, & \mbox{ for }\check{M}(x,0,\phi
)\rightarrow M_{0}; \\
\mathbf{\Omega }=\check{M}^{-1}\left( \frac{{}}{(x+1)^{3}e^{2\mathring{q}%
(x^{2}+3)}[\mathring{q}x^{2}+\mathring{q}x+1]}\right) ^{1/2} &
\Longrightarrow \mathring{\Omega}, & \mbox{ for }\check{M}(x,0,\phi
)\rightarrow M_{0},%
\end{array}
\label{obs1}
\end{equation}
where the dipole moment $a_{2}:=\mathring{q}$ \ from (\ref{legpoliy}) is
used, and the nonassociative polarized mass $\check{M}(x,0,\phi )$ $\ $is
determined by formula (\ref{deflegpoly}). All such formulas are valid
locally, in some \ neighborhood of the horizon $y=1$ and have physical
meaning for real values which impose some constraints on the range of
coordinate $x$ (considering a prescribed value $\mathring{q}$).

Analyzing formulas (\ref{obs1}) we conclude that possible nonassociative
real R-flux distortions do not change nearly the horizon the disk energy of
a locally anisotropic BH (or BE for polarizations of type (\ref{ellipsconf}%
)) but may result in string constant $\kappa $-polarizations on $\phi $ of
the angular momentum and angular velocity, $\mathbf{L}$ and $\mathbf{\Omega}$%
, via $\check{M}(x,0,\phi ).$ So, in principle, nonassociative R-flux
modifications can be observed in certain thin disk accretion processes by
additional rotation on $\phi $ effects which can be of ellipsoid type
polarization. For a fixed value $\phi _{0}$ and vanishing $\mathring{q},$ we
have monotonically decreasing \ function $\mathbf{\Omega }(\phi _{0},r)$ of
the radius. Such a BH, or BE, is surrounded by a mass distribution
"embedded" into a nonassociative deformed gravitational vacuum, when after
some distance the behaviour of $\mathbf{\Omega }(\phi ,r)$ manifests the
existence both of an external matter and effective R-flux source. There is a
extremum when the influence of the surrounding matter becomes strong when
the local solution is no longer valid. Here we note that for $\mathring{q}<0$
such an extremum can appear within a valid range of radial coordinates with
real $\mathbf{E,L,}$ and $\mathbf{\Omega .}$ For $\mathring{q}>0,$ such an
extremum is usually outside the valid range even the additional dependence
on $x$ and $\phi $ in $\check{M}$ may open some new possibilities comparing
to the case $\check{M}=M_{0}.$

Here we note that the inner edge of the standard thin disk model in \cite%
{faraji20} is assumed to be at the Innermost Stable Circular Orbit (ISCO; it
is also called the marginally stable orbit). Let us analyze the location of
the ISCO in the nonassociative distorted Schwarzschild spacetime. The
reflection symmetry states such a conditions for existence of geodesics in
the equatorial plane: $a_{2l-1}=0,$ for $l>0,$ but this does not give any
new in the study of quadrupoles for prime configurations (\ref{legpoliy}).
Similarly, the nonassociative gravitational polarization for
quasi-stationary solutions does not result in contributions to the effective
potential of R-flux modified Schwarzschild BH because%
\begin{eqnarray*}
\ ^{Eff}V &=&\frac{x-1}{x+1}e^{2\mathring{\psi}(x,0)}\left( 1+\frac{\mathbf{L%
}^{2}}{\mathbf{M}^{2}}\frac{e^{2\mathring{\psi}(x,0)}}{(x+1)^{2}}\right) \\
&\simeq &\frac{x-1}{x+1}e^{2\mathring{\psi}(x,0)}\left( 1-\frac{e^{\mathring{%
q}(x^{2}+3)}[-\mathring{q}x^{3}+\mathring{q}x+1]}{2x^{3}\mathring{q}+x(1-2%
\mathring{q})-2}\frac{e^{2\mathring{\psi}(x,0)}}{(x+1)}\right) ,\mbox{ for }(%
\ref{deflegpoly})\mbox{ and }(\ref{obs1}),
\end{eqnarray*}%
which is equivalent to formula (49) in \cite{faraji20}. The main results of
the thin disk models for distorted Schwarzschild BH are described and
plotted in section V with figures 1-8 of that work. Comparing with the usual
Schwarzschild spacetimes in GR, we should mention that formulas for $\mathbf{%
E,L,}$ $\mathbf{\Omega }$ and $\ ^{Eff}V$ characterizing nonassociative BHs
and BEs are only valid locally, i.e. in the vicinity of the horizon. So, we
have only considered the inner part of the respective disks in such (non)
associative and nonholonomically deformed spacetimes. The choice of the
quadrupole (additionally to ellipsoidal R-flux distortions) imposes
respective limits o which points we can extend the thin disk solutions.
Finally, we conclude that the solutions for static BHs in GR extend to
certain quasi-stationary ones when observable thin disk effects are $\kappa $%
-polarized on $\phi $ with nontrivial contributions to the angular momentum
and angular velocity, $\mathbf{L}$ and $\mathbf{\Omega ,}$ via nonlinear
gravitational locally anisotropic polarizations of mass $\check{M}(x,0,\phi
).$

\section{Summary and Conclusions}
\label{sec5}
\subsection{Main results}
 In this work, we constructed and discussed new classes of exact
and parametric solutions for four dimensional, 4-d, black holes, BHs, and
black ellipsoids, BEs, with distortions encoding nonassociative star
deformations and R-flux effective sources from string theory. Such
nonassociative vacuum solutions are described by generic off--diagonal
symmetric metrics when certain nonsymmetric metric components $\ _{\star
}^{\shortparallel }\mathfrak{a}_{\mu _{s}\nu _{s}}$ (\ref{aux40aa}) contain
nontrivial contributions for higher shells in corresponding 8-d
nonassociative phase spaces. Our geometric method of constructing solutions
in modified gravity theories and GR (see \cite{partner02} and references
therein, on AFCDM and applications) was extended to a level bearing direct
relevance to observable nonassociative contributions using for relativistic
thin disk models around such compact objects. We computed the most important
quantities (energy, angular momentum, and angular velocity) for the most
important physical quantities characterizing the thin disk model. The
results of this paper prove that the main differences of such formulas in nonassociative
gravity and GR are consequences of different types of nonholonomic structures
resulting in angular anisotropies on $\phi $ (in prolate coordinates);
generic off-diagonal terms; and, for respective symmetries of generating
functions) ellipsoidal type deformations, of BH horizons and thin accretion
disks.

In the face-on case, we found that using nonholonomic frames, analytic
approximations with effective cosmological constants for 8-d phase spaces,
via nonlinear symmetries and parametric decompositions on string constant of
nonassociative geometric objects we obtain effective real sources encoding
contributions of star R-flux deformations. Such 8-d and 4-d nonassociative
vacuum Einstein equations were originally proposed in \cite%
{blumenhagen16,aschieri17}. We have disentangled the roles of phase and
spacetime nonassociative vacuum gravitational dynamics using nonholonomic
dyadic decompositions and restricting the class of effective sources
encoding star deformations. Such 4-d gravitational models can be studied
independently up to a level when we have to compute nonsymmetric metrics
coefficients, involve generalized phase space nonlinear symmetries and
analyze explicitly certain higher dimension contributions. The BH \& BE solutions constructed in this work define certain nonassociative generalizations of some classes of solutions in noncommutative and string gravity  \cite{vacaru09a}.

\subsection{Concluding remarks and perspectives}
Our work does not attempt to perform a complete study of BH solutions in
nonassociative gravity with R-fluxes. Instead, we use simple toy models
which provide intuition for possible geometric effects of nonassociative
distortions for the 4-d Schwarzschild solution and imprints on
related thin disks accretion effects. The techniques on generating exact and
parametric nonassociative quasi-stationary phase space solutions elaborated
in \cite{partner02} can be applied to classes of of solutions in 8-d and
10-d phase spaces found in \cite{bubuianu19,vacaru16b} and construct new
classes of nonassociative phase space and string BH and BE solutions,
generalizing the Tangherlini and higher dimension Kerr metrics.

Here we note that only for some very special effective ellipsoid / spheroid
horizons, the BHs and BEs can be characterized by corresponding
Bekenstein-Hawking thermodynamic models. For more general classes of
quasi-stationary and locally anisotropic solutions, we have to elaborate on
nonassociative generalizations of relativistic geometric flow theory and
Grigory Perelman's entropic functionals and statistical/ information
thermodynamics \cite{bubuianu19a}. In particular, we can treat the 4-d
sector of nonassociative stationary vacuum gravitational solutions as
relativistic Lorentz-Ricci solitons. This will allow to compute
corresponding thermodynamical variables determined by R-flux distortions and
encoding star product deformations and off-diagonal (non) symmetric effects.
To study nonassociative (non) symmetric metric contributions, geometric and
information flows (see related associative and commutative results in \cite%
{bubuianu17}) is a project for our future research.

\vskip6pt

\textbf{Acknowledgments:} This work develops for nonassociative geometry and gravity some research programs on geometry and physics supported during 2006-2015 by senior fellowships at the Perimeter Institute and Fields Institute (Ontario, Canada), CERN (Geneva, Switzerland) and Max Planck Institut f\"{u}r Physik / Werner Heisenberg Institut, M\"{u}nchen (Germany). SV is grateful to professors V. G. Kupriyanov, D. L\"{u}st, N. Mavromatos, J. Moffat, D. Singleton and P. Stavrinos for respective hosting of short/
long terms visits, seminars, and/or discussing important ideas and preliminary results.


\begin{thebibliography}{99}
\bibitem{shafer95} R. D. Schafer, An Introduction to Nonassociative Algebras (Dover Publications, New York, 1995)

\bibitem{baez02} J. C. Baez, The octonions, Bull. Ammer. Math. Soc. 39 (2002) 145-202
[Erratum: 42 (2005) 2013] arXiv: math-ra/0105155

\bibitem{blumenhagen16} R. Blumenhagen and M. Fuchs, Towards a theory of
nonassociative gravity, JHEP 1601 (2016) 039; arXiv: 1604.03253

\bibitem{aschieri17} P. Aschieri, M. Dimitrijevi\'{c} \'{C}iri\'{c}, R. J.
Szabo, Nonassociative differential geometry and gravity with non-geometric
fluxes, JHEP 02 (2018) 036; arXiv: 1710.11467

\bibitem{szabo19} R. J. Szabo, An introduction to nonassociative physics,
Published in: PoS CORFU2018 (2019) 100; arXiv: 1903.05673

\bibitem{jordan32} P. Jordan, \"{U}ber Eine Klasse Nichassociativer Hyperkomplexer Algebre, Nachr. Ges. Wiss. G\"{o}ttingen (1932) 569-575

\bibitem{jordan34} P. Jordan, J. von Neumann and E. Wigner, On algebraic generalization of the quantum mechanical formalism, Ann. Math. 35 (1934) 29-64

\bibitem{kurdgelaidze} D. F. Kurdgelaidze, The foundation of nonassociative
classical field theory, Acta Phys. Hung. 57 (1985) 79

\bibitem{okubo} S. Okubo, Introduction to Octonion and other Non-associative
Algebras in Physics (Cambridge Univ. Press, 1995)

\bibitem{castro1} C. Castro Perelman, The noncommutative and nonassociative
geometry of octonionic spacetime, modified dispersion relations and grand
unification, J. Math. Phys, 48 (2007) 073517

\bibitem{castro2} C. Castro Perelman, Octonionic ternary gauge field
theories revisited, IJGMMP, (2014) 1450013

\bibitem{mylonas12} D. Mylonas, P. Schupp, and R. J. Szabo, Membrane
sigma-models and quantization of non-geometric flux backgrounds, JHEP 09
(2012); arXiv: 1207.0926

\bibitem{mylonas13} D. Mylonas, P. Schupp, and R. J. Szabo, Non-geometric
fluxes, quasi-Hopf twist deformations and nonassociative quantum mechanics,
J. Math. Phys. 55 (2014) 122301; arXiv: 1312.1621

\bibitem{kupriyanov15} V. G. Kupriyanov and D. V. Vassilevich,
Nonassociative Weyl star products, JHEP 1509 (2015) 102; arXiv: 1506.02329

\bibitem{kupriyanov18} V. G. Kupriyanov, Non-associative star products and
quantization of non-geometric backgrounds in string and M-theory, PoS
CRFU2017 (2018) 200; arXiv: 1804.10161

\bibitem{gunaydin} M. G\"{u}naydin, D. L\"{u}st and E. Malek,
Non-associativity in non-geometric string and M-theory background, the
algebra of octonions, and missing momentum models, JHEP 1611 (2016) 027,
arXiv: 1607.06474

\bibitem{bouwknegt04} P. Bouwknegt, K. Hannabuss, and V. Mathai,
Nonassociative tori and applications to T-duality, Commun. Math. Phys. 264
(2006) 41-69; arXiv: hep-th/o412092

\bibitem{alvarez06} L. Alvarez-Gaume, F. Meyer, and M. A. Vazquez-Mozo,
Comments on noncommutative gravity, Nucl. Phys. B753 (2006) 92-127; arXiv:
hep-th/0605113

\bibitem{luest10} D. L\"{u}st, T-duality and closed string non-commutative
(doubled) geometry, JHEP 12 (2010) 084; arXiv: 1010.1361

\bibitem{blumenhagen10} R. Blumenhagen and E. Plauschinn, Nonassociative
gravity in string theory? J. Phys. A44 (2011) 015401; arXiv: 1010.1263

\bibitem{condeescu13} C. Condeescu, I. Florakis, C.\ Kounnas, and D. L\"{u}%
st, Gauged supergravities and non-geometric Q/R-fluxes from asymmetric
orbifold CFT's, JHEP 10 (2012) 057; arXiv: 1307.0999

\bibitem{blumenhagen13} R. Blumenhagen, M. Fuchs, F. Ha\ss ler, D. L\"{u}st,
and R.\ Sun, Non-associative deformations of geometry in duble field theory,
JHEP 04 (2014) 141; arXiv: 1312.0719

\bibitem{kupriyanov19} V. G. Kupriyanov, $L_{\infty }$-Bootstrap approach to
non-commutative gauge theories, Fortsch. Phys. 67 (2019) 1910010; arXiv:
1903.02867

\bibitem{kupriyanov19a} V. G. Kupriyanov, Non-commutative deformations of
Chern-Simons theory, Eur. Phys. J. C80 (2020) 42; arXiv: 1905.08753

\bibitem{partner01} S. Vacaru, E. V. Veliev, and L. Bubuianu, Nonassociative
nonholonomic geometry of phase spaces with star R-flux deformations and
(non) symmetric metrics, Fortschr. Physik 69 (2021) 2100029

\bibitem{partner02} E. V. Veliev, L. Bubuianu, and S. I. Vacaru, Decoupling
and integrability of nonassociative vacuum phase space gravitational
equations with star and R-flux parametric deformations, Fortschr. Physik 69
(2021) 2100030

\bibitem{drinf} V. G. Drinfeld, Quasi-Hopf algebras, Alg. Anal. 1 N6 (1989)
114-148


\bibitem{vacaru96b} S. Vacaru, Superstrings in higher order extensions of
Finsler superspaces, Nucl. Phys. B, 434 (1997) 590-656; arXiv: hep-th/9611034

\bibitem{vacaru03} S. Vacaru, Exact solutions with noncommutative symmetries
in Einstein and gauge gravity, J. Math. Phys. 46 (2005) 042503; arXiv:
gr-qc/0307103


\bibitem{vacaru09a} S. Vacaru, Finsler black holes induced by noncommutative
anholonomic distributions in Einstein gravity, Class. Quant. Grav. 27 (2010)
105003 (19pp); arXiv: 0907.4278 [math-ph]




\bibitem{bubuianu18a} L. Bubuianu and S. Vacaru, Axiomatic formulations of
modified gravity theories with nonlinear dispersion relations and
Finsler-Lagrange-Hamilton geometry, Eur. Phys. J. C 78 (2018) 969


\bibitem{bubuianu19} L. Bubuianu and S. Vacaru, Black holes with MDRs and
Bekenstein-Hawking and Perelman entropies for Finsler-Lagrange-Hamilton
spaces, Annals of Physics, NY, 404 (2019) 10-38; arXiv: 1812.02590

\bibitem{misner} C. W. Misner, K. S. Thorn and J. A. Wheeler, Gravitation
(Freeman, 1973)

\bibitem{hawking73} S. W. Hawking and C.F. R. Ellis, The Large Scale
Structure of Spacetime (Cambridge University Press, 1973)

\bibitem{wald82} R. W. Wald, General Relativity (Universtiy of Chicago
Press, Chicago, IL, 1984)

\bibitem{kramer03} D. Kramer, H. Stephani, E. Herdlt, and M. A. H.
MacCallum, Exact Solutions of Einstein's Field Equations, 2d edition
(Cambridge University Press, 2003)

\bibitem{bubuianu17} L. Bubuianu and S. Vacaru, Deforming black hole and
cosmological solutions by quasiperiodic and/or pattern forming structures in
modified and Einstein gravity, Eur. Phys. J. C 78 (2018) 393; arXiv:
1706.02584

\bibitem{bubuianu19a} I. Bubuianu, S. I. Vacaru, and E. V. Veliev, Entropy
functionals and thermodynamics of relativistic geometric flows, stationary
quasi-periodic Ricci solitons, and gravity, Ann. Phys. NY 423 (2020) 168333;
arXiv: 1903.04920




\bibitem{vacaru03be2} S. Vacaru, Perturbations and stability of black
ellipsoids, Int. J. Mod. Phys. D 12 (2003) 461-478; arXiv: gr-qc/0206016



\bibitem{vacaru16b} S. Vacaru and K. Irwin, Off-diagonal deformations of
Kerr metrics and black ellipsoids in heterotic supergravity, Eur. Phys. J. C
77 (2017) 17; arXiv: 1608.01980

\bibitem{chandr02} S. Chandrasekhar, The Mathematical Theory of Black Holes.
(Oxford Univ. Press, Oxford, 2002)

\bibitem{abram13} M. A. Abramowicz and P. Chris Fragile, Foundations of
black hole accretion disk theory, Living Reviews in Relativity 16(1) (2013) 1

\bibitem{faraji20} Shokoufe Faraji and E. Hackmann, Thin accretion disk
around the distorted Schwarzschild black hole, Phys. Rev. D 101 (2020)
023002, arXivl: 2010.02786v2

\bibitem{geroch82} R. Geroch and J. B. Hartle, Distorted black holes, J.
Math. Phys. 23 (1982) 680-692

\bibitem{breton97} N. Bret\'{o}n, T. E. Denisova, and V. S. Manko, A Kerr
black hole in the external gravitational field, Phys. Lett. A 230 (1997) 7-11

\bibitem{quevedo90} H. Quevedo, Multipole moments in general relativity
static and stationary vacuum solutions, Fortschr. Physik 38 (1990) 733-840

\bibitem{novikov73} I. D. Novikov and K. S. Thorne, Astrophysics of black
holes. In: C. Dewitt and B. S. Dewitt, editors, Black Holes (Les Astres
Occlus, 1973), pag. 343-450

\bibitem{compere17} G. Comp\`{e}re and R. Oliveri, Self-similar accretion in
thin discs arround near-extremal black holes, MNRAS (468) 2017 4351-4361
\end{thebibliography}
\end{document}